\documentclass{article}       
\usepackage{epriv}                
\usepackage[dvips]{epsfig}

\newcommand{\al}{\alpha}
\newcommand{\de}{\delta}
\newcommand{\G}{\Gamma}
\newcommand{\g}{\gamma}
\newcommand{\si}{\sigma}
\newcommand{\e}{\epsilon}
\newcommand{\ta}{\theta}
\newcommand{\equ}[2]{\begin{equation} \label{#1} #2 \end{equation} }

\begin{document}


\title{Inclusive Single- and Dijet Rates in Next-to-Leading 
	  Order QCD for $\g^*p$ and $\g^*\g$ Collisions}

\author{B.~P\"otter}

\institute{II. Institut f\"ur Theoretische
  Physik,\thanks{Supported by Bundesministerium f\"ur Forschung und 
  Technologie, Bonn, Germany, under Contract 05~7~HH~92P~(0), and by
  EEC Program {\it Human Capital and Mobility} through Network {\it
    Physics at High Energy Colliders} under Contract CHRX--CT93--0357
  (DG12 COMA).} Universit\"at Hamburg, 
  Luruper Chaussee 149, D-22761 Hamburg, Germany; 
  e-mail: poetter@mppmu.mpg.de}


\maketitle

\begin{abstract}
We present one- and two-jet inclusive cross sections for 
$\g^*\g$ scattering and virtual photoproduction in $ep$ collisions. The 
hard cross sections are calculated in next-to-leading order QCD. Soft and
collinear singularities are extracted using the phase-space-slicing
method. The initial state singularity of the virtual photon depends 
logarithmically its' virtuality. This logarithm is large and has to be
absorbed into the parton distribution function of the virtual
photon. We define for this purpose an $\overline{\mbox{MS}}$
factorization scheme similar to the real photon case. We numerically
study the dependence of the inclusive cross sections on the transverse
energies and rapidities of the outgoing jets and on the photon
virtuality. The ratio of the resolved to the direct cross section in
$ep$ collisions is compared to ZEUS data.
\end{abstract}

\section{Introduction}

The topic of this work is the structure of the virtual photon as it
can be determined in jet production in high energy collisions. In
particular, we will study electron proton scattering as is explored at
HERA and the scattering of virtual on real photons as is possible
at $e^+e^-$ colliders.

In the parton model \cite{1a} a hadron is thought to consist of
point-like particles that can be identified mainly with the valence
quarks. The valence quarks are surrounded by a sea of virtual quarks
and are bound by gluons. These particles obey the laws of Quantum
Chromodynamics (QCD) which is a non-abelian SU(3) gauge theory
\cite{1b}. Hadrons are usually probed in high energy scattering
experiments. These experiments involve contributions from a wide range
of scales. An important property of QCD is its asymptotic freedom
\cite{1c}, which states that the coupling between quarks and gluons
vanishes for asymptotically small distances. Factorization theorems allow
a separation of the short and the long distance contributions of the
high energy scattering (for reviews on this subject see \cite{1d,
1e}). This permits the application of perturbative QCD to calculate the
hard part of the cross section. The contributions from long distances
are parametrized by the parton distribution functions (PDF's).

At the HERA collider at DESY, the scattering of leptons on protons
produces jets with large transverse energies $E_T$. The ZEUS and H1
collaborations have observed an important fraction of events at small
virtualities $P^2\simeq 0$ of the exchanged photon \cite{3, 4}. 
The lepton is only weakly deflected in these so-called photoproduction
events, so that it escapes unobserved in the beam direction. The
momentum spread and the slight off-shellness of the photons that are 
radiated by the lepton, is described by the Weizs\"acker-Williams 
formula \cite{39}, where the photons are assumed to be real
($P^2=0$). The exchange of the other electroweak gauge boson, 
$Z^0$, is largely suppressed for photoproduction and can be 
neglected. The transverse energy $E_T$ serves as the large scale in
photoproduction, which allows a perturbative calculation of the hard
part of the scattering. In leading order (LO) two different processes
can be identified in the hard cross section. In the {\em direct}
interaction, the photon couples as a point-like particle to the
partons from the proton, leading to the Compton scattering and the
photon-gluon fusion subprocesses. In the {\em resolved} interaction,
the photon acts as a source of partons, which can interact with the
partons from the proton. The resolved photon is described by the photon
PDF. For quasi-real photons with virtuality $P^2\simeq 0$ the parton
content is constrained reasonably well by data from deep-inelastic
$\g^*\g$ scattering \cite{2a, 34a}. Both LO processes produce two
outgoing jets with large $E_T$. Studies of photoproduction events with
two jets  in the final state at HERA have shown that both, the direct
and the resolved processes are present for photons with very small
virtuality $P^2\simeq 0$ \cite{3, 4}. Comparisons between theoretical
predictions for dijet photoproduction rates and the data from
\cite{3, 4} have been done in \cite{5, 6, 6f}.

For the comparison between the data on jet cross sections and the
theoretical predictions in \cite{5, 6, 6f}, the hard part of the
scattering has been calculated in next-to-leading order (NLO) QCD. In
NLO, one encounters inital and final state singularities, due to
collinear and soft radiation of partons in the initial or final
state. There are two reasons for performing NLO calculations, which
are far more cumbersome than the LO ones. First, one wishes to reduce
the unphysical scale dependences. Second, only in NLO can one sensibly
implement a jet algorithm, which is needed for a comparison between
theory and experiment. However, the above discussed distinction
between direct and resolved photoproduction becomes ambiguous in
NLO. When two-jet events are observed in an experiment, a disposition
of energy near the beam pipe of the detector in the forward region of
the photon  can be attributed either to the photon remnant of a
resolved photon or to a collinear final state particle from the direct
interaction. The collinear particle in the NLO direct cross section
produces a large contribution that has to be subtracted and combined
with the LO resolved term. This introduces a dependence of the photon
PDF on the factorization scale $M_\g$. The factorization scale
determines the part of the NLO direct contribution, which has to be
absorbed into the resolved contribution. The $M_\g$ dependences of
the remaining NLO direct and the LO resolved contribution coming
from the photon PDF cancel to a large extent. This cancellation has
been demonstrated and analyzed numerically in \cite{6f, 6e, 6b} for
real photoproduction (see \cite{6a} for related work).

Information complementary to the $ep$ collision experiments from HERA
can be obtained from $e^+e^-$ colliders. Assuming the two leptons
to emit quasi-real photons that are both described by the 
Weizs\"acker-Williams approximation, one effectively has $\g\g$
scattering. Both photons can be point-like or act as a
source of partons. Three cases can be distinguished, according to the
different contributions to the cross section \cite{13b}. The
interaction of a direct with a resolved photon is denoted as the 
single resolved (SR) contribution. Interactions, where both photons
are resolved are called double-resolved (DR) contributions. These two
cases are also encountered in $ep$ scattering, where the one resolved
photon has to be substituted by the proton. In addition to these
possibilities, also both photons can interact directly in $\g\g$
scattering, which gives the direct (D) contribution. The region of high 
center of mass energies is of special interest for obtaining information 
beyond the low $E_T$ region that is determined by soft physics. This
has been measured at LEP \cite{14} and TRISTAN \cite{15}. Comparison of
the data in \cite{15} with theoretical predictions using similar
methods as those employed in \cite{6e} can be found in \cite{6f,16,17} 
for real photons.

Recently, data has been presented by the ZEUS \cite{7} and
the H1 \cite{8} collaborations for electron-proton collisions
involving photons with small, but not-vanishing $P^2$ that allow a
test of the virtual photon structure. So far, there has only been one
measurement of the virtual photon structure function from the PLUTO
collaboration at the PETRA $e^+e^-$ collider \cite{9}. In \cite{10} we
made a comparison of theoretical NLO predictions for $\g^*p$ inclusive
jet production with data from \cite{7}, by extending the methods used
in \cite{5, 6}. This extension will be described in detail in this
work. Some theoretical studies of inclusive $\g^*p$ cross sections in
LO have been presented in \cite{11, 12, 13}. We also
include the case of $\g^*\g$ scattering that will become important at
LEP2 \cite{19}, which is an extension of the work from \cite{16, 17}. 

Since the partonic subprocesses of $ep$ and $\g^*\g$ scattering are
very similar, we will take over the notation 'SR' and 'DR' from the
$\g\g$ case to $ep$ scattering to simplify the discussion. In $ep$
scattering, the SR component denotes the contribution, where the
virtual photon is directly interacting with the partons from the
proton, whereas in the DR component the resolved photon interacts with
the partons from the proton. 

The extension from real to virtual photoproduction is done by taking the
Weizs\"acker-Williams formula to describe the momentum spread of the
virtual photon, but keeping $P^2$ fixed, not integrating over the
region of small $P^2$ and not assuming $P^2=0$. This is described by
the unintegrated Weizs\"acker-Williams formula. In the hard process,
the matrix elements for finite $P^2$ have to be taken. The matrix
elements and the initial and final state singularities for the SR
contribution with $P^2\ne 0$ have been calculated in \cite{23, 23b} in
connection with deep-inelastic $ep$ scattering (DIS) at
HERA, where $P^2$ is large. Since we consider $P^2$ to be finite, the
photon initial state singularities encountered in real photoproduction
do not occur. Instead, when integrating over the phase-space of the final
state particles, a logarithm of the type $\ln (P^2/E_T^2)$ occurs. In
DIS this logarithm is small, since $P^2$ is of the order of $E_T^2$, and
thus has not to be considered separately. In virtual photoproduction
though, $P^2$ is small and the logarithm gives a large
contribution. This large term has to be subtracted from the SR hard
cross section, where the virtual photon is direct, and combined with
the resolved virtual photon from the DR contribution. This introduces
a dependence of the virtual photon PDF on the factorization scale
$M_\g$, just as in the case of real photons. The cancellation of 
the $M_\g$ scale dependences of the NLO direct and resolved
contributions must hold also for virtual photoproduction with $P^2\ne
0$. This has been worked out in \cite{10} and will be studied
numerically in this work. 

The D contribution is needed for the direct interaction of one real and one
virtual photon in $\g^*\g$ reactions. The initial and final state
corrections for the D contribution are calculated here for the first
time. For the real photon, the singularities are handled as discussed
in \cite{6f,17}. For the virtual photon, the procedure is equivalent
to the one described for the SR contribution in $ep$ scattering.

The theoretical calculation of the resolved cross sections requires
the parton distribution functions of real and virtual photons.
Several parametrizations of the parton contents of the real photon are
available in the literature by now \cite{20, 36, 35} and seem to be 
consistent with dijet production data in $ep$ scattering \cite{6f, 6e}. 
For virtual photons theoretical models have been constructed that 
describe the evolution
with the scale $Q^2$ of the parton distributions and the input
distributions at scale $Q_0$ with changing $P^2$ \cite{11, 21,
22}. However, these virtual photon PDF's are not available in a form
that parametrizes the $Q^2$ evolution in NLO. Only the LO
parametrizations for the virtual photon are given in \cite{21, 22}.

The outline of this work is as follows. In section 2 we will discuss
the general structure of factorization and renormalization for NLO
corrections.  Section 3 contains a calculation of the LO and NLO
partonic cross sections for the D, SR, and DR contributions with a
virtual photon. We recall the virtual corrections to the Born cross
sections and present the inital and final state singularities, using 
the phase-space-slicing method. The DR contribution has been calculated in
\cite{6f} and will be considered only briefly. The parton
distribution function of the virtual photon is discussed in section
4. Section 5 contains numerical results for inclusive
single- and dijet production in $ep$ scattering. Several numerical
tests will be presented and the available data is compared with our
theoretical predictions. Section 6 gives theoretical results for
$\g^*\g$ collisions with the kinematics of LEP2. Finally we present a
summary and an outlook in section 7. The appendix contains the
analytic results for virtual, initial, and final state corrections.

\section{General Structure of the Hadronic Cross Sections}

The key to using perturbative QCD is the idea of factorization. It
states that a cross section is a convolution of different factors that
each depend only on physics relevant at one momentum scale. In this
section we explain how factorization shows up in the 
hadronic cross sections we use in this work. Especially we will
discuss the divergences appearing in a NLO calculation of the
perturbative hard cross section and explain how these divergences are
factorized and absorbed by a redefinition of the PDF's involved in
each process. The general procedure described in this section will be
applied to the specific partonic NLO cross sections that are
calculated in section 3.

\subsection{Factorization of Hard and Soft Regions}

\begin{figure}[hhh]
\unitlength1mm
\begin{picture}(121,32)
\put(15,-33){\psfig{file=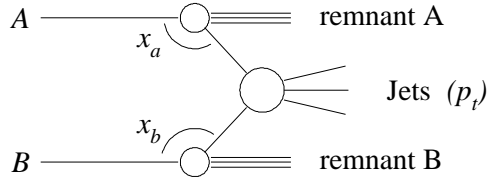,width=10.4cm}}
\end{picture}
\caption{\label{fack}Factorization of hard and soft
    processes in the hadronic cross section.}
\end{figure}

The physical cross sections considered in this work have a general
structure, where the long-distance and short-distance parts are
separated. The hadronic cross section $d\si^H$ of a process is given
by a convolution of the hard cross section $d\si_{ab}$ 
and the PDF's $f_{a/A}(x_a)$ and $f_{b/B}(x_b)$:
\equ{welle}{ d\si^H = \sum_{a,b} \int dx_adx_b f_{a/A}(x_a) d\si_{ab}
  f_{b/B}(x_b) \quad . }
In general, the PDF $f_{i/A}(x)$ of a hadron $A$ gives the probability 
of finding a parton $i$ (quark or gluon) with momentum fraction $x$
within the hadron. It cannot be calculated  perturbatively and
has to be fixed by measurement. The partons from the PDF are thought
to interact in a hard process involving a large scale that allows to
make use of the asymptotic freedom of QCD \cite{1c}, i.e.\ the
vanishing of the coupling between the partons for asymptotically small
distances. For a large scale $\mu$, the QCD coupling constant $g(\mu )$ 
behaves as $g(\mu )\sim 1/\ln (\mu /\Lambda_{QCD})$ and a
perturbative expansion of the hard cross section in the strong
coupling constant can be applied. The hard process in (\ref{welle}) is
described by the {\em partonic} cross section
\equ{well}{ d\si_{ab} = \frac{1}{2x_ax_bs} |{\cal M}_{ab}|^2
  d\mbox{PS}^{(n)} \quad , } 
where $2x_ax_bs$ is the flux factor, $|{\cal M}_{ab}|^2$ are the
partonic matrix elements and $d\mbox{PS}^{(n)}$ represents the phase
space of the $n$ final state particles of the subprocess. In the final
state we are interested in jets for which suitable jet definitions
have to be defined in order to go from the partonic level to
observable quantities. We will come back to this in section 5. The
general structure of the cross sections discussed in this work is
indicated in Fig.\ \ref{fack}.
For the case of $ep$ scattering, $A$ will be a lepton, that radiates a
virtual photon and $B$ will be a proton. The remnant $A$ will stem
from the resolved virtual photon, whereas the remnant $B$ comes from
the proton. In the case of $\g^*\g$ scattering, $A$ is a virtual and $B$
a real photon. They can both have a hadronic structure, leading to the
remnants $A$ and $B$. The different subprocesses, encountered in these
two cases are explained in section 3.

\subsection{NLO Corrections}

The matrix elements of the partonic cross section (\ref{well})
can be calculated by summing up all Feynman diagrams
to a given order. We will be interested in processes with two initial
and at least two final state particles. The calculation of the LO
contributions is straightforward. In NLO several difficulties are
encountered. We have to distinguish the virtual corrections to 
the $2\to 2$ partonic processes, which contain self energy and vertex
corrections, and the real corrections, which stem from the radiation
of an additional real parton from the $2\to 2$ processes, leading
to $2\to 3$ processes. Both these contributions contain characteristic
divergences. 

\begin{figure}[hhh]
\unitlength1mm
\begin{picture}(121,25)
\put(25,-42){\psfig{file=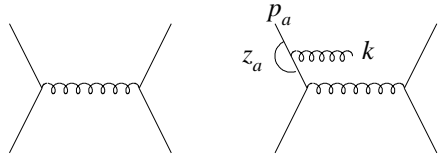,width=7.0cm} }
\end{picture}
\caption{\label{rb}Partonic cross section:
      Born graph and real gluon emission.}
\end{figure}

As an example for a $2\to 3$ process the Feynman diagrams for the amplitudes
of the Born subprocess $q\bar{q}\to q\bar{q}$ and the
${\cal O}(\al_s)$ correction containing a real gluon emission are drawn in
Fig.\ \ref{rb}.
We consider a parton with momentum $p_a$ emitted from a resolved
photon. Taking $k$ to be the momentum of the
outgoing gluon, the Feynman diagram contains a propagator of the form
\equ{propagator}{ G \sim \frac{1}{(p_a-k)^2} \equiv \frac{1}{M^2} \quad . }
In the limit of massless quarks, the propagator diverges in certain
regions of phase space. The denominator 
\equ{}{ M^2 = (p_a-k)^2 = -2p_ak = -2|p_a||k|(1-\cos\ta ) \quad , }
where $\ta$ is the angle between the gluon and the parton, vanishes
if $\cos\ta =1$ ({\em collinear} divergence) and if
$|k|=0$ ({\em soft} divergence). Both, the collinear and soft
divergences are infra-red (IR) divergences that can be
regularized in the dimensional regularization scheme \cite{24, 25}. In
this scheme $n=4-2\e$ dimensions are chosen for the phase space
integration, so that the singularities appear in poles like $1/\e$ and
$1/\e^2$. After the poles have been removed the limit $\e\to 0$ is
taken and the  four-dimensional result is obtained. The singularity shown
in Fig.\ \ref{rb} is due to the radiation of a gluon in the initial
state and is thus called an {\em initial state singularity}. In addition to
these, {\em final state singularities} occur, when a parton is
collinear or soft in the final state.

The virtual corrections involve loop integrals over internal momenta,
that lead to ultra-violet (UV) and IR divergences. These divergences
can be extracted, as the real corrections, in the dimensional
regularization scheme as poles in $1/\e$ and $1/\e^2$. The UV
divergences are removed completely by adding a counter term to the QCD
Lagrangian, where the singularities are absorbed by a renormalization
of the quark charge, quark field, and gluon field. 

\begin{figure}[hhh]
\unitlength1mm
\begin{picture}(121,25)
\put(26,-28){\psfig{file=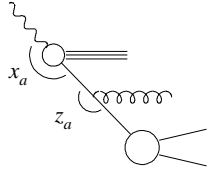,width=8cm} }
\end{picture}
\caption{\label{facto}The factorization theorem for
      the singular part of the partonic cross section.}
\end{figure}

When the virtual and the real contributions are added, the IR
divergences cancel partly, leaving only initial state singularities. 
It can be proven that in the remaining hard cross section
the short distance finite parts and the long distance singular
parts factorize \cite{1d, 1e}. One defines a bare partonic cross
section $d\si$, that is calculable in perturbative QCD, a renormalized
finite partonic cross section $d\bar{\si}$ and transition functions
$\G_{i\leftarrow j}$ so that \cite{sterm}
\equ{fact-theo}{   d\si_{ij}(s) = \sum_{k,l} \int dz_adz_b 
   \G_{i\leftarrow k}(z_a,\mu_A) d\bar{\si}_{kl}(z_az_bs,\mu_A,\mu_B) 
   \G_{j\leftarrow l}(z_b,\mu_B)  \quad . }
The variables \mbox{$z_a,z_b\in [0,1]$} give the momentum 
fraction of $p_a,p_b$ in the propagator after a parton is
radiated as can be seen in Fig.\ \ref{rb}. The singular terms are
absorbed into the transition functions in such a way that the
renormalized cross section is finite. This absorption depends on the 
scales $\mu_A$ and $\mu_B$, which are the factorization scales for
the hadrons $A$ and $B$, respectively. To obtain a hadronic cross
section, which is free of divergent parts, one needs renormalized
PDF's $\bar{f}$, which are defined as 
\begin{eqnarray}
 \bar{f}_{iA}(\eta_a,\mu_A) &\equiv & \int\limits_0^1\int\limits_0^1
 dxdz f_{jA}(x) \G_{i\leftarrow j} (z_a,\mu_A) \de (\eta_a-x_az_a)
 \nonumber \\ 
 &=& \int\limits_{\eta_a}^1 \frac{dz_a}{z_a} \  f_{jA}\left(
 \frac{\eta_a}{z_a} \right) \G_{i\leftarrow j}(z_a,\mu_A) 
	\label{part-dens} \quad .
\end{eqnarray}
As one sees, the factorization scale dependence of the transition
functions leads to a scale dependence of the renormalized PDF's. 
The factorization of the hard and soft parts in the partonic cross
section is pictured in Fig.\ \ref{facto} for the case of a resolved
photon with the subprocess depicted in Fig.\ \ref{rb}.
The factorization scales $\mu_A$ and $\mu_B$ define what is to be
understood as the hard and the soft part of the cross
section. Referring to equation (\ref{propagator}), $M^2$ gives the
off-shellness of the propagator in the initial state of the partonic
cross section. Interactions with $M^2< \mu_A$ are described with  help
of the PDF of hadron $A$, whereas for $M^2\ge \mu_F$ one can apply
perturbative QCD and calculate the partonic cross section.

Using the definitions of the renormalized quantities, the IR safe
hadronic cross section reads
\equ{master2}{ d\si^H(s) =  \sum_{k,l} \int d\eta_ad\eta_b
  \bar{f}_{kA}(\eta_a,\mu_A) d\bar{\si}_{kl}(\eta_a\eta_bs,\mu_A,\mu_B)
  \bar{f}_{lB}(\eta_b,\mu_B) \quad , }
where the variables $\eta_a,\eta_b\in [0,1]$ are defined as $\eta_a=
x_az_a$ and  $\eta_b = x_bz_b$. The connection between the IR safe and
the bare hadronic cross sections can be easily seen, by inserting the
definition of the renormalized PDF's (\ref{part-dens}) into 
(\ref{master2}) and performing the integrations over the delta
functions, making use of the definition (\ref{fact-theo}). The
factorization scale
dependences of the renormalized partonic cross section and the PDF's
cancel to a large extend. The transition functions that connect the
renormalized partonic cross section and the PDF  are, however, not
unique in NLO and arbitrary finite parts can be shifted from the PDF's
to the renormalized partonic cross section. Therefore one has to
define a factorization scheme to be used for a consistent
calculation. Commonly used schemes are the DIS \cite{24a} and the
$\overline{\mbox{MS}}$ \cite{24b} schemes.

To extract the renormalized from the unrenormalized quantities, one
assumes $d\bar{\si}, d\si$ and the transition functions to have
perturbative expansions in $\al_s$ \cite{sterm}:
\begin{eqnarray}
  d\bar{\si}(s) &=& \sum_n^\infty \left(
  \frac{\al_s}{2\pi}\right)^n d\bar{\si}^{(n)}(s) \label{ex1} \\
  d\si (s) &=& \sum_n^\infty \left(
  \frac{\al_s}{2\pi}\right)^n d\si^{(n)}(s) \label{ex2} \\
  \G_{i\leftarrow k} (z) &=& \de_{ik}\de (1-z) + \sum_{n=1}^\infty \left(
   \frac{\al_s}{2\pi}\right)^n \G_{i\leftarrow k}^{(n)}(z) \label{ex3}
\end{eqnarray}
The LO contributions are understood to be the $n=0$th order
contributions. For the DR partonic cross section it is of order ${\cal
O}(\al_s^2)$, for the SR contribution is of order ${\cal
O}(\al\al_s)$ and for the D contribution it is of order ${\cal
O}(\al^2)$. Inserting the expansions (\ref{ex1})--(\ref{ex3}) into
(\ref{fact-theo}) gives up to ${\cal O}(\al_s)$
\begin{eqnarray}
  d\si^{(0)}_{ij}(s) + \frac{\al_s}{2\pi} d\si^{(1)}_{ij}(s) &=&
  d\bar{\si}^{(0)}_{ij}(s) + \frac{\al_s}{2\pi}
  \bigg[ d\bar{\si}_{ij}^{(1)}(s,\mu_A,\mu_B) \nonumber \\ &+&
   \sum_k \int dz_1\ 
   \G_{i\leftarrow k}^{(1)}(z_1,\mu_A) d\bar{\si}_{kj}^{(0)}(z_1s)
   \nonumber \\ &+& \sum_k\int dz_2\ d\bar{\si}_{ik}^{(0)}(z_2s) 
  \G_{k\leftarrow j}^{(1)}(z_2,\mu_B) \bigg]  \quad .
\end{eqnarray}
Comparing the left hand and the right hand side in LO gives
$d\bar{\si}^{(0)} = d\si^{(0)}$. The NLO correction is obtained by
comparing  the left hand and right hand side to order $\al_s$ and
rearranging the terms:
\equ{nlo}{   d\bar{\si}_{ij}^{(1)} = d\si_{ij}^{(1)} - 
  \sum_k \int dz_1 \G_{i\leftarrow k}^{(1)} d\si_{kj}^{(0)}  - 
  \sum_k \int dz_2 \G_{i\leftarrow k}^{(1)} d\si_{kj}^{(0)} \quad . }
Thus the prescription for subtracting the singular parts from the bare
cross section is simple: the singularities are removed by a
convolution of the finite Born cross section with the singular 
${\cal O}(\al_s)$ transition functions.

As we have seen, the PDF's acquire a dependence on the factorization
scales. The evolution of the PDF's with the scale are predicted in
perturbative QCD by the DGLAP evolution equations \cite{27, 28,
29}. This is the following set of integro-differential equations:
\equ{dglap}{ \frac{df_i(x,Q^2)}{d\ln Q^2} = \sum_i \int\limits_x^1
  \frac{dz}{z} P_{i\leftarrow j}(z,\al_s(Q^2))f_j\left(
  \frac{x}{z},Q^2\right) \quad . } 
Here, $Q^2$ is a general scale, and $P_{ij}(z)$ are the splitting
functions that represent a process in which a parton with momentum
fraction $x$ radiates a parton with momentum fraction $(1-z)x$ and
continues with momentum fraction $xz$. The splitting functions can be
expanded in powers of the strong coupling constant with $\al_s \equiv
g^2/4\pi$:
\equ{}{  P_{i\leftarrow j}(z,\al_s) = \frac{\al_s}{2\pi} 
  P_{i\leftarrow j}^{(0)}(z) + \left( \frac{\al_s}{2\pi}\right)^2 
  P_{i\leftarrow j}^{(1)}(z) + \ldots }
The evolution equations (\ref{dglap}) predict the PDF's at a higher
scale once they are fixed at some input scale $Q_0^2$.

\subsection{Factorization for the Photon}

In the previous section we have described, how the absorption of the
singularities works in the case, when $A$ and $B$ are hadrons or
resolved photons. For the direct photon one additional complication
has to be taken into account.

As mentioned in the introduction, in LO the photon gives rise to
direct and resolved contributions in the hadronic cross section. In
NLO, the creation of a collinear $q\bar{q}$-pair in the initial state
leads to initial state singularities in the direct contribution. The
only place for these singularities to be absorbed is the PDF of the
resolved photon. This leads to a point-like term in evolution
equations of the the photon PDF \cite{witt, bar}. A subtraction procedure
for the real photon that is consistent with the evolution of the
photon PDF has been worked out in \cite{32b}.

For the virtual photon, one actually has no real singularity, since
the virtuality $P^2$ regularizes the divergence. Integrating over the
phase space of the $q\bar{q}$-pair in the initial state leads to a
logarithmic dependence on $P^2$, namely $\ln (P^2/Q^2)$, where $Q^2$
is the hard scale of the process. This logarithm becomes large for
$P^2\ll Q^2$ and is absorbed into the PDF in much the same way as
described in \cite{32b}. This leads to an inhomogeneous term in
the PDF of the virtual photon, which differs somewhat from the
point-like term in the case of the real photon. This will be described
in more detail in section 4, where the construction of the virtual
photon PDF is explained.

\section{Partonic Cross Sections}

In this more technical section, we proceed with a computation of the
perturbatively calculable partonic cross sections. The partonic cross
sections contributing to $ep$ and $\g^*\g$ scattering are very similar
and will therefore be treated together in this section. Both
hadronic cross sections contain single resolved (SR) contributions, in
which the virtual photon couples directly to the subprocess and double
resolved (DR) contributions, in which the virtual photon is
resolved. In $ep$ scattering the virtual photon and its partonic
content interact with the partons of the proton, whereas in $\g^*\g$
scattering they interact with the parton content of the real photon. 
In addition to the SR and DR contribution, in $\g^*\g$ scattering one
encounters the direct (D) interaction of both photons. 

After an introduction to the notation of the various relevant subprocesses, 
we give the formul{\ae} for the Born and virtual contributions. Then we
explain the phase-space slicing method as a tool to separate singular
regions of phase space in the partonic cross section, so that we can
calculate the singular parts of the real final and initial state
corrections. The results from \cite{23} for the SR contributions are
recalled for completeness and consistency. The DR contributions can be
found in \cite{6f} and will be considered only briefly.

\subsection{Notation}

For the calculation of hadronic cross sections in sections 5 and 6, we
will have to calculate the matrix elements for the various partonic cross
sections in LO and in NLO. In LO the D contribution is of ${\cal
O}(\al^2)$, the SR contribution is of ${\cal O}(\al\al_s)$ and
the DR contribution is of ${\cal O}(\al_s^2)$. In NLO the D, SR
and DR contributions are of one order higher in $\al_s$. 
Since the partonic cross sections have to be convoluted with the PDF's
of the photon, which are of order $\al /\al_s$ in the high
energy limit \cite{witt}, the different contributions will turn out to
be of the same order. The matrix elements $|{\cal M}|^2$ from equation
(\ref{well}) are obtained by taking the trace of the hadron tensor
that corresponds to each subprocess. We define $H\equiv
-g_{\mu\nu}H^{\mu\nu}$. The Born contributions are labeled $H_B$, the
virtual corrections are $H_V$, and the real corrections are $H_R$.

\begin{table}[bbb]
\renewcommand{\arraystretch}{1.6}
\caption{\label{born-con}Definition of the LO matrix elements.}
\begin{center}
\begin{tabular}{|c||c|c|} \hline
 D and SR Processes &  
 \multicolumn{2}{c|}{DR Contributions} \\ \hline\hline 
 $B_1=H_B(\g^*\g \to q\bar{q})$ & 
 \makebox[3.6cm][c]{$B_4=H_B(qq'\to qq')$}  &
 \makebox[3.6cm][c]{$B_9=H_B(q\bar{q}\to gg)$ } \\ \hline 
 $B_2=H_B(\g^*q\to qg)$ & $B_5=H_B(q\bar{q}'\to q\bar{q}')$ &
 $B_{10}=H_B(qg\to qg)$ \\ \hline
 $B_3=H_B(\g^*g\to q\bar{q})$ & $B_6=H_B(qq\to qq)$ &
 $B_{11}=H_B(gg\to q\bar{q})$ \\ \hline
 & $B_7=H_B(q\bar{q}\to q'\bar{q}')$ &
 $B_{12}=H_B(gg\to gg)$ \\ \hline
 & $B_8=H_B(q\bar{q}\to q\bar{q})$ &  \\ \hline
\end{tabular}
\end{center}
\renewcommand{\arraystretch}{1}
\end{table}
\begin{table}[hhh]
\renewcommand{\arraystretch}{1.6}
\caption{\label{tab2}Definition of the NLO matrix elements for the 
  D and SR processes. }
\begin{center}
\begin{tabular}{|c||c|c|} \hline
 Virtual Corrections $2\to 2$ &  
 \multicolumn{2}{c|}{Real Corrections $2\to 3$ } \\ \hline\hline 
 $V_1=H_V(\g^*\g \to q\bar{q})$ & 
 \makebox[3.6cm][c]{$H_1=H_R(\g^*\g \to q\bar{q}g)$}  &
 \makebox[3.6cm][c]{$H_4=H_R(\g^*q\to qq\bar{q})$  } \\ \hline 
 $V_2=H_V(\g^*q\to qg)$  
 & $H_2=H_R(\g^*q \to qgg)$ & $H_5=H_R(\g^*g\to q\bar{q}g)$ \\ \hline 
 $V_3=H_V(\g^*g\to q\bar{q})$ & $H_3=H_R(\g^*q\to qq'\bar{q}')$ & \\ \hline
\end{tabular}
\end{center}
\renewcommand{\arraystretch}{1}
\end{table}
\begin{table}[hhh]
\renewcommand{\arraystretch}{1.6}
\caption{\label{tab1}Definition of the NLO matrix elements for the 
  DR contribution.}
\begin{center}
\begin{tabular}{|c||c|c|} \hline
 Virtual Corrections $2\to 2$ &  
 \multicolumn{2}{c|}{Real Corrections $2\to 3$} \\ \hline\hline 
 $V_4=H_V(qq'\to qq')$ & 
 \makebox[3.6cm][c]{$H_6=H_R(qq'\to qq'g)$}  &
 \makebox[3.6cm][c]{$H_7=H_R(q\bar{q}'\to q\bar{q}'g)$} \\ \hline 
 $V_5=H_V(q\bar{q}'\to q\bar{q}')$ & $H_8=H_R(qq\to qqg)$
 & $H_9=H_R(q\bar{q}\to q'\bar{q}'g)$  \\ \hline 
 $V_6=H_V(qq\to qq)$ & $H_{10}=H_R(q\bar{q}\to q\bar{q}g)$
 & $H_{11}=H_R(qg\to qq'\bar{q}')$ \\ \hline
 $V_7=H_V(q\bar{q}\to q'\bar{q}')$ & $H_{12}=H_R(qg\to qq\bar{q})$
 & $H_{13}=H_R(q\bar{q}\to ggg)$ \\ \hline 
 $V_8=H_V(q\bar{q}\to q\bar{q})$
 & $H_{14}=H_R(qg\to qgg)$ & $H_{15}=H_R(gg\to q\bar{q}g)$ \\ \hline 
 $V_9=H_V(q\bar{q}\to gg)$
 & $H_{16}=H_R(gg\to ggg)$ & \\ \hline 
 $V_{10}=H_V(qg\to qg)$ & & \\ \hline
 $V_{11}=H_V(gg\to q\bar{q})$ & & \\ \hline
 $V_{12}=H_V(gg\to gg)$ & & \\ \hline
\end{tabular}
\end{center}
\renewcommand{\arraystretch}{1}
\end{table}

First, we have collected the definition of the LO Born matrix elements
in Tab.\ \ref{born-con}. The matrix elements for incoming anti-quarks
are the same as those for quarks and only give a factor 2 in the
sum over all contributions in the hadronic cross section. This holds
also for the contributions in the other tables. Of special interest in
this work will be the NLO corrections to the Born matrix elements for
processes involving a virtual photon. These are the D and SR
contributions, that are collected in Tab.\ \ref{tab2}. The SR
contributions with one virtual photon have been studied by several
authors \cite{23, 23b}, the D and SR contributions for real photons
can be found in \cite{6f}, whereas  the D contributions with one
virtual photon have only been studied in this work, yet.

\begin{figure}[hhh]
\unitlength1mm
\begin{picture}(121,65)
\put(10,-20){\psfig{file=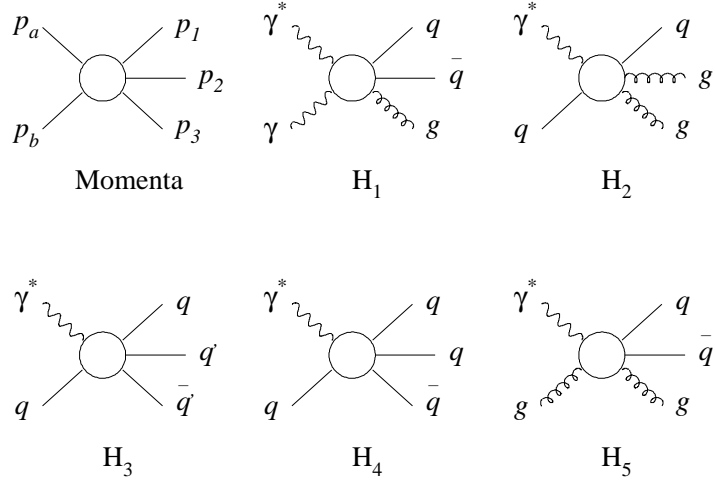,width=10cm}  }
\end{picture}
\caption{\label{procs}Notation of the different
      processes involving one virtual photon. }
\end{figure}

Because of their importance, we show in Fig.\ \ref{procs} the classes
of matrix elements as collected in Tab.\ \ref{tab2} explicitly. We also
show the definition of the momenta which will be used throughout this work
for all three, i.e. D, SR and DR, contributions. In Tab.\ \ref{tab1}
the matrix elements of the NLO processes for the DR case are
collected. These matrix elements have been calculated in \cite{ellsex}
and the integrations over the singular regions of phase space where
performed in \cite{6f}. 

\subsection{The Two-Body Processes}

For the $2\to 2$ processes we use the Mandelstam variables 
\begin{eqnarray}
 s &=& (p_a+p_b)^2 = (p_1+p_2)^2 \quad , \nonumber \\ 
 t &=& (p_a-p_1)^2 = (p_b-p_2)^2 \quad , \\
 u &=& (p_a-p_2)^2 = (p_b-p_1)^2 \quad . \nonumber 
\end{eqnarray}
Note, that in the D and SR case, $p_a=q$ with $P^2\equiv -q^2$. In the
D case $p_b$ is a real photon, in the SR and DR cases $p_b$ is a
massless parton. The partonic cross section is given
by the flux factor, the two-particle phase space and the matrix
elements:
\equ{}{ d\si (ij\to \mbox{jets})= \frac{1}{2s} H(ij\to \mbox{jets})
  d\mbox{PS}^{(2)} \quad . }
The two-particle phase space is given by
\equ{}{ d\mbox{PS}^{(2)} = \frac{1}{\G (1-\e )} \left( \frac{4\pi}{s}
   \right)^\e [z(1-z)]^{-\e} \frac{dz}{8\pi} \quad , }
where $z\equiv (p_bp_1)/(p_ap_b)$. Expressed by the
Mandelstam variables, the phase space reads
\equ{2bph1}{ d\mbox{PS}^{(2)} = \frac{1}{\G (1-\e )} \left( \frac{4\pi}{stu}
   \right)^\e (s+P^2)^{-1+2\e} \frac{dt}{8\pi} \quad , }
if the particle $p_a=q$ has mass $-P^2$. This is valid for the D and
SR case. In the DR case all partons are massless, so we substitute 
$P^2=0$ and the phase space reduces to 
\equ{2bph2}{  d\mbox{PS}^{(2)} = \frac{1}{\G (1-\e )} \left( \frac{4\pi s}{tu}
   \right)^\e \frac{dt}{8\pi s} \quad . }
The Born matrix elements for the D and SR case read, using the
notation of Tab.\ \ref{born-con}, 
\begin{eqnarray}
 B_1 &=& (16\pi^2\al^2)\ (Q_i^4 8N_C) \ T_\g (s,t,u) \quad , \\
 B_2 &=& -(16\pi^2\al\al_s)\ (Q_i^2 2C_F) \ T_\g (u,t,s) \quad , \\
 B_3 &=& (16\pi^2\al\al_s)\ (Q_i^2) \ T_\g (s,t,u) \quad , 
\end{eqnarray}
where the definition of $T_\g (s,t,u)$ can be found in the appendix, 
section 8.1. The Born matrix elements for the DR case are given by 
\begin{eqnarray}
 B_4 &=& (16\pi^2\al_s^2)\ {\textstyle \frac{1}{4N_C^2}}\ T_1(s,t,u)
 \quad , \\  
 B_5 &=& (16\pi^2\al_s^2)\ {\textstyle \frac{1}{4N_C^2}}\ T_1(u,t,s)
 \quad , \\ 
 B_6 &=& (16\pi^2\al_s^2)\ {\textstyle \frac{1}{2}
 \frac{1}{4N_C^2}}\ \left[ T_1(s,t,u)+T_1(s,u,t)+T_2(s,t,u) \right]
 \quad , \\ 
 B_7 &=& (16\pi^2\al_s^2)\ {\textstyle \frac{1}{4N_C^2}}\ T_1(u,s,t)
 \quad , \\ 
 B_8 &=& (16\pi^2\al_s^2)\ {\textstyle
 \frac{1}{4N_C^2}}\ \left[ T_1(u,t,s)+T_1(u,s,t)+T_2(u,t,s) \right]
 \quad , \\
 B_9 &=& (16\pi^2\al_s^2)\ {\textstyle \frac12
 \frac{1}{4N_C^2}}\ T_3(s,t,u) \quad , \\
 B_{10} &=& -(16\pi^2\al_s^2)\ {\textstyle\frac{1}{8N_C^2C_F}}\ 
 T_3(t,s,u) \quad , \\
 B_{11} &=& (16\pi^2\al_s^2)\ {\textstyle\frac{1}{16N_C^2C_F}}\ 
 T_3(s,t,u) \quad , \\
 B_{12} &=& (16\pi^2\al_s^2)\ {\textstyle \frac12 \frac{1}{16N_C^2C_F}}\ 
 T_4(s,t,u) \quad .
\end{eqnarray}
The factors $1/2$ in some of the expressions are symmetry factors for
two identical particles in the final state. The definitions of the
matrix elements $T_1,\ldots ,T_4$ can again be found in the appendix
8.1. 

The virtual corrections for the SR case are calculated by
multiplying the one-loop diagrams for $\g^*\g\to q\bar{q}$, 
$\g^*q\to gq$ and $\g^*g\to q\bar{q}$ with the corresponding Born
diagrams, which leads to an extra factor $\al_s$ in the matrix
elements. The virtual corrections $V_2$ and $V_3$ are well known for
quite some years from the processes $e^+e^-\to q\bar{q}g$  \cite{31,
32}. They are achieved by crossing from the known matrix elements
\cite{23}. The contribution $V_1$ for the D case can be inferred from
the SR case by considering the contribution $H_V(\g^*g\to
q\bar{q})$. Only the parts which have no gluon self-coupling are taken
into account and the color factors have to be adjusted
appropriately. To compare the singular structure of the virtual
corrections with those from the real corrections, we write down the
final result for the D and SR case in the form 
\begin{eqnarray}
 V_1 &=& 16\pi^2 \frac{\al^2\al_s}{2\pi} 
 \left(\frac{4\pi\mu^2}{s} \right)^\e \frac{\G (1-\e )}{\G (1-2\e )} 
 E_3 \quad , \\
 V_2 &=& 16\pi^2\frac{\al\al_s^2}{2\pi} 2(1-\e )
 \left(\frac{4\pi\mu^2}{s} \right)^\e \frac{\G (1-\e )}{\G (1-2\e )} 
   \left[ C_F^2 E_1 - \frac{1}{2}N_CC_F E_2 \right. \nonumber \\
 &+& \left. \frac{1}{\e} \left( \frac{1}{3}N_f-\frac{11}{6}N_C \right)
 C_FT_\g (s,t,u) \right] \quad , \\
 V_3 &=& 16\pi^2 \frac{\al \al_s^2}{2\pi} 
 \left(\frac{4\pi\mu^2}{s} \right)^\e \frac{\G (1-\e )}{\G (1-2\e )} 
 \left[ C_F E_3 - \frac{1}{2}N_C E_4 \right. \nonumber \\
 &+& \left. \frac{1}{\e} \left( \frac{1}{3}N_f-\frac{11}{6}N_C \right)
  T_\g (s,u,t) \right] \quad . 
\end{eqnarray}
Terms of order ${\cal O}(\e )$ have been neglected. 
The expressions $E_1,\ldots ,E_4$ are given in the appendix 8.2. In
the DR case, the Born processes $B_4,\ldots ,B_{12}$ have to be
multiplied by the corresponding one-loop processes. The results can be
found in \cite{6f}. They are given by $V_4, \ldots ,V_{12}$ in
\cite{18}.

\subsection{Phase-Space-Slicing Method}

As discussed in the section 2, the partonic $2\to 3$ corrections are
singular in certain regions of phase space. One possibility is to
integrate the hard cross section over the complete phase space in $n$
dimensions and to obtain analytical results for the integrated matrix
elements. We choose a somewhat different method in this work. We
separate the singular regions of phase space from the finite regions
by inserting an invariant mass cutt-off $y_c$ into the integration,
symbolically 
\equ{sepp}{ \int\limits_0^1 d\mbox{PS}^{(3)} |{\cal M}_{2\to 3}|^2 = 
  \int\limits_0^{y_c} d\mbox{PS}^{(3)} |{\cal M}_{2\to 3}|^2 +
  \int\limits_{y_c}^1 d\mbox{PS}^{(3)} |{\cal M}_{2\to 3}|^2 \quad . }
The first integral on the right hand side of this equation contains
the singular region of phase space and is integrated analytically in
$n=4-2\e$ space time dimensions. If the cut-off $y_c$ is chosen
appropriately small, this singular phase space separates into a 
$2\to 2$ phase space $d\mbox{PS}^{(2)}$ that will be kept and a
remaining part $d\mbox{PS}^{(r)}$ that is integrated out together with
the matrix elements. For small $y_c$ various approximations can be
applied to the matrix elements, so that the integration of the matrix
elements over $d\mbox{PS}^{(r)}$ is simplified considerably as will
become evident in sections 3.4--3.6. The non-singular second term on the right
hand side of (\ref{sepp}) is integrated numerically, opening the
possibility to adopt a wide range of jet definitions and experimental
cuts. This flexibility allows a detailed comparison between theory and
experiment. Of course, the dependences of the first and the second
contribution on the parameter $y_c$ should compensate, leaving a
result independent of $y_c$, since the cut-off has no physical
significance. This also provides a strong test of the results, which
will be described in section 5. The method described here is referred
to as the phase-space-slicing (PSS) method \cite{kramer}.

An important step in the application of the PSS method is the
separation of the two-body phase space from the singular part, 
$d\mbox{PS}^{(3)}\to d\mbox{PS}^{(2)}d\mbox{PS}^{(r)}$. This
separation is different for the three cases encountered in this work,
which are singularities in the final state, in the initial state for
massless particles and in the initial state for a massive virtual
photon. In the next two subsections we will provide the formul{\ae} that
serve as a basis for the calculation of the singular parts of the
partonic cross sections in sections 3.4--3.6. 

\subsubsection{Singularities in the Final and in the Massless Initial State}

\begin{figure}[bbb]
\unitlength1mm
\begin{picture}(121,35)
\put(10,-23){\psfig{file=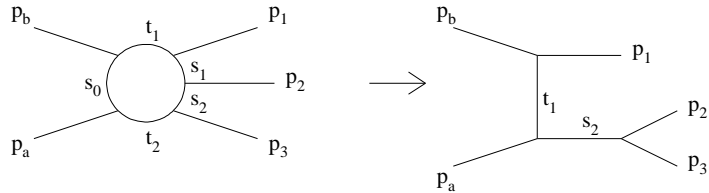,width=12.0cm}  }
\end{picture}
\caption{\label{psc-splitt}Separation of the three particle phase space.}
\end{figure}

\begin{figure}[hhh]
\unitlength1mm
\begin{picture}(121,60)
\put(22,-5){\psfig{file=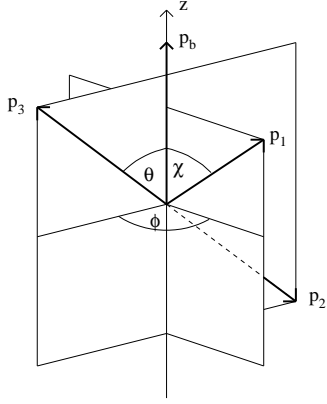,width=7.0cm}  }
\end{picture}
\caption{\label{k-system}The three particle final
      state in the c.m.\ system of the particles $p_2$ and $p_3$.}
\end{figure}

We consider the splitting for a phase space containing one massive
particle $p_a$ with mass $P^2=-p_a^2$. In general, the
$2\to 3$ phase space in $n$ dimensions is given by \cite{26a}
\equ{3pps}{
  d\mbox{PS}^{(3)} = \frac{d^{n-1}p_1}{2E_1(2\pi )^{n-1}}
 \frac{d^{n-1}p_2}{2E_2 (2\pi )^{n-1}} 
 \frac{d^{n-1}p_3}{2E_3(2\pi )^{n-1}} (2\pi )^n
 \de^n(p_b+q-p_1-p_2-p_3) \quad . }
It is useful to introduce the following irreducible set of invariants:
\begin{eqnarray}
 s_0 = 2p_ap_b -P^2, \qquad \qquad & & \qquad \qquad t_1 = -2p_bp_1,
 \nonumber \\
 s_1 = 2p_1p_2, \qquad \qquad \qquad & & \qquad \qquad t_2 = -2p_ap_3 -
 P^2, \label{5freunde} \\
 s_2 = 2p_2p_3.  \qquad \qquad \qquad & & \nonumber 
\end{eqnarray}
These five invariants are pictured in Fig.\ \ref{psc-splitt} on the
left. 
The separation pictured on the right of Fig.\ \ref{psc-splitt} is
achieved by inserting \cite{26}
\equ{eins}{  1 = \int \frac{ds_2}{2\pi} \int \frac{d^{n-1}p_{23}}{(2\pi
    )^{n-1}2E_{23}} \de^{(n)}(p_{23}-p_2-p_3) (2\pi )^n }
into (\ref{3pps}), where the definition $s_2 = p_{23}^2=(p_2+p_3)^2$
is used and $E_{23}$ is the energy of this intermediate particle. One
obtains
\begin{eqnarray} 
  d\mbox{PS}^{(3)} &=& \frac{ds_2}{2\pi } \left\{\frac{d^{n-1}p_1}{2E_1(2\pi
  )^{n-1}} \frac{d^{n-1}p_{23}}{2E_{23}(2\pi )^{n-1}}
  \de^n(p_a+p_b-p_1-p_{23}) (2\pi )^n \right\} \nonumber \\
 & \times & \left\{\frac{d^{n-1}p_2}{2E_2(2\pi )^{n-1}}
 \frac{d^{n-1}p_3}{2E_3 (2\pi )^{n-1}} \de^n(p_{23}-p_2-p_3) (2\pi )^n
  \right\}  \label{splitt-phase}  \quad . 
\end{eqnarray}
To perform the integration over the delta functions in
(\ref{splitt-phase}) it is useful to define the kinematical variables in
the c.m.\ system of the outgoing partons $p_2$ and $p_3$. The angles
of the partons $p_1$ and $p_2, p_3$ with respect to the parton $p_b$ 
are shown in Fig.\ \ref{k-system}.
To parametrize the angles, the variables $b\equiv\frac{1}{2}(1-\cos\ta )$
and $z_1\equiv (p_bp_1)/(p_ap_b)$ are used. After integrating over the delta
functions and expressing the variables by $z_1, b$ and the irreducible
invariants introduced in (\ref{5freunde}), the three particle phase space in
$n=4-2\e$ dimensions reads
\equ{ppmaster}{
 d\mbox{PS}^{(3)} = \frac{ds_2}{2\pi} \frac{dz_1}{64\pi^2}\frac{db}{N_b}
 [b(1-b)]^{-\e} \frac{d\phi}{N_\phi} \frac{\sin^{-2\e}\phi }{\G (2-2\e )} 
  \left[ \frac{16\pi^2}{s_2z_1(s_0(1-z_1)-s_2)}
  \right]^\e \quad , } 
with the normalization constants 
\equ{}{  N_b = \int\limits_0^1 db [b(1-b)]^{-\e}  = \frac{\G^2(1-\e
   )}{\G(2-2\e )} \quad \mbox{and} \quad 
 N_\phi = \int\limits_0^\pi d\phi \sin^{-2\e}\phi = \frac{4^\e\pi\G
   (1-2\e )}{\G^2(1-\e )} \quad . }
For singularities in the final state the invariants $s_2$ and $b$ will
vanish, whereas for the singularities in the initial state of the massless
particle $p_b$ the invariant $t_1$ and therefore $z_1$ will vanish.

\subsubsection{Singularities in the Initial State of a Virtual Photon}

For this case, the mass $P^2=-p_a^2$ of the photon with momentum $p_a$
serves as a regulator for the integration. Therefore, the phase space
(\ref{3pps}) can be calculated in $n=4$ dimensions. We introduce the
five invariants 
\begin{eqnarray}
 s_0 = 2p_ap_b -P^2, \qquad \qquad & & \qquad \qquad t_1 = -2p_bp_3,
 \nonumber \\
 s_1 = 2p_1p_2, \qquad \qquad \qquad & & \qquad \qquad t_2 = -2p_ap_1 -
 P^2,  \label{quatsch} \\
 s_2 = 2p_2p_3,  \qquad \qquad \qquad & & \nonumber 
\end{eqnarray}
and separate the phase space analogously to the case discussed above by
inserting (\ref{eins}) into (\ref{3pps}). Again we move to the
c.m. frame of the particles $p_2$ and $p_3$ and define the variable $b\equiv
\frac{1}{2}(1-\cos\theta )$, but now $\theta$ is defined
as the angle between $p_a$ and $p_3$. The singularities occur for
$t_2\to 0$, which we parametrize by the variable
$z_2\equiv (p_ap_3)/(p_ap_b)$. It is obvious from the definition of
$t_2$ that the variable $z_2$ vanishes only for $P^2\to 0$. After
integrating over the delta functions, the result is simply
\equ{ppmastur}{ d\mbox{PS}^{(3)} = \frac{d\phi}{\pi}
	\frac{ds_2}{2\pi}\frac{dz_2}{8\pi} \frac{db}{8\pi} \quad . }

\subsection{Final State Singularities}

The singularities in the final state appear when the invariant $s_2$
in equation (\ref{5freunde}) becomes on-mass-shell. We define the
variable $r\equiv s_2/s_0$ and consider the limit $r\to 0$. We start
by considering the D and SR cases, for which we define the two-body 
Mandelstam variables as
\begin{eqnarray}
  s &=&(p_b+q)^2      = 2p_bq-P^2, \quad , \nonumber \\
  t &=&(p_b-p_1)^2    =-2p_bp_1 \quad , \label{bornbar} \\
  u &=&(p_b-p_{23})^2 =-2p_bp_{23} \quad . \nonumber 
\end{eqnarray}
In the limit  $r\to 0$ the five invariants (\ref{5freunde}) reduce to
these variables, with $s_0\to s$ and 
$t_1\to t$. The definitions of $s, t$ and $u$ are only unique in the
limit $r\to 0$. In the limit $s_2\to 0$ the phase space (\ref{ppmaster}) 
separates according to $d\mbox{PS}^{(3)} = d\mbox{PS}^{(2)}
d\mbox{PS}^{(r)}$ with 
\equ{1ppss}{  d\mbox{PS}^{(r)} = \frac{\G (1-\e )}{\G (2-2\e )}
 \frac{d\phi}{N_\phi} \sin^{-2\e}\phi  \left( \frac{4\pi}{s} \right)^\e
 \frac{s}{16\pi^2} G_F(r) dr r^{-\e}\frac{db}{N_b} [b(1-b)]^{-\e} }
where 
\equ{}{  G_F(r) = \left[ 1- \frac{r}{(1-z)} \right]^{-\e} = 1 +
  {\cal O}(r) \quad .} 
The two-body phase space $d\mbox{PS}^{(2)}$ is given by equation
(\ref{2bph1}). The limits of integration in
$d\mbox{PS}^{(r)}$ are given by $r\in [0,-t/(s+P^2)], b\in [0,1]$ and 
$\phi\in [0,\pi ]$. The invariant $s_2$ is integrated up to 
$s_2\le y_cs_0$, which restricts the range of $r$ to 
$0\le r \le \mbox{min}[-t/(s+P^2),y_c]\equiv y_F$. 

We have now achieved the separation of the phase space and have to
integrate the matrix elements over the region $d\mbox{PS}^{(r)}$. 
Therefore the $2\to 3$ matrix elements $H_1,\ldots H_5$ are expressed
by the variables $s, t, u, r, b$ and $\phi$ in the limit $r\to 0$,
which leads to approximated matrix elements with final state 
singularities $H_{F1},\ldots ,H_{F5}$ \cite{23}. A difficulty arises 
for those squared matrix elements that contain real gluons. In that
case more than one invariant can vanish in a propagator, so that
the different classes of singularities, such as initial and final
state singularities, are not properly separated. The separation is
achieved by partially fractioning the matrix elements. 

After the matrix elements have been partially fractioned and
approximated in the limit $r\to 0$, the integration over the singular
region of phase space yields
\equ{int1}{ \int d\mbox{PS}^{(r)} H_{Fi} = 8\pi
   \left(\frac{4\pi\mu^2}{s} \right)^\e
  \frac{\G (1-\e )}{\G (1-2\e )}  (1-\e ) F_i + {\cal O}(\e ) \quad . }
The final results for $F_1,\ldots ,F_5$ are listed in appendix
8.3. They contain the IR collinear and soft singularities that cancel
against those of the virtual corrections. It is essential that the
singular terms are proportional to the LO matrix elements $T_\g$ and
that the variables $s, t$ and $u$ defined in (\ref{bornbar}) 
correspond to the two-body variables in the above discussed limit. 
The contributions $F_2,\ldots ,F_5$ have been calculated in
\cite{23}. The final state correction $F_1$ of the D contribution can be
derived from $F_5$ by keeping only the abelian part and adjusting the
color factor. One can compare the singularities for the case of direct
real photons as stated in \cite{18} with the ones given in appendix
8.3. The real photon contributions $F_1,\ldots ,F_5$ in \cite{18}
follow from the contributions $F_1,\ldots ,F_5$ in this work by taking
the limit $P^2\to 0$, so that $t+u=-s$. 

Turning to the resolved photon case, the phase space is obtained from
the formula (\ref{1ppss}) by substituting $q\to p_a$ with
$p_a^2=0$, so that $s=2p_ap_b$. The two-body variables in the resolved
case are given by 
\equ{}{ s=2p_ap_b, \qquad t =-2p_bp_1, \qquad u=-2p_bp_{23} \quad .}
Note, that the two-body phase space for the DR case is given by
equation (\ref{2bph2}) of section 3.2. 
Expressing the matrix elements for the resolved processes $H_6,\ldots
,H_{16}$ as classified in Tab. \ref{tab1} through the variables 
$s, t, u, r, b$ and $\phi$ yields the matrix elements containing final
state singularities $H_{F6},\ldots ,H_{F16}$ \cite{6f, 18}. The
integral over the singular phase space is the same as in equation
(\ref{int1}). The final state corrections for the resolved case
$F_6,\ldots ,F_{16}$ will not be stated here again since they can be
found in \cite{18}.

\subsection{Initial State Singularities for Massless Particles}

We turn to the discussion of the initial state singularities for
particles with zero mass. This includes the photon initial state
singularity for the real photon for the D case, the parton initial
state singularities in the SR case and the parton initial state
singularities in the DR case. We start with the SR case, from which
the others can be inferred. 

\subsubsection{Parton Initial State Singularities in the SR Case}

In the SR case parton initial state singularities arise for the
incoming particle $p_b$. The incoming particle $p_a$ is the virtual
photon with $p_a=q$ and $P^2=-q^2$. Using the notation of section
3.3.1, the singularities appear when the invariant $t_1$ becomes
on-mass-shell, i.e.\ for $z_1\to 0$. The invariant $s_2$ does not
vanish in the case of initial state singularities but rather defines
the partonic c.m.\ energy of the corresponding two-body process. We
define the new variable 
\equ{}{ z_b \equiv \frac{p_2p_3}{qp_b} = \frac{s_2}{s_0+P^2} \in
	[\eta_b,1] \quad , }  
that gives the fraction of the momentum $p_b$ that participates in the
subprocess after a particle has been radiated in the initial
state. The variable $\eta_b$ is given by $\eta_b=x_bz_b$. We 
define the two-body variables as
\begin{eqnarray}
  s &=& (p_2+p_3)^2  = 2p_2p_3 \quad , \nonumber \\
  t &=& (z_bp_b-p_2)^2 =-2z_bp_bp_2 \quad , \label{tbsr} \\
  u &=& (z_bp_b-p_3)^2 =-2z_bp_bp_3 \quad . \nonumber 
\end{eqnarray}
In the limit $z\to 0$ the variable $s_2$ reduces to $s$. In the same
limit the phase space (\ref{ppmaster}) separates according to
$d\mbox{PS}^{(3)} = d\mbox{PS}^{(2)} d\mbox{PS}^{(r)}$, where
\begin{eqnarray}
 d\mbox{PS}^{(r)} &=& \frac{1}{\G (1-\e )} \frac{d\phi}{N_\phi}
 \sin^{-2\e}\phi \left( \frac{4\pi}{s} \right)^\e \frac{s}{16\pi^2}  
 G_I(z_1) \nonumber \\
 &\times & dz_1 z_1^{-\e} \frac{dz_b}{z_b} \left(
 \frac{1-z_b}{z_b} - \frac{P^2}{s} \right)^{-\e} 
  \left( 1+ \frac{P^2(1-z_b)}{z_b(z_bs-(1-z_b)P^2)} \right)^{1-\e}
    \label{voz1}
\end{eqnarray}
with 
\equ{}{ G_I(z_1) = \left[ 1- z_1 \frac{s-z_bP^2}{s(1-z_b)-z_bP^2}
	\right]^{-\e} = 1 + {\cal O}(z_1) \quad . }
The two-body phase space is given by equation (\ref{2bph1}). 
The integration over $d\mbox{PS}^{(r)}$ with $z_1\in [0,-u/(s+P^2)]$,
$z_b\in [\eta_b,1]$ and $\phi\in [0,\pi ]$ is restricted to the singular
region of $z_1$ in the range 
$0\le z_1\le\mbox{min}\{-u/(s+P^2),y_c\}\equiv y_I$. 

Expressing the matrix elements for the direct photon case, listed in
Tab.\ \ref{tab2}, with the variables $s, t, u, z_1, z_b,
b$ and $\phi$ and taking the limit $z_1\to 0$, one obtains the matrix
elements $H^b_{I2}, \ldots ,H^b_{I5}$ \cite{23} that contain initial state
singularities on the parton side B. These are integrated according to 
\equ{int22}{ \int d\mbox{PS}^{(r)} H^b_{Ii} = \int\limits_{\eta_b}^1
  \frac{dz_b}{z_b} 8\pi \left(\frac{4\pi\mu^2}{s} \right)^\e
   \frac{\G (1-\e )}{\G (1-2\e )}  (1-\e ) I^b_i + {\cal O}(\e ) \quad . } 
where results for $I^b_2,\ldots ,I^b_5$ are written down in appendix
8.4. They can also be found in \cite{23}. Apart from the two-body
variables $s, t, u$ and the cut-off $y_I$, they still
depend on the integration variable $z_b$. The results for the $I_i^b$
contain IR singularities proportional to $1/\e^2$ that cancel
against the corresponding singularities in the virtual
corrections. The remaining singular parts are proportional to $1/\e$
and to the Altarelli-Parisi kernels in four dimensions. These are
removed by a redefinition of the PDF's that are the source of particle
$p_b$, which can be a hadron or a resolved photon. The redefinition,
as explained in section 2.2, is achieved by equation 
(\ref{part-dens}) for the PDF's, introducing the factorization scale
($M_b$) dependence through the transition functions 
$\G_{i\leftarrow j}^{(1)}$: 
\equ{}{ \bar{f}_{iB}(\eta_b,M_b^2) = \int\limits_{\eta_b}^1
  \frac{dz_b}{z_b} \ \left( \de_{ij}\de (1-z_b) + \frac{\al_s}{2\pi}
  \G_{i\leftarrow j}^{(1)}(z_b,M_b^2) \right) f_{jB}\left(
  \frac{\eta_b}{z_b} \right) \quad .}  
Here $f_{jB}\left( \frac{\eta_b}{z_b} \right)$ is the PDF of hadron B
in LO before absorption of the collinear singularities. The NLO
transition functions are given by
\equ{nlo-trans}{  \G_{i\leftarrow j}^{(1)}(z_b,M_b^2) =
  -\frac{1}{\e}P_{i\leftarrow j}(z_b) 
  \frac{\G (1-\e )}{\G (1-2\e )}\left( \frac{4\pi\mu^2}{M_b^2}
  \right)^\e + C_{ij}(z_b) }
with $C_{ij}=0$ in the $\overline{\mbox{MS}}$ scheme. The renormalized
partonic cross section $d\bar{\si}(\g^*i\to \mbox{jets})$ for parton
initial state singularities is calculated from the unrenormalized
cross section $d\si$ by 
\equ{}{ d\bar{\si}(\g^*i\to \mbox{jets}) = d\si (\g^*i\to
  \mbox{jets}) - \frac{\al_s}{2\pi} \sum_j \int dz_b 
  \G_{i\leftarrow j}^{(1)}(z_b,M_b^2) d\si^B(\g^*j\to \mbox{jets}) \quad . }
The $d\si^B$'s denote the Born level partonic cross sections that
can be found in section 3.2. The factor $4\pi\mu^2/M_b^2$ in
equation (\ref{nlo-trans}) is combined with the factor $4\pi\mu^2/s$ in
equation (\ref{int22}) and leads to a factorization scale dependent
term of the form 
\equ{}{ -\frac{1}{\e}P_{i\leftarrow j}(z_b) \left[ 
  \left(\frac{4\pi\mu^2}{s}\right)^\e - 
    \left(\frac{4\pi\mu^2}{M_b^2}\right)^\e \right]
 = - P_{i\leftarrow j}(z_b) \ln \left( \frac{M_b^2}{s}\right) \quad . }
In this way, the subtracted partonic cross section will depend on the
scale $M_b^2$, as does the PDF of the hadron B, $f_{iB}$.

\subsubsection{Real Photon Initial State Singularities in the D Case}

In the D case the direct real photon can split into
a $q\bar{q}$ pair that gives rise to a collinear singularity if the
partons are emitted parallel. The calculation proceeds along the
lines that have been described for the parton initial state
singularities in the previous section 3.5.1. The two-body variables are
defined as in equations (\ref{tbsr}). The phase space separation as
well as the formula (\ref{int22}) for the integration over the
singular regions remains unchanged. For the singular matrix element
$H^b_{I1}=H^b_I(\g^*\g\to q\bar{q}g)$ we have  
\equ{}{ \int d\mbox{PS}^{(r)} H^b_{I1} = \int\limits_{\eta_b}^1
  \frac{dz_b}{z_b} 8\pi \left(\frac{4\pi\mu^2}{s} \right)^\e
  \frac{\G (1-\e )}{\G (1-2\e )}  (1-\e ) I^b_1 + {\cal O}(\e ) \quad ,}
where $I^b_1$ is stated in the appendix 8.4. As remarked for the
virtual and final state corrections, the initial state correction for
the D case can be infered from the SR case with an incoming gluon
instead of the real photon by dropping the non-abelian terms and
adjusting the color factor. The singularity appearing in $I^b_1$ is
proportional to the splitting function $P_{q\leftarrow\g}(z_b)$ given
in appendix 8.1. This function appears in the evolution equation of
the PDF of the real photon as an inhomogeneous (so-called point-like)
term, as will be explained in more detail in section 4. Therefore, the
photon initial state singularities can be absorbed into the real
photon PDF, according to the procedure given in \cite{32b}. We
define the renormalized PDF $\bar{f}_{qe}$  of a quark $q$ in the
electron as
\equ{}{ \bar{f}_{qe}(\eta_b,M_b^2) = \int\limits_{\eta_b}^1 
	\frac{dz_b}{z_b} \left(\de_{q\g}\de (1- z_b) + 
   \frac{\al_s}{2\pi} \G_{q\leftarrow\g}^{(1)}(z_b,M_b^2) \right) f_{\g e} 
	\left( \frac{\eta_b}{z_b} \right) \quad . }
The NLO transition functions are given by
\equ{}{  \G_{q\leftarrow\g}^{(1)}(z_b,M_b^2) =
  -\frac{1}{\e}P_{q\leftarrow \g}(z_b) 
  \frac{\G (1-\e )}{\G (1-2\e )}\left( \frac{4\pi\mu^2}{M_b^2}
  \right)^\e + C_{q\g}(z_b) }
with $C_{q\g}=0$ in the $\overline{\mbox{MS}}$ scheme. In the
discussed order, $P_{g\leftarrow \g}(z)=0$. The partonic
cross section $d\bar{\si}(\g^*\g\to \mbox{jets})$ for the photon
initial state singularity is calculated from the unrenormalized cross
section $d\si$ by 
\equ{}{ d\bar{\si}(\g^*\g\to \mbox{jets}) = d\si (\g^*\g\to
  \mbox{jets}) - \frac{\al_s}{2\pi}  \int dz_b 
  \G_{q\leftarrow\g}^{(1)}(z_b,M_b^2) d\si^B(\g^*q\to \mbox{jets}) \quad . }
The cross section $d\si^B$ contains the LO virtual photon-parton
scattering matrix element $B_2$ given in section 3.2. The
dependence of the real photon PDF on the factorization scale enters in
the same way as discussed for the SR case, section 3.4.1.

\subsubsection{Parton Initial State Singularities in the DR Case}

The calculation of the initial state singularities for the DR case is
very similar to the calculations shown for the SR case, only now both
incoming partons are massless. The calculations for the partons $p_a$
and $p_b$ yield identical results and thus we have to consider these
singularities only once. The singularities occur in the region 
$z_1\to 0$.  The formula for the phase space 
separation is obtained from equation (\ref{voz1}) 
by substituting $q\to p_a$ with $p_a^2=P^2=0$, so that
$s=2z_bp_ap_b$. With these substitutions, the phase space (\ref{voz1})
reduces to
\equ{}{ d\mbox{PS}^{(r)} = \frac{1}{\G (1-\e )} \frac{d\phi}{N_\phi}
 \sin^{-2\e}\phi \left( \frac{4\pi}{s} \right)^\e
 \frac{s}{16\pi^2}  G_I(z_1) dz_1 z_1^{-\e} \frac{dz_b}{z_b} \left(
 \frac{1-z_b}{z_b}\right)^{-\e}  }
with 
\equ{}{ G_I(z_1)= \left[ 1- \frac{z_1}{1-z_b}\right]^{-\e} = 1 +
  {\cal O}(z_1) \quad . }
The two-body phase space in the case of initial state
singularities is given by (\ref{2bph2}).
Expressing the resolved matrix elements $H_6,\ldots ,H_{16}$ with the
variables $s, t, u, z_1,z_b,b$ and $\phi$ in the limit
$z_1\to 0$ leads to the resolved matrix elements containing initial
state singularities $H_{J_6},\ldots ,H_{J_{16}}$ \cite{6f, 18}. These
are integrated similar to equation (\ref{int22}), leading to the
results $J_6, \ldots ,J_{16}$. These are not stated here, they can be
found in \cite{6f, 18}.  The initial state singularities on the proton
side $J_6, \ldots ,J_{16}$ are given by $R_6^a,\ldots R_{16}^a$ in
\cite{18}. The cancellation of the poles from the real and virtual
corrections proceeds as in the SR case
discussed above. The remaining poles in $1/\e$ are proportional to the
Altarelli-Parisi kernels and are absorbed into the PDF's of the
hadrons A and B that emit the particles $p_a$ and $p_b$.

\subsection{Initial State Singularities for the Virtual Photon}

The initial state singularities described in the previous section were
extracted and handled in the dimensional regularization scheme,
i.e.\ in $d=4-2\e$ dimensions. In the case of the real photon this is
necessary, because the real photon is massless. The singular
terms are proportional to a simple pole in $\e$ multiplied by the
splitting function $P_{q\leftarrow \g}$. These initial state
singularities are absorbed into the PDF of the real photon.
The NLO correction for the direct virtual photon becomes
singular only in the limit $P^2\to 0$. After integrating the phase
space up to the invariant cut-off $y_c$, the logarithm $\ln (P^2/s)$
will occur that becomes large in the limit of small $P^2$. The
logarithm has to be absorbed into the PDF of the virtual photon,
instead of the $1/\e$ poles in the real photon case. 

To show the subtraction of the logarithm explicitly, we start by
defining the two-body variables for the virtual photon initial state
singularities. They are given by 
\begin{eqnarray}
  s &=& (p_2+p_3)^2 = 2p_2p_3 \quad , \nonumber \\
  t &=& (p_b-p_2)^2 =-2p_bp_2 \quad , \\
  u &=& (p_b-p_3)^2 =-2p_bp_3 \quad . \nonumber 
\end{eqnarray}
We define the variable $z_a$ as
\equ{}{ z_a \equiv \frac{p_2p_3}{qp_b} = \frac{s_2}{s_0+P^2} \in
	[\eta_a,1]  }	
with $\eta_a=x_az_a$. It gives the momentum fraction of the
three-body c.m.\ energy that participates in the two-body process. The
definition of the three-body variables is given in section 3.3.2, equation
(\ref{quatsch}). As mentioned above, the mass $P^2$ regularizes
the initial state singularities of the virtual photon. The singular
terms appear in the case $t_2\to 0$ in (\ref{quatsch}), which
corresponds to $z_2\to 0$ for $P^2\to 0$. For $z_2\to 0$ the phase
space (\ref{ppmastur}) separates according to $d\mbox{PS}^{(3)} =
d\mbox{PS}^{(2)} d\mbox{PS}^{(r)}$, with
\equ{}{ d\mbox{PS}^{(r)} = \frac{s}{16\pi^2} \frac{d\phi}{\pi}	
	\frac{dz_a}{z_a}dz_2 \qquad \mbox{and} \qquad d\mbox{PS}^{(2)} = 
	\frac{1}{8\pi}\ \frac{dt}{s+P^2} \quad . } 
The limits of integration are given by $z_2\in [0,-t/(s+P^2)]$, 
$z_a\in [\eta_a,1]$ and $\phi \in [0,\pi ]$. Since the integration of
$z_2$ is restricted to the singular region we define the integration
range for $z_2$ by $0\le z_2 \le \mbox{min}\{-t/(s+P^2),y_c\}\equiv y_J$. 

Expressing the matrix elements for the direct photon case, listed in
Tab.\ \ref{tab2}, with the help of the variables $s, t, u, z_2, z_b, b$
and $\phi$ and taking the limit $z_2\to 0$, we obtain matrix 
elements $H^a_{I1}, \ldots ,H^a_{I5}$ that contain the initial state
singularities on the virtual photon side. These are integrated according to 
\equ{int2}{ \int d\mbox{PS}^{(r)} H^a_{Ii} = \int\limits_{\eta_a}^1
  \frac{dz_a}{z_a} 8\pi I^a_i  \quad . } 
The results for $I^a_1,\ldots ,I^a_5$ are collected in appendix
8.5. They contain the singularities for the D and SR
contributions. Apart from $s, t, u$ and $y_J$, the results also depend
on the integration variable $z_a$. All five expressions $I^a_1,\ldots
,I^a_5$ contain the term 
\equ{m(P2)}{ M(P^2) = \frac{1}{2N_C} P_{q\leftarrow\g}(z_a) \ln\left(
  1 + \frac{y_Js}{z_aP^2} \right)  } 
which is large for $P^2\ll s$ and singular for $P^2=0$ as
expected. The large contribution has to be subtracted and absorbed
into the PDF of the virtual photon. Here, we have the same freedom as
in the case of the real photon, as has been described above for the D
case. Finite parts can be shifted from the PDF to the direct cross
section and vice versa. 

However, the virtual photon PDF's used later on in this work are
constructed in a scheme similar to the $\overline{\mbox{MS}}$ scheme
for real photons and we have to use the same scheme to
obtain consistent results. Therefore, we subtract those terms that
will yield the $\overline{\mbox{MS}}$ scheme of the real photon in the
limit $P^2\to 0$. In order to make the comparison with 
the case of the real photon possible, we state here the singular parts
of the expressions $I^a_1,\ldots I^a_5$, that appear for the real
photon in $d=4-2\e$ dimensions. They are given by \cite{6e}
\equ{}{  M = -\frac{1}{\e} \frac{1}{2N_C} P_{q\leftarrow \g}(z_a)
  + \frac{1}{2N_C}P_{q\leftarrow \g}(z_a)\ln\left(
  \frac{(1-z_a)}{z_a} y_J\right) + \frac{1}{2} \quad . }
The characteristic singularity proportional to $1/\e$ is subtracted by
absorbing the transition function
\equ{}{ \G^{(1)}_{q\leftarrow\g}(z_a,M_\g^2) = 
  -\frac{1}{\e}P_{q\leftarrow\g}(z_a) \frac{\G (1-\e )}{\G (1-2\e )}
  \left( \frac{4\pi\mu^2}{M_\g^2} \right)^{\e} }
into the PDF of the real photon. This subtraction produces a
factorization scale dependence of the photon PDF and gives the finite
contributions to the cross section. The expression remaining after the
absorption is, in the $\overline{\mbox{MS}}$ scheme:
\equ{mmsb}{ M_{\overline{MS}} = -\frac{1}{2N_C} 
  P_{q\leftarrow\g}(z_a) \ln\left( \frac{M_\g^2z_a}{y_Js(1-z_a)}
  \right) + \frac{1}{2} \quad .}
In order to obtain the same finite terms in $M(P^2)$ from equation
(\ref{m(P2)}) in the limit
$P^2\to 0$ for the virtual photon case, we absorb the transition
function  
\equ{llog}{ \G^{(1)}_{q\leftarrow\g^*}(z_a,M_\g^2,P^2) = \ln\left(
	\frac{M_\g^2}{P^2(1-z_a)} \right) P_{q\leftarrow\g}(z_a) -N_C } 
into the PDF of the virtual photon. This leaves the finite term
\equ{}{ M_{\overline{MS}}(P^2) = -\frac{1}{2N_C}
    P_{q\leftarrow\g}(z_a) \ln\left(\frac{M_\g^2z_a}{(z_aP^2+y_Js)(1-z_a)} 
    \right) + \frac{1}{2} \quad , }   
that reduces to the expression $M_{\overline{MS}}$ in (\ref{mmsb}) for
real photons in the limit $P^2\to 0$. We therefore call this form of
factorization the $\overline{\mbox{MS}}$ factorization for virtual
photons.

\section{Parton Distribution Function of the Photon}

As mentioned in the introduction and worked out in section 3, the
photon produces a $q\bar{q}$-pair in the initial state of the NLO
direct contribution that leads to a large logarithm for virtual photons
and a singularity for 
real photons. These terms have to be absorbed into the PDF of the virtual
and real resolved photons, respectively, leading to a point-like term
in the evolution equations of the photon PDF's. In this section, we
wish to introduce the PDF of the real and the virtual photon. 

After a
general discussion of the origin of the photon structure, we define
the structure function and the PDF of the real photon. The evolution
equations will be explained and the differences to the proton will be
pointed out. A discussion of the formalism for the virtual photon PDF
resembling the formalism for the real photon follows. Finally, we
compare two parametrizations of the virtual photon PDF which we use in
our computations.

\subsection{Origin of the Photon Structure}

The photon is the elementary gauge boson of QED. However one knows
from soft low energy $\g p$ reactions, that a photon can behave like a
hadron. If the time of the
interaction between the proton and the photon is much smaller than the
fluctuation time $t_f$ of the $q\bar{q}$-pair, the pair will interact
with the proton rather than with the photon itself and will give rise
to a hadronic structure of the photon. The fluctuation time for high
energy photons with virtuality $P^2$ can be estimated from the
uncertainty principle by \cite{1}  
\equ{}{ t_f = \frac{2q_0}{P^2+m_{q\bar{q}}^2} \quad , }
where $q_0$ is the energy of the photon and $m_{q\bar{q}}$ is the mass
of the pair. As $P^2$ increases, $t_f$ becomes smaller, giving back the
photon its structureless character. Thus, one can identify  
the direct photon, which interacts directly as a structureless
object, and the resolved photon, which has a hadronic structure.

At this point it is important to further distinguish the possible
configurations of the $q\bar{q}$-pair which yield different
contributions to the resolved photon. The photon can create a large
size, asymmetric configuration with small transverse momentum $k_T$
that gives rise to soft non-perturbative effects and a small size,
symmetric configuration with large $k_T$ that yields hard, perturbative
interactions \cite{33}. The soft part will behave more like a
hadron and it will therefore be called the {\em hadronic} part of the
resolved photon, whereas the hard part behaves more like a point-like
photon and will therefore be called the {\em point-like} part. A
possible physics interpretation of the soft part of the resolved
photon is the fluctuation of the $q\bar{q}$-pair into a
vector-meson with $m_{q\bar{q}}\simeq m_V$, which is described by the
vector-meson dominance (VMD) model \cite{2}. The coupling of the
photon to the vector-meson $\frac{4\pi}{f_V^2}$ has been predicted by
the VMD model, giving
\equ{}{ \frac{f_\rho^2}{4\pi} : \frac{f_\omega^2}{4\pi} :
	\frac{f_\varphi^2}{4\pi} = 9:1:7 \quad . }
These ratios have been confirmed by measurements of the reaction
$e^+e^-\to \mbox{hadrons}$.

\begin{figure}[hhh]
\unitlength1mm
\begin{picture}(121,32)
\put(20,-29){\psfig{file=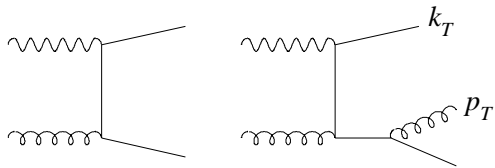,width=10.1cm} }
\end{picture}
\caption{\label{yg-fusion}Direct and resolved photon
	processes for photon-gluon fusion.}
\end{figure}

We have introduced the distinction between direct and resolved
photons, but this distinction is unambiguous only in LO. In NLO the
direct and resolved parts of the photon become intermixed through the
point-like part of the resolved photon. A possibility to distinguish
between the direct and the resolved photon interaction has been
suggested by Levy \cite{33}. Consider photon-gluon fusion as
shown in Fig. \ref{yg-fusion}. 
The diagram on the left is usually denoted as a direct process in LO.
The diagram on the right describes the fluctuation of the
photon into a $q\bar{q}$ pair with a given $k_T$ followed by the
interaction of one of the partons with a gluon, which produces a final
state with some $p_T$. If $k_T\ll p_T$, the process can be called a
resolved interaction. For $k_T\gg p_T$ the $p_T$ is too small for the
final state partons to form two separate jets, so the diagram looks
like the diagram on the left und thus can be considered as a direct
interaction. For large $P^2$ it is more likely that $k_T\gg p_T$ so
that the direct component is dominant. For low $P^2$ the resolved part
will be more dominant. 

\subsection{Parton Distribution Functions of the Real Photon}

\begin{figure}[bbb]
\unitlength1mm
\begin{picture}(121,31)
\put(15,-33){\psfig{file=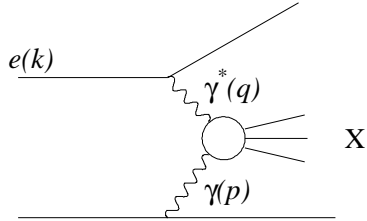,width=10.1cm} }
\end{picture}
\caption{\label{ey-eX}Single-tag DIS $e\g$ experiment.}
\end{figure}

The structure of the real photon has been analyzed in the process
$e\g\to eX$, depicted in Fig.\ \ref{ey-eX}. 
We define $x\equiv Q^2/(Q^2+W^2)$ with $Q^2=-q^2$ and $P^2=-p^2$, 
where $W$ is the c.m.\ energy of the $\g^*\g$ system and $y\equiv
(pq)/(pk)$. Denoting the longitudinal polarization state of the
photon as $l$ and the transversal one as $t$, we can define photon
structure functions
\begin{eqnarray}
   F_1^\g &=& \frac{Q^2}{4\pi\al} \frac{1}{2x} \si_{tt} \quad , \\
   F_2^\g &=& \frac{Q^2}{4\pi\al} (\si_{tt}+\si_{lt}) \quad .   
\end{eqnarray}
Using these definitions, the cross section for $e\g$ scattering can
be written as
\equ{egx}{ \frac{d\si (e\g\to eX)}{dxdy} = \frac{4\pi\al s}{Q^4} \left[
	(1-y)F_2^\g + xy^2F_1^\g \right]  \quad . }
This is in complete analogy to the DIS $ep$ reaction. The difference to the 
case of $ep$ scattering lies in the fact that the photon structure
function can be calculated perturbatively in the limit of large
$Q^2$. This is not possible for the proton structure function. 
The photon structure function is computable in the quark-parton model (QPM) 
from the box diagram $\g^*\g\to q\bar{q}$ and gives in LO
in the limit $m_{q_i}^2\ll Q^2$, where $m_{q_i}$ are the quark masses,
\cite{35.6, 35.7, 35.8}
\equ{}{ F_2^{\g ,pl} = \sum_{i=1}^{N_f} x q_i^\g (x,Q^2) }
with
\equ{qpmr}{ q_i^\g (x,Q^2)= 3e_{q_i}^2\frac{\al}{2\pi} \left\{
	[x^2+(1-x)^2] \ln \frac{Q^2(1-x)}{m_{q_i}^2x} + 8x(1-x)-1
	\right\} \quad . }
The function $q_i^\g$ can be interpreted as the PDF of the quark in
the photon, in analogy to the proton case. 

The QPM result can be modified substantially by QCD effects, such as
multiple gluon radiation ladder diagrams. In the limit of large $Q^2$
these kind of corrections where shown to be exactly calculable in LO
\cite{witt} and NLO \cite{bar}. The result is of the form
\equ{}{ F_2^{\g ,asymp} = \al \left[ \frac{a(x)}{\al_s(Q^2)} + b(x)
	\right] \quad , }
where $a(x)$ is the LO and $b(x)$ the NLO result. 
Unfortunately $F_2^{\g ,asymp}$ becomes negative for small values of $x$,
which cannot be true, since the photon structure function 
is measurable. The problem cannot
be cured by adding the VMD contributions that have been mentioned in
the previous section, since this non-perturbative contribution is
expected to be well-behaved. Therefore it is not possible to compute 
the photon structure function by perturbation theory alone. 

To handle the problems of the photon structure function, Gl\"uck and
Reya \cite{gr} have suggested to formally add all contributions to the
photon structure, namely the QPM, their QCD corrections and the VMD
contributions into a single photon structure function 
$F_2^\g = \sum_i x q_i^\g (x,Q^2)$ and fix the quark
distributions at some input scale $Q_0^2$ in analogy to the proton
case. Then the photonic parton densities at different values of $Q^2$
follow from the inhomogeneous evolution equations, that are in LO
\begin{eqnarray}
  \frac{dq_i^\g}{dt} &=& h^{box} + \frac{\al_s}{2\pi} \int\limits_x^1 
	\frac{dz}{z} \left[ P_{q\leftarrow q}\left(\frac{x}{z}\right)q_i^\g
	+ P_{g\leftarrow q}\left(\frac{x}{z}\right)g^\g \right] \\ 
  \frac{dg^\g}{dt} &=& \frac{\al_s}{2\pi} \int\limits_x^1 
	\frac{dz}{z} \left[ P_{q\leftarrow g}\left(\frac{x}{z}\right)
     q_i^\g + P_{g\leftarrow g}\left(\frac{x}{z}\right)g^\g \right] \quad . 
\end{eqnarray}
Here, $t\equiv \ln (Q^2/\Lambda^2)$ and the inhomogeneity  is given by
\equ{}{ h^{box} = 3e_{q_i}^2 \frac{\al}{2\pi} [x^2+(1-x)^2] \quad . }
The solution of the homogeneous equations is similar to the solution
of the DGLAP equations for hadrons and can therefore be called
$F_2^{\g ,had}$. The particular solution of the inhomogeneous equation
is due to the inhomogeneity that stems from the point-like coupling of
the photons to the quarks and can therefore be called $F_2^{\g ,pl}$. 
The general solution of the inhomogeneous evolution equations for the
photon is thus given by  
\equ{had-point}{ F_2^\g = F_2^{\g ,had} + F_2^{\g ,pl} \quad . }
This equation allows one to speak about the hadronic and the
point-like part of the photon structure function.

\begin{figure}[hhh]
\unitlength1mm
\begin{picture}(121,122)
\put(10,126){\epsfig{file=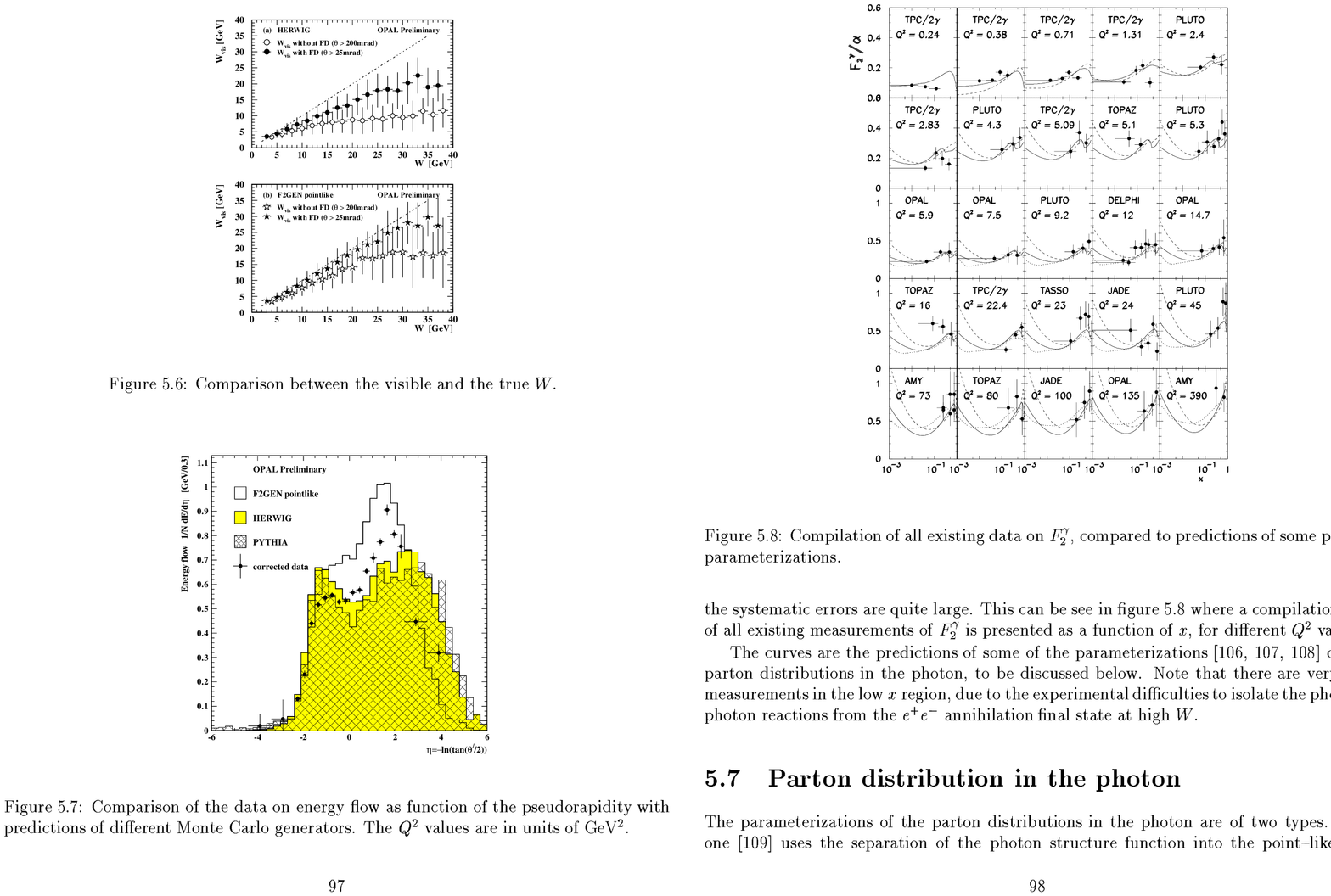,width=12.9cm,
	bbllx=42pt,bblly=514pt,bburx=347,bbury=746pt,angle=270,clip=}}
\end{picture}
\caption{\label{f2y-data}Compilation \cite{34a} of all
	existing data on $F_2^\g$ in comparison to predictions of
	the PDF parametrizations in \cite{20,36,22}.}
\end{figure}

The measurement of the photon structure function is not as easy as for
the proton case. Due to limited detector acceptance, the measured
hadronic energy is not equal to the total hadronic energy, so that the
photon energy is not determined well, which leads to large systematic
errors. In addition, the structure function $F_2^\g$ is small and the
cross section (\ref{egx}) is suppressed by $1/Q^4$, which leads to
large statistical errors. In spite of these difficulties, the photon
structure has been measured for various values of $Q^2$. In
Fig. \ref{f2y-data} all existing data for $F_2^\g$ is presented as a
function of $x$ for different values of $Q^2$ (taken from \cite{34}).

One important difference in the behaviour of the proton and the photon
structure functions is that $F_2^\g$ manifests strong scaling
violation even in LO without gluon radiation included. It is positive in
the whole $x$ region. Furthermore, $F_2^\g$ should be large at large
$x$ due to the point-like part of the photon structure function, while
the structure function of the proton is small at large $x$. Predictions 
from several parametrizations for the PDF of the real photon (solid
curve \cite{20}, dashed curve \cite{36}, dotted curve \cite{22}) are
also shown in Fig.\ \ref{f2y-data}.

\subsection{Parton Distribution Functions of the Virtual Photon}

We now turn to a discussion of the virtual photon structure. Some old data
from the PLUTO collaboration \cite{9} exist, which show the structure
function of a target photon with virtuality $P^2\simeq 0.4$ GeV$^2$ at
$Q^2\simeq 5$ GeV$^2$. At HERA information about the structure of the
virtual photon has been obtained by two different methods \cite{7, 8}. 
One was by tagging photons with a mean virtuality of $P^2\simeq
10^{-5}$ GeV$^2$ with the electron calorimeter of the luminosity
system. Another method was to use the beam-pipe calorimeter to tagg photons
with a range in virtuality of $0.1<P^2<0.6$ GeV$^2$. The ratio of the
resolved to the direct contribution can be plotted as a function of
the photon virtuality $P^2$, where an experimental definition of the
direct and the resolved part of the cross section has to be given. We
will come back to this data in section 5, where we calculate one- and
two-jet cross sections in NLO for $ep$ scattering under HERA
conditions. We will compare the ratios as defined in the experiment
with our theoretical predictions. 

Here, we concentrate on the construction of the PDF's for a
virtual photon. First, we state the LO QPM result for virtual photons,
which substitutes the result (\ref{qpmr}) for real photons. The
virtuality $P^2$ serves as a regulator in this case and in the limit
$\Lambda^2\ll P^2\ll Q^2$ one obtains \cite{35.6, 35.7, 35.8} 
\equ{qpm}{ q_i^{\g^*} (x,Q^2)= 3e_{q_i}^2\frac{\al}{2\pi} \left\{
	[x^2+(1-x)^2] \ln \frac{Q^2}{x^2P^2} + 6x(1-x)-2
	\right\} \quad . }
The evolution equations in $Q^2$ for virtual photons and the resulting
PDF's are exactly calculable in perturbative QCD in a limited range of
photon virtuality, $\Lambda^2\ll P^2\ll Q^2$ 
\cite{35.7, 35.8}. The PDF's of the real photon are known for the
region $P^2\ll \Lambda^2$, as described in section 4.2. At HERA,
though, the intermediate region $P^2\simeq \Lambda^2$ is of special
interest, as has been noted above. The aim of Gl\"uck, Reya and
Stratmann (GRS) in \cite{21} and Schuler and Sj\"ostrand (SaS) in 
\cite{22} was to construct PDF's for virtual photons, that are valid
in the whole $P^2$-region, i.e.\ $0\le P^2 \le Q^2$. We explain
the constructions of these PDF's in the following.

\subsubsection{The PDF's of GRS}

Gl\"uck, Reya and Stratmann have used for their construction a VMD
inspired interpolation between the PDF's of real photons and those
valid at $P^2\gg \Lambda^2$ \cite{21}, since the PDF's 
$f_i^\g (x,Q^2,P^2)$ obey evolution equations similar to those of the real
photon. The question therefore reduces to finding appropriate boundary
conditions at $Q^2=P^2$. Defining
\equ{}{ \eta (P^2) \equiv \frac{m_\rho^4}{(m_\rho^2+P^2)^2} \quad ,}
where $m_\rho^2=(0.77)^2$ GeV$^2$ refers to some effective mass in the
vector-meson propagator, PDF's valid for all values of $P^2$ are
defined as
\equ{}{ f^{\g^*}(x,Q^2=P^2,P^2) = \eta (P^2)f^{\g^*}_{had}(x,P^2) + 
	[1-\eta (P^2)]f^{\g^*}_{pl}(x,P^2) \quad . }
In this formula, the perturbatively calculable pointlike part
$f^{\g^*}_{pl}$ is given by the function $q_i^{\g^*}(x,Q^2=P^2)$ from
equation (\ref{qpm}) and $g^{\g^*}(x,Q^2=P^2)=0$ in NLO. For a LO
construction, $f^{\g^*}_{pl}(x,Q^2=P^2)=0$. These results are for the
DIS$_\g$ scheme \cite{20}, which is connected to the
$\overline{\mbox{MS}}$-scheme via the transformations
\begin{eqnarray} 
   q^{\g^*}_{DIS_\g} &=& q^{\g^*}_{\overline{MS}} +
   q^\g_i(x,Q^2=m_i^2)  \nonumber \\
   g^{\g^*}_{DIS_\g} &=& g^{\g^*}_{\overline{MS}} \quad , \label{ms-disg}
\end{eqnarray}
where $q^\g_i(x,Q^2)$ is given by (\ref{qpmr}). These transformations
are valid only for the NLO distributions, whereas the LO distributions are
equal in both schemes. The reason for introducing the DIS$_\g$ scheme
was that the differences between the LO and the NLO result are
small. The hadronic, non-perturbative input is given by 
\equ{}{ f^{\g^*}_{had}(x,P^2) = \kappa \frac{4\pi\al}{f_\rho^2} 
	\times \left\{ {f^\pi(x,P^2) \quad , \quad P^2>\mu^2 
	\atop f^\pi(x,\mu^2) \quad , \quad 0\le P^2\le\mu^2 }
	\right. }
where $\mu_{LO}^2=0.25$ GeV$^2$ and $\mu_{NLO}^2=0.3$ GeV$^2$. The
function $\kappa (4\pi\al /f_\rho^2)f^\pi(x,\mu^2)$ is just the
prescription for the boundary conditions at input scale $Q^2=\mu^2$
for real photons. As one observes,
\equ{}{ \eta (P^2=0)=1 \qquad \mbox{and} \qquad 1-\eta (P^2\gg
	\Lambda^2) \to 1 \quad . } 
Thus, the usual real photon PDF is regained for $P^2=0$, whereas
the perturbatively calculable part dominates for $P^2\gg \Lambda^2$. 
The number of flavors is set to $N_f=3$. The heavy quark sector 
($c, b, \ldots$) is supposed to be added as predicted by perturbation
theory of fixed order with no active $c$ and $b$ quarks in the proton
and photon PDF's. In LO this amounts to adding the processes $\g^*g\to
c\bar{c}$ and $\g^*g\to b\bar{b}$ to the cross section, keeping $m_c,
m_b\ne 0$. 

\begin{figure}[hhh]
  \unitlength1mm
  \begin{picture}(122,115)
    \put(8,-3){\epsfig{file=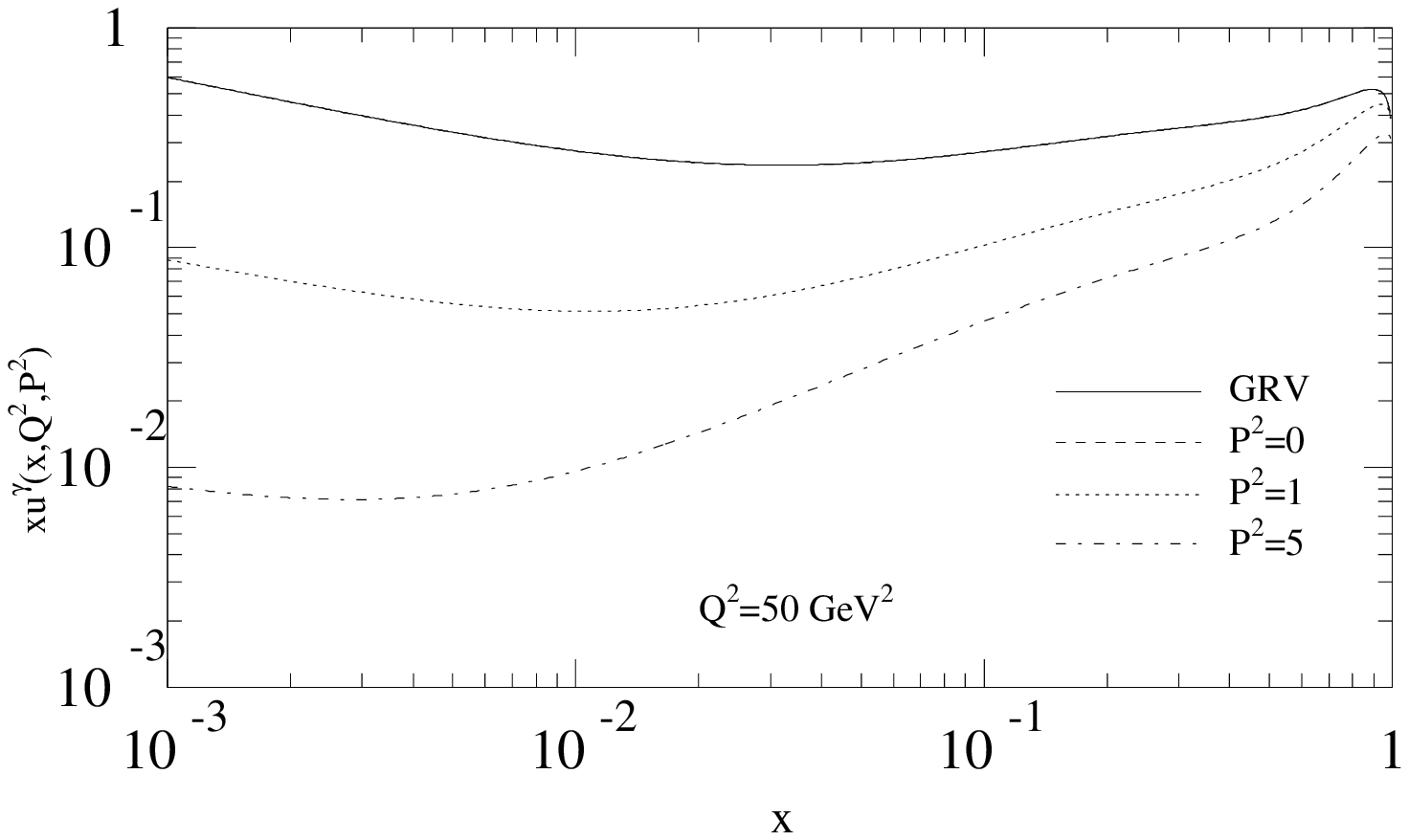,width=11cm}}
    \put(8,-61){\epsfig{file=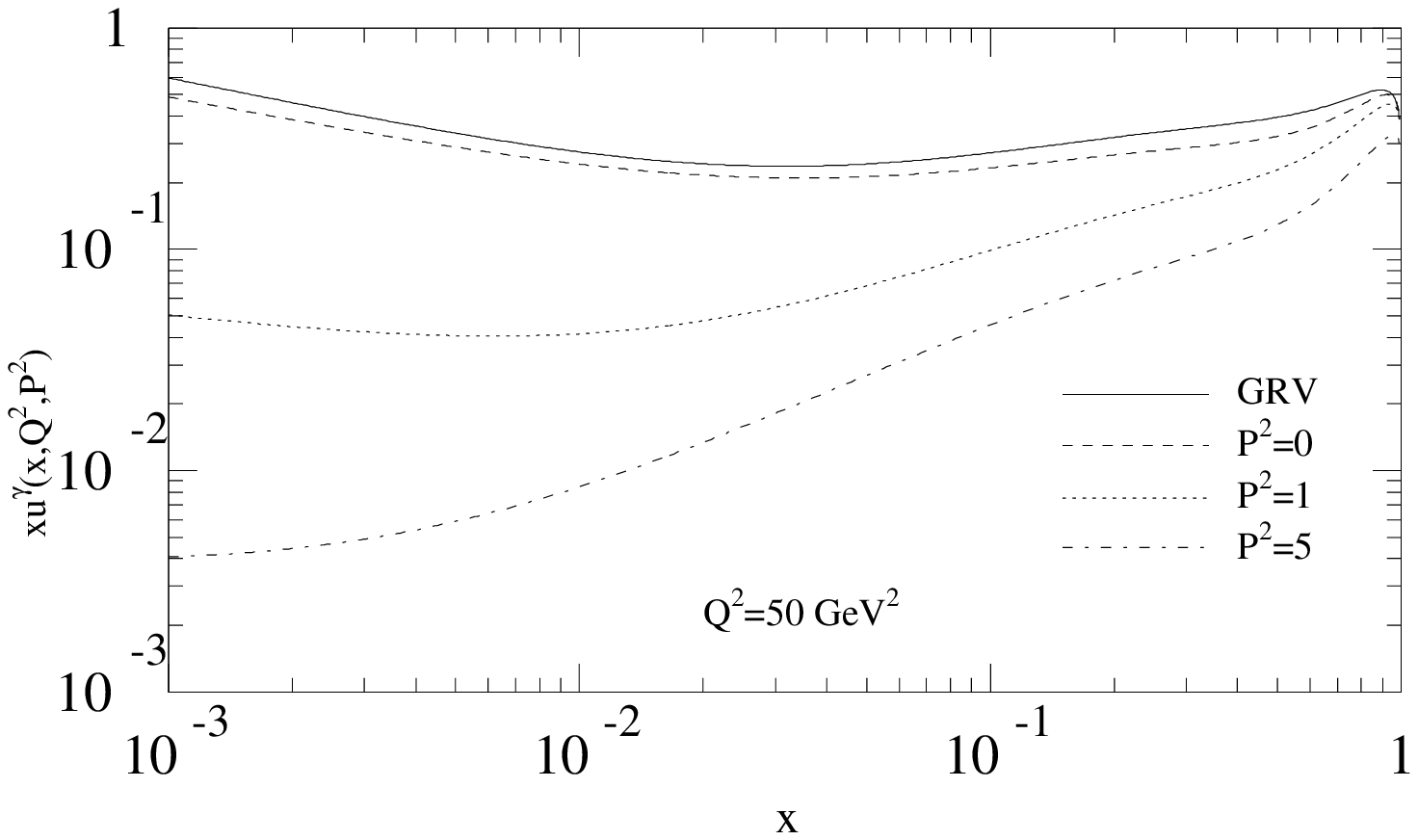,width=11cm}}
    \put(62,108){\footnotesize (a)}
    \put(62,49){\footnotesize (b)}
  \end{picture}
\caption{\label{u-dis}(a) The GRS LO 
	prediction in the $\overline{\mbox{MS}}$ scheme for the
	up-distribution of a virtual photon at $Q^2=50$ GeV$^2$ and
	various fixed values of $P^2$. For comparison, the LO 
	prediction of GRV for a real photon \cite{20} is shown, which
	lies exactly on top of the PDF of GRS for $P^2=0$; (b) SaS1M PDF's.}
\end{figure}

In \cite{11} GRS have provided PDF's of the virtual photon in a
parametrized form in LO that can be conveniently used for numerical
calculations. The input scale is $Q_0=0.5$ GeV and the restriction
$P^2\le Q^2/5$ is implemented as to fulfill the condition $P^2 \ll
Q^2$. We show the $x$-distribution for the up-quark at a scale of
$Q^2=50$ GeV$^2$ for three different values of $P^2$, namely $P^2=0, 1$
and $5$ GeV$^2$ in Fig.\ \ref{u-dis} a. As one observes, the  
distribution decreases with increasing $P^2$. For $P^2=0$ the real
photon PDF of Gl\"uck, Reya, Vogt \cite{20} is reproduced exactly and
the curves fall on top of each other. The use of the
$\overline{\mbox{MS}}$ scheme for the LO distributions has to be
explained. The authors GRS and SaS both give distributions in the
DIS$_\g$ scheme, but their schemes differ slightly. SaS actually also
give $\overline{\mbox{MS}}$ distributions. In order to make
the results comparable, we treat the distributions of GRS formally as
in NLO and use the transformation equations (\ref{ms-disg}) also for
the GRS LO distributions.

It should be mentioned, that GRS have calculated NLO distributions in
\cite{11}. Distinct differences occur for larger $P^2$ and
$x>10^{-3}$ which is mainly  due to the different NLO perturbative
boundary condition at $P^2=Q^2$, which does not exist for the real
photon structure function.
\begin{figure}[hhh]
  \unitlength1mm
  \begin{picture}(122,115)
    \put(8,-3){\epsfig{file=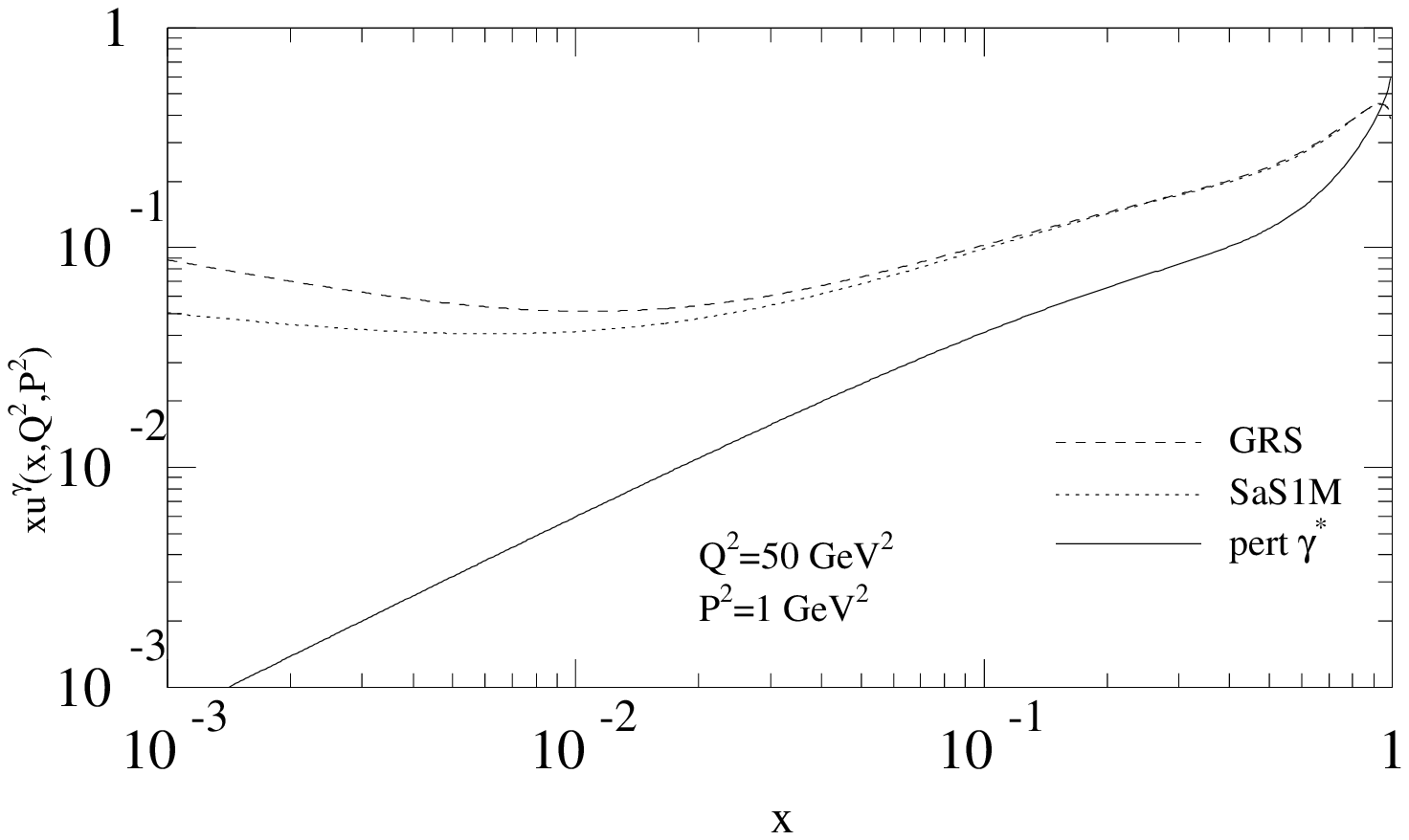,width=11cm}}
    \put(8,-61){\epsfig{file=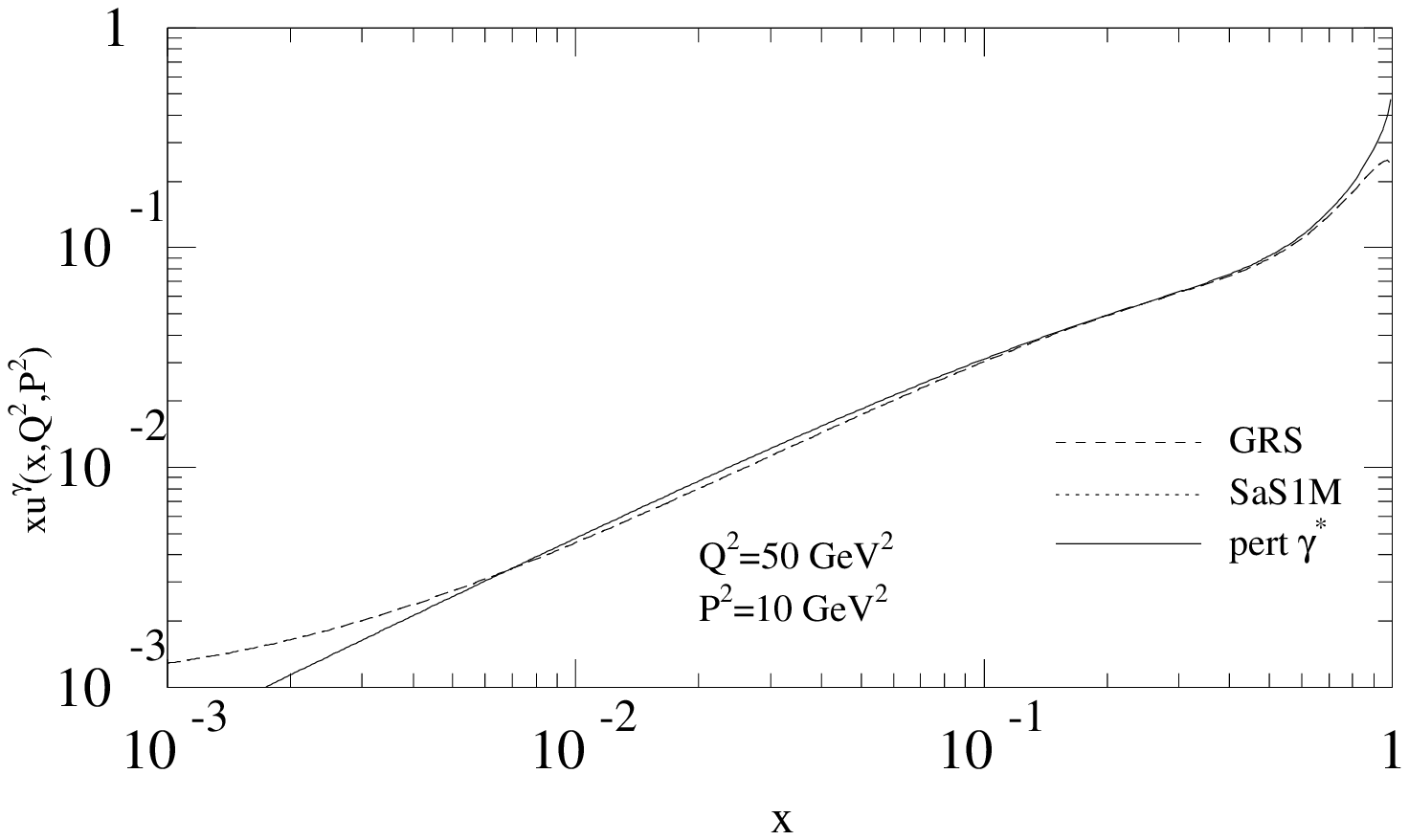,width=11cm}}
    \put(62,108){\footnotesize (a)}
    \put(62,49){\footnotesize (b)}
  \end{picture}
\caption{\label{u-pt}(a) Comparison
	between the GRS and SaS1M LO predictions for the
	up-distribution of a virtual photon at $Q^2=50$ GeV$^2$ and
	$P^2=1.0$ GeV$^2$ and the purely perturbative contribution
	in the $\overline{\mbox{MS}}$ scheme; (b) $P^2=10$
	GeV$^2$. The distribution for GRS lies exactly atop of the
	SaS1M curve.}
\end{figure}

\subsubsection{The PDF's of SaS}

Schuler and Sj\"ostrand represent the solution of the inhomogeneous
evolution equations of the real photon as a sum of a perturbative and
a non-perturbative term \cite{22}
\equ{rpdfs}{ f^\g (x,Q^2) = \sum_V \frac{4\pi\al}{f_V^2} 
	f^{\g ,VMD}(x,Q^2;Q_0^2) + \frac{\al}{2\pi}\sum_i 2e_{q_i}^2
	\int\limits_{Q_0^2}^{Q^2} \frac{dk^2}{k^2} 
	f^{\g ,q\bar{q}}(x,Q^2;k^2) \quad . }
Here, $Q_0^2\ge \Lambda^2$ is the input scale for the non-perturbative 
solution $f^{\g ,VMD}$ of the homogeneous evolution equations, which
can be interpreted as a fluctuation of the real photon into vector-mesons.
The second term represents the anomalous perturbative solutions of the $\g\to
q\bar{q}$ fluctuations, where $k^2$ is the virtuality of the
$q\bar{q}$-pair, which has a continuous spectrum. As noted above, the
evolution equations of the PDF's of the virtual photon can be exactly
calculated in the range  $Q_0^2\ll P^2\ll Q^2$. For real photons in
the region of $P^2\ll Q_0^2$, the PDF's are given by equation
(\ref{rpdfs}). To obtain results valid for the whole $P^2$ region, SaS
make use of the dispersion relation
\equ{}{ f^{\g^*}(x,Q^2,P^2) =\frac{\al}{2\pi}\sum_i 2e_{q_i}^2
	\int\limits_0^{Q^2} \frac{dk^2}{k^2} 
	\left(\frac{k^2}{k^2+P^2} \right)^2 f^{\g ,q\bar{q}}(x,Q^2;k^2)
	\quad . } 
This model provides the correct behaviour for both $P^2\to 0$ and
the above described perturbative region. Now, the region of low $k^2$
can be associated with the discrete set of vector-mesons, so that by
introducing the cut-off $Q_0^2$ in the $k^2$-integration SaS obtain
\cite{22}
\begin{eqnarray}
 f^{\g^*}(x,Q^2,P^2) &=& \sum_V \frac{4\pi\al}{f_V^2} 
	\left(\frac{m_V^2}{m_V^2+P^2} \right)^2	
	f^{\g ,VMD}(x,Q^2;\tilde{Q}_0^2) \nonumber \\
  &+& \frac{\al}{2\pi}\sum_i 2e_{q_i}^2
	\int\limits_{Q_0^2}^{Q^2} \frac{dk^2}{k^2} 
	\left(\frac{k^2}{k^2+P^2} \right)^2 f^{\g ,q\bar{q}}(x,Q^2;k^2)
	\quad . 
\end{eqnarray}
These parton distributions are the solutions of the inhomogeneous
evolution equations of the virtual photon. Note, that the input scale
for the VMD PDF's has been shifted from $Q_0^2\to \tilde{Q}_0^2$ with
$Q_0^2< \tilde{Q}_0^2$. This is motivated by a study of the evolution 
equations in \cite{35.5}, which shows that the evolution for virtual
photons starts later in $Q^2$.

The authors SaS have provided the PDF's of the virtual photon in a
para\-me\-trized form in LO for $N_f=4$ in \cite{22} for two different
input scales, namely $Q_0=0.6$ GeV and $Q_0=2$ GeV. We will use the
lower scale in this work. In contrast to the GRS parametrization, the
$c$ quark is included as a massless flavor in the PDF that undergoes
the usual evolution as the other massless quarks except for a shift of
the starting scale $Q_0$. We show the up-quark distribution in
comparison to the ones obtained by GRS in Fig.\ \ref{u-dis} b for the
same scale and $P^2$ values as in Fig.\ \ref{u-dis} a. They show
roughly the same behaviour and deviate only in the small $x$
region. For $P^2=0$, the GRV \cite{20} distribution of the real photon
is  recovered more or less. We have used the SaS1M parametrization,
which is given in the $\overline{\mbox{MS}}$ scheme, to make the
SaS results comparable with the GRS distributions. 

In section 3.5 we have calculated the part of the hard cross section
for an incoming virtual photon that couples directly to the
subprocess, leading to the logarithmic terms $\ln (P^2/Q^2)$. The
logarithm is absorbed into the PDF of the virtual photon. As we
suspect from the above discussion, the PDF of the virtual photon 
should reduce approximately to the perturbatively calculable contribution
(\ref{llog}) in the region of large $P^2$. Thus, comparing the
contribution (\ref{llog}) directly with the, say, up-quark distribution
for virtual photons should lead to results of the same order of
magnitude for large $P^2$. They cannot give exactly the same
results, since the perturbative part of the $u$-quark distribution is
evolved with help of the evolution equations. In Fig.\ \ref{u-pt} a, b
we show the purely perturbative contribution in comparison to the
$u$-quark distributions GRS and SaS1M at $Q^2=50$ GeV$^2$ for the two
values $P^2=1$ GeV$^2$ and $P^2=10$ GeV$^2$ in the 
$\overline{\mbox{MS}}$-scheme. As one observes, the
perturbative solution and the $u$-distribution coincide  rather well
for the larger $P^2$ value, especially in the large $x$ range, with a
slight enhancement of the perturbative curve near $x=1$.

\section{Electron-Proton Scattering at HERA}

We come to an analysis of inclusive jet-rates in electron-proton
scattering for slightly off-shell photons. After the introduction of
the hadronic cross section, using the calculated partonic cross
sections of section 3, we explain the matching of theoretical and
experimental jet definitions. Afterwards, some numerical tests are
discussed, and finally numerical results for one- and two-jet inclusive
cross sections are given. A comparison with present HERA data is
shown.

\subsection{Hadronic Cross Section}

We write electron-proton scattering for the production of two jets as
\equ{}{ e(k)+P(p) \to e(k') + \mbox{Jet}_1(E_{T_1},\eta_1) +
    \mbox{Jet}_2(E_{T_2},\eta_2) + X \quad .}
Here, $k$ and $p$ are the momenta of the incoming electron and proton,
respectively, and $k'$ is the momentum of the outgoing electron. The
jets in the final state are characterized by their transverse momenta
$E_{T_i}$ and rapidities $\eta_i$, which are observables in an
experimental setup. The interaction of the electron with the proton is
mediated by an electroweak vector boson with four-momentum $q\equiv (k-k')$
and virtuality $P^2\equiv (-q^2)$. The process is dominated by a photon,
especially for the small virtualities under consideration. We therefore
concentrate on the photon and neglect contributions from the other
electroweak bosons. The phase space of the electron can be
parametrized by the variables $y\equiv (pq)/(pk)$ and $P^2$.
In the case of small virtualities $P^2\ll q_0^2$, where
$q_0$ is the energy of the virtual photon, $y$ gives the momentum
fraction of the electron energy $E_e$, carried away by the virtual
photon, so $y \simeq q_0/E_e$. The c.m.\ energy
of the hadronic system is given by $s_H=(p+k)^2$, whereas the c.m.\
energy of the photon-proton subsystem is $W^2=(p+q)^2$. 

We have discussed the factorization of hard and soft regions of the 
electron-proton cross section in section 2. The hadronic cross section
$d\si^H$ may be written as a convolution of the hard scattering
process $d\si_{e/k}$ with the PDF of the proton $f_{k/P}(x_b)$, where
$x_b$ is the momentum fraction of the parton from the proton:
\equ{}{ d\si^H(s_H) = \sum_k \int dx_b \ f_{k/P}(x_b)
  d\si_{e/k}(x_bs_H) \quad .} 
The hard process is given by the squared matrix element $|{\cal
M}|^2$, which has to be divided by the flux factor $2s_Hx_b$ and 
multiplied by the phase space of $n$ final state particles of the
subprocess and the electron, $d\mbox{PS}^{(n+1)}$:
\equ{hard1}{ d\si_{e/k} = \frac{1}{4s_Hx_b} |{\cal M}|^2
  d\mbox{PS}^{(n+1)} \quad . } 
The matrix element $|{\cal M}|^2$ separates into the hadron tensor
$H^{\mu\nu}$ and the lepton tensor $L_{\mu\nu}=4(k_\mu k'_\nu -k'_\mu
k_\nu - g_{\mu\nu}kk')$:
\equ{}{ |{\cal M}|^2 = \frac{4\pi\al}{P^4}
  L_{\mu\nu}H^{\mu\nu} \quad . }
The constant $\al$ is the electromagnetic coupling constant. The
separation of the phase space into a part depending only on the
electron $dL$ and a part depending only on the final state particles
of the subprocess $d$PS$^{(n)}$ is easily achieved by inserting a
delta function for the intermediate virtual photon and gives 
\equ{}{ d\mbox{PS}^{(n+1)} = dL d\mbox{PS}^{(n)} \qquad \mbox{with}
  \qquad dL = \frac{P^2}{16\pi^2} \frac{d\phi}{2\pi}
  \frac{dydP^2}{P^2} \quad . }
Here $\phi$ is the azimuthal angle of the outgoing electron. This degree
of freedom can be integrated out, yielding
\equ{}{  \frac{1}{4P^2} \int \frac{d\phi}{2\pi} L_{\mu\nu}H^{\mu\nu} = 
   \frac{1+(1-y)^2}{2y^2} H_g + \frac{4(1-y) + 1+(1-y)^2}{2y^2} H_L
   \quad , } 
with the definitions $H_g \equiv -g^{\mu\nu}H_{\mu\nu}$ and $H_L\equiv
(4P^2)/(s_Hy)^2 p^\mu p^\nu H_{\mu\nu}$. Since we
will consider the range of small photon virtualities $P^2$ throughout
this work, the contribution $H_L$ proportional to $P^2$ will be
neglected. We approximate the spectrum of the virtual photons by
\equ{1/p2}{ \frac{df_{\g /e}(y)}{dP^2} =
  \frac{\al}{2\pi} \frac{1+(1-y)^2}{y} \frac{1}{P^2}  \quad ,}
which is the unintegrated Weizs\"acker-Williams \cite{39}
formula. For later use we define the virtuality $P^2_{eff}=0.058$ GeV$^2$. By
inserting $P^2_{eff}$ into the unintegrated Weizs\"acker-Williams
formula, we obtain the value for the Weizs\"acker-Williams formula
integrated in the region $P^2_{min} \le P^2 \le P^2_{max} = 4$
GeV$^2$ using the minimum photon virtuality
$P^2_{min}:=\frac{m_e^2y^2}{1-y}$. In this way, we can reproduce the
$P^2 \simeq 0$ results.

Defining the partonic cross section 
\equ{}{ d\si_{\g /k} = \frac{1}{4x_bys_H} H_{\g /k} d\mbox{PS}^{(n)}
	\quad , }
the hard scattering (\ref{hard1}) integrated over the angle $\phi$ can
be written as 
\equ{}{ d\si_{e/k} = d\si_{\g /k}df_{\g /e}(y)dy \quad . }

\begin{figure}[hhh]
\unitlength1mm
\begin{picture}(121,48)
\put(15,-30){\psfig{file=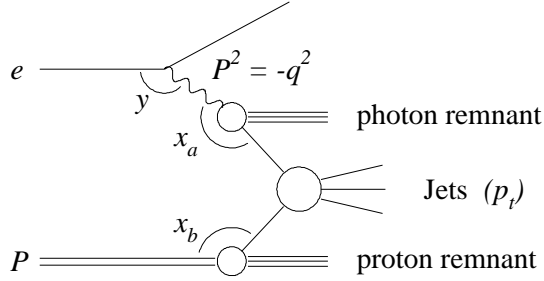,width=10.4cm} }
\end{picture}
\caption{\label{fac1}Factorization of hard and soft
    contributions in electron-proton scattering.}
\end{figure}

As discussed in the previous section, a photon with moderate
virtuality interacts with a proton not only as a point-like particle,
but also via its hadronic content. The hadronic structure of the
photon is described by a PDF $f_{\g /l}(x_a)$, introducing the new
variable $x_a$ that gives the momentum fraction of the parton from the
photon. To simplify the notation, the case of a direct photon is
included into the PDF of the photon via the delta function
$f_{\g\g}(x_a)=\de (1-x_a)$. Summarizing the above results, the
hadronic cross section $d\si^H(s_H)$ may be written as a convolution
of the hard scattering $d\si_{k/l}$ with the PDF of the photon 
$f_{\g /l}(x_a)$ and of the proton  $f_{k/P}(x_b)$, multiplied by the
photon spectrum $df_{\g /e}(y)$: 
\equ{fac2}{ d\si^H = \sum_{k,l} \int \! dx_a dx_b dy \ df_{\g /e}(y) \
  f_{\g /l}(x_a) d\si_{l/k} f_{k/P}(x_b) \quad . } 
The hard cross section $d\si_{kl}$ now describes the interactions of
the partons from the photon (and the photon itself) with the partons
from the proton and is given by the trace of the hadron tensor,
multiplied with the phase space of the final state particles, divided
by the flux factor:
\equ{}{ d\si_{l/k} = \frac{1}{4x_ax_bys_H} H_{l/k} d\mbox{PS}^{(n)} \quad .} 
The factorization (\ref{fac2}) is visualized in Fig.\ \ref{fac1}.

For electron-proton scattering, the SR case, in which the virtual
photon couples directly to the partons from the proton, and the DR
case, in which the photon serves as a source of partons, are present.
As mentioned in section 2, the matrix elements cannot be integrated
over the whole region of phase space in NLO. A cut-off has to be 
introduced that separates the singular from the finite regions. The
results for the integration over the singular regions were given in
section 3 and the factorization of singular terms has been
discussed. Although the calculation of the NLO matrix elements is
straightforward, especially using algebraic programs like REDUCE
\cite{40}, the results for the full matrix elements are too cumbersome
to be stated here. For both, the SR and the DR cross
section, we have a set of two-body contributions and a set of
three-body contributions. Each set is completely finite, as all
singularities have been canceled or absorbed into PDF's. Each part
depends separately on the phase space slicing parameter $y_c$. The
analytic calculations are valid only for very small $y_c$, since terms
${\cal O}(y_c)$ have been neglected in the analytic integrations. As
explained in section 3.3, the two separate pieces have no
physical meaning. When the two-body and three-body contributions are
superimposed to yield a suitable inclusive cross section, as for
example the inclusive one- or two-jet cross section, the dependence on
the cut-off $y_s$ will cancel. This has been checked explicitly and
will be demonstrated in section 5.4. 

We now come to the kinematics of electron-proton scattering. We first
concentrate on the SR cross section, for which the kinematics is most
easily treated in the c.m.\ system of the virtual photon and the
proton, where for the three-vectors ${\bf p}+{\bf q}={\bf 0}$. We
denote the momenta of the final state particles as $p_1$ and $p_2$,
which can be expressed by their transverse momenta
$E_{T_1}=E_{T_2}=E_T$ and their rapidities $\eta_1$ and $\eta_2$ in
the $\g^*P$ c.m.\ system by $p_i=E_T(\cosh\eta_i,0,0,\sinh\eta_i)$
(remember that the azimuthal angle has been integrated out).
From energy and momentum conservation one obtains
\begin{eqnarray}
  W &=& E_T(e^{-\eta_1}+e^{-\eta_2}) \label{kin1} \quad , \\
  y &=& \frac{W^2+P^2}{s_H} \quad , \\
  x_b &=& 1 + \frac{2W}{W^2+P^2}\ E_T (\sinh\eta_1 + \sinh\eta_2) \quad . 
\end{eqnarray}
The phase space, including the integration over $x_b$ and $y$, can
be expressed as
\equ{kin2}{ d\mbox{PS}^{(2)}dx_bdy = \frac{W^2}{W^2+P^2}
  \frac{2E_T}{s_H} \frac{dE_T}{(2\pi )^2} \ d\eta_1d\eta_2 \quad . } 
The Mandelstam variables $s, t$ and $u$ are defined as
\begin{eqnarray}
 s &=& (p_b+q)^2 = (p_1+p_2)^2 \quad , \nonumber \\ 
 t &=& (q-p_1)^2 = (p_b-p_2)^2 \quad , \\
 u &=& (q-p_2)^2 = (p_b-p_1)^2 \quad . \nonumber
\end{eqnarray}
In the DR case, the rapidities of the final state partons
$\eta_1'$ and $\eta_2'$ are expressed in the c.m.\ system of the two
partons and have to be boosted into the photonic c.m.\ system via
\equ{}{ \eta_i = \eta_i' + \frac12 \ln x_a \qquad \mbox{for} \quad
i=1,2 \quad . } 
Inserting these transformed rapidities into the above equations
(\ref{kin1})--(\ref{kin2}) the correct formul{\ae} in the case of
the resolved photon for $x_a, x_b, y$ and $W$, now containing
$\eta_i'$, are obtained. For $x_a=1$, which defines the SR case, the
two systems are identical and $\eta_i'=\eta_i$.

For a comparison with HERA data the rapidities and transverse momenta
have to be transformed from the photonic c.m.\ system to the HERA
laboratory system. The calculation of the SR and DR cross sections
proceed as for real photoproduction, i.e.\ the transverse momentum
($q_T$) of the virtual photon and other small terms proportional to $P^2$
are neglected so that the virtual photon momentum is in the
direction of the incoming electron and $q_0=E_ey$. The transformation
from the c.m.\ system into the HERA laboratory system is as
for real photoproduction:
\equ{}{ \eta^{lab}_i = \eta_i + \frac{1}{2} \ln\frac{E_p}{yE_e} \quad . }

\subsection{Snowmass Jet Definition}

The factorization of hard and soft regions in the hadronic cross section
has been discussed so far for the initial state. The non-perturbative
and not calculable regions are parametrized through the PDF's of the
hadron or the resolved photon. A similar problem occurs in the final
state. The partons that are emitted from the subprocess cannot be
observed directly due to the confinement of color charge. The
hadronization of partons into single hadrons in the final state can be
described, similarly to hadrons in the initial state, by fragmentation
functions. Another possibility is the observation of a shower built
from a large number of hadrons without resolving the specific type of
hadrons emitted. One then has to define jets in order to identify the
hadron showers with individual partons or their combinations from the
subprocess. The combination of hadrons  into jets is done by cluster
algorithms, where jet definitions can be implemented. The jet
definitions should fulfill a number of criteria \cite{41}, such as
they should be simple to implement in theory and experiment, be well
defined and yield finite cross sections in any order of perturbation
theory and give cross sections that are more or less insensitive to
the hadronization processes.

Looking at the theoretical side of inclusive two-jet cross sections,
in LO there is a one-to-one correspondence between the parton from the
subprocess and the jet in the final state. Therefore the theory
is not sensitive to any specific cluster algorithm used in the
experiment. This is not sensible, because the experimental results
depend strongly on the used algorithm. Only in NLO can one implement
certain jet definitions on the theoretical side, because the jet can
obtain a substructure due to the nearly collinear radiation of a
parton in the final state.

Several jet definitions have been proposed to date, one of the first
being the $(\e ,\delta )$ criterion of Sterman and Weinberg
\cite{42} (see also \cite{kramer}). We will adopt the jet definition of 
the Snowmass 
meeting \cite{44}. According to this definition, two partons $i$ and $j$
are recombined, if $R_{i,J}<R$, where 
$R_{i,J}=\sqrt{(\eta_i-\eta_J)^2+(\phi_i-\phi_J)^2}$ and  $\eta_J, \phi_J$
are the rapidity and the azimuthal angle of the combined jet
respectively, defined as
\begin{eqnarray}
 E_{T_J} &=& E_{T_1} + E_{T_2} \quad , \\
 \eta_J  &=& \frac{E_{T_1}\eta_1 + E_{T_2}\eta_2}{E_{T_J}} \quad , \\
 \phi_J  &=& \frac{E_{T_1}\phi_1 + E_{T_2}\phi_2}{E_{T_J}} \quad .
\end{eqnarray}
The cone-radius $R$ is chosen as in the experimental analysis. Thus, two 
partons are considered as two separate jets or as a single jet depending 
on whether they lie outside or inside the cone with radius $R$ around the
jet momentum. In NLO, the final state may consist of two or three
jets. The three-jet sample contains all three-body contributions,
which do not fulfill the cone condition. The rapidities used for the
cone constraint are evaluated in the HERA laboratory system.

\subsection{Numerical Input}

We now describe the input for the numerical calculations. We
have chosen the CTEQ3M proton structure function \cite{45} which is
a NLO parametrization in the $\overline{\mbox{MS}}$ scheme, with
$\Lambda^{(4)}_{\overline{MS}} = 239$ MeV. This $\Lambda$ value
is also used to calculate $\al_s$ from the two-loop formula 
\equ{}{ \al_s(\mu ) = \frac{12\pi}{(33-2N_f)\ln
      \frac{\mu^2}{\Lambda^2}} \left( 1 - 
      \frac{6(153-19N_f)}{(33-2N_f)^2} \frac{\ln (\ln	
      \frac{\mu^2}{\Lambda^2})}{\ln \frac{\mu^2}{\Lambda^2}} \right)
	\quad . }
We use this formula for both the LO and NLO calculations. In the case
of photoproduction, the scale $\mu$ is set equal to the transverse
momentum of the jets, since this is the only hard scale present in the
interactions. Here, $P^2\ll E_T^2$, so that we also set $\mu =E_T$. 
Equivalently, the factorization scales are chosen to be $M_\g=M_p=E_T$. 

For the PDF's of the virtual photon we choose either the GRS \cite{21} set
or the SaS1M set \cite{22}. Both sets are given in parametrized form
for all scales $M_\g^2$ so that they can be applied without repeating
the computation of the evolution. As mentioned in section 4, both sets
are given only in LO, i.e. the boundary conditions for $P^2=M_\g^2$
and the evolution equations are in LO. Since neither of the two PDF's is 
constrained by empirical data from  scattering on a virtual photon
target we consider these LO distribution functions as sufficient for our
exploratory studies on jet production and treat them as if they
were obtained in NLO.  As noted in section 4, the heavy quarks
are supposed to be added as predicted by fixed order perturbation
theory  with no active heavy quarks  in the PDF's of the proton and
the photon. Since in this section we are primarily interested in studying
the sum of the direct and resolved contributions and the influence of
the consistent subtractions of the NLO direct part we refrain from
adding the LO or NLO cross sections for direct heavy quark production
as suggested in \cite{11, 21}. So, our investigations in connection
with the GRS parametrization of the virtual photon PDF are for a model
with three flavors only. For consistency we take also $N_f=3$ in the
NLO corrections and in the two-loop formula for $\alpha_s$. Of course,
the proton PDF has been obtained for $N_f=4$. In comparison to the GRS
parametrization, we studied the relevant cross sections also with  
the virtual photon PDF's of SaS \cite{22}, which are for $N_f=4$.

The cross sections we have computed are for kinematical conditions as
in the HERA experiments, for which positrons of $E_e = 27.5$ GeV  
which collide with protons of $E_p = 820$ GeV. To have the equivalent
conditions as in the ZEUS analysis we choose the constraints $y_{min}=
0.2$ and $y_{max}=0.8$ for the variable occuring in the unintegrated
Weiz\"acker-Williams approximation. The cone radius is set to $R=1$.

\subsection{Numerical Tests}

\begin{figure}[hhh]
  \unitlength1mm
  \begin{picture}(122,115)
    \put(8,-6){\epsfig{file=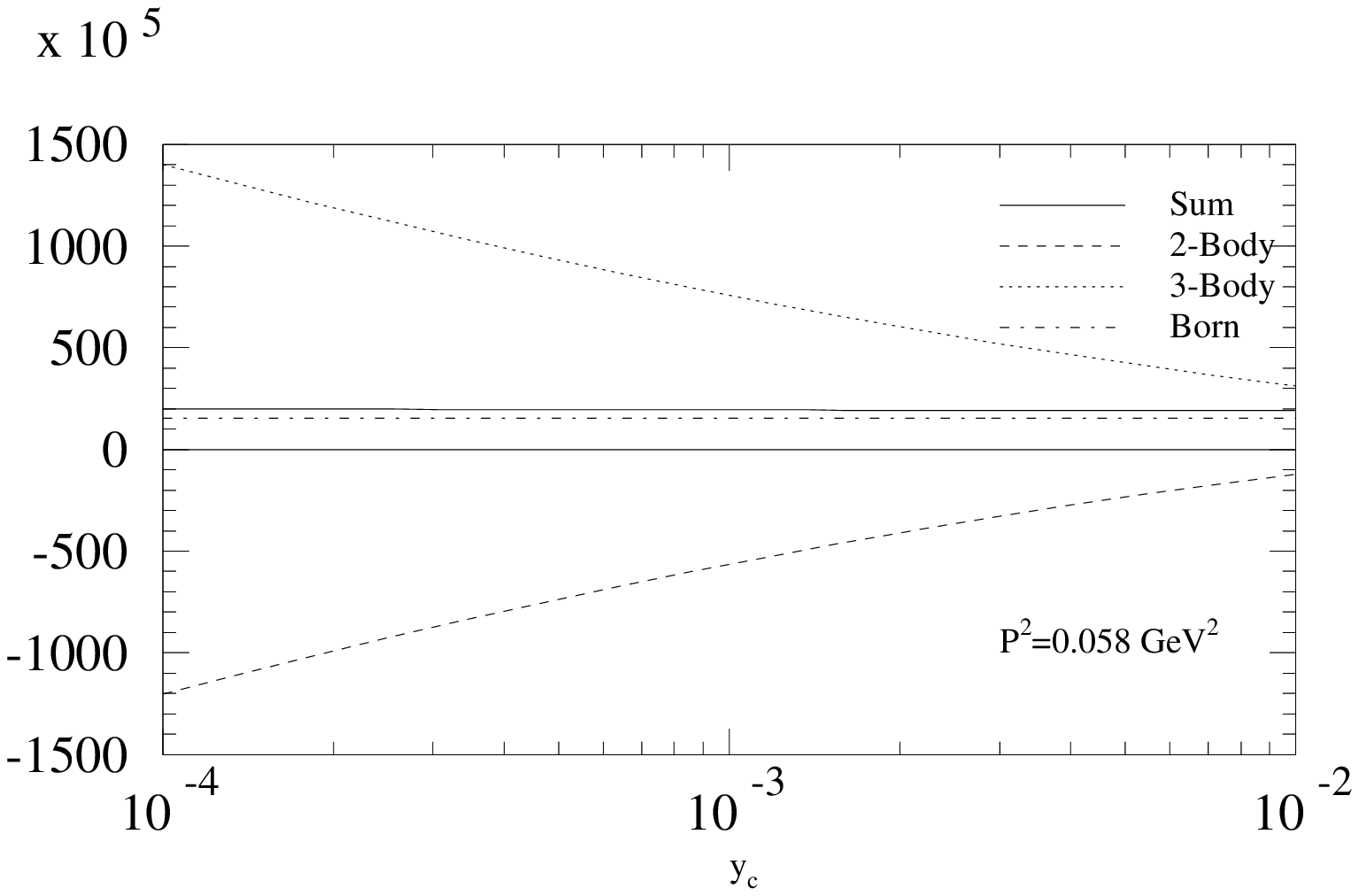,width=11cm}}
    \put(8,-64){\epsfig{file=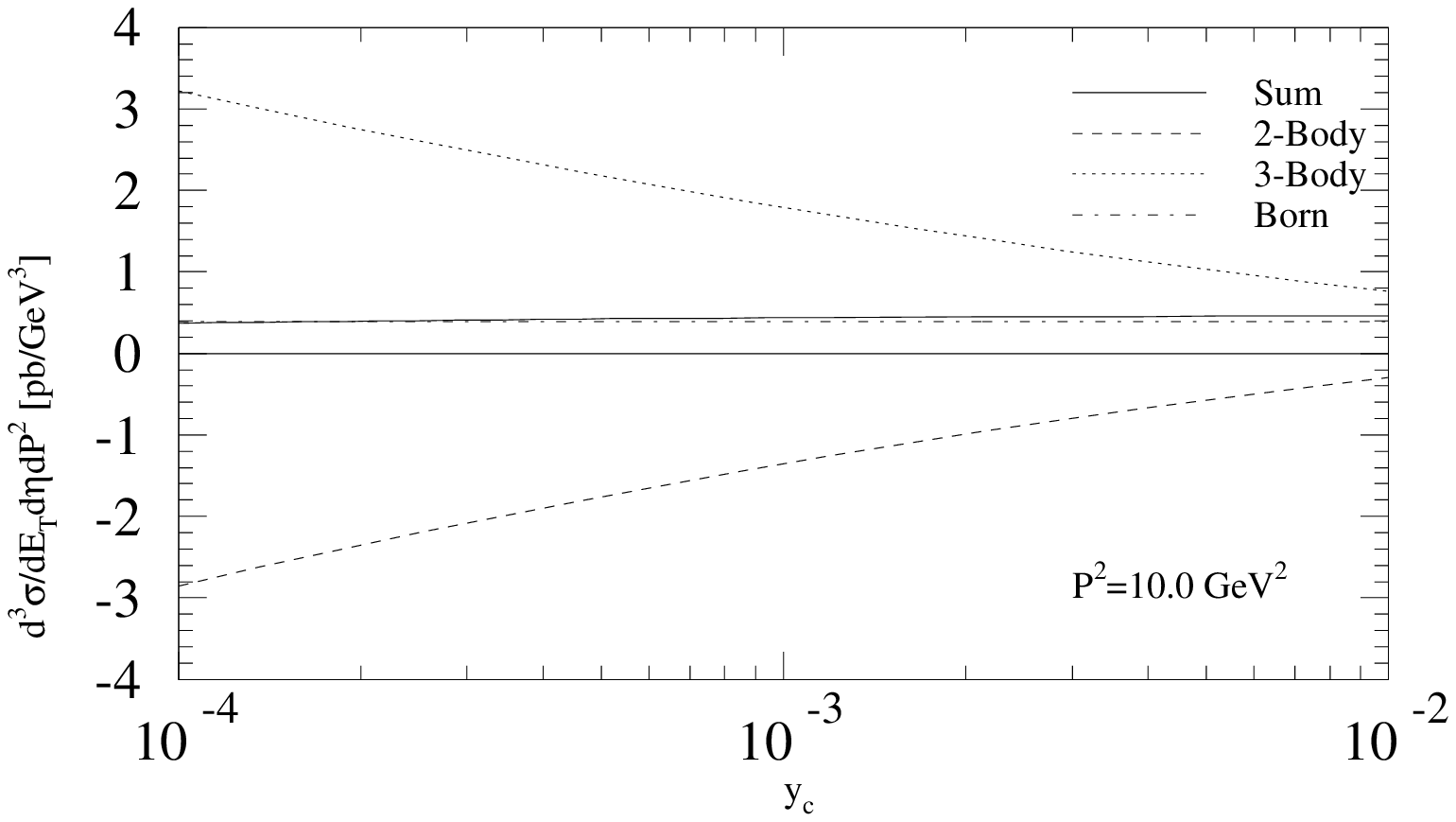,width=11cm}}
    \put(62,105){\footnotesize (a)}
    \put(62,46){\footnotesize (b)}
  \end{picture}
\caption{\label{cut}(a) Single-jet
        inclusive cross section integrated over the physical $\eta$ region
	for $E_T=20$ GeV and for the virtuality $P^2=0.058$ GeV$^2$ as
	a function of the slicing parameter $y_c$. The solid line gives
        the sum of the two- and the three-body contributions; (b)
	$P^2=10.0$ GeV$^2$.}
\end{figure}

Since the separation of the two-body and three-body contributions
with the slicing parameter $y_c$ is a purely technical device in order to
distinguish the phase space regions where the integrations are done
analytically from those where they are done numerically, the sum of
the two- and three-body contributions should be independent from
$y_c$. The dependence of the two-body contributions on the slicing
parameter is logarithmic, giving rise to $(\ln y_c)$- and
$(\ln^2y_c)$-terms. The parameter $y_c$ has to be
quite small to guarantee that the approximations in the
analytical calculations are valid. Typically, $y_c$ is of the order of
$10^{-3}$, forcing the two-body contributions to become negative,
whereas the three-body cross sections are large and positive. In Fig.\
\ref{cut} a, b we have checked for two different values of $P^2$, by
varying $y_c$ between $10^{-4}$ and $10^{-2}$, that the superimposed
two- and three-body contributions are independent of $y_c$ for the
inclusive single-jet cross sections integrated over the whole
kinematically allowed $\eta$ region for fixed $E_T=20$ GeV. Only the
SR contribution is tested, since the insensitivity of the DR
contributions on $y_c$ has been checked in \cite{6f, 18}. 

Furthermore, we have explicitly checked that the SR one- and two-jet 
cross sections for virtual photons are in perfect agreement with the
ones from real photoproduction given in \cite{6e, 6f} by integrating
the virtuality numerically over the region of small $P^2$ with
$P^2_{min}\le P^2 \le 4$ GeV$^2$. The main contribution to the cross
section comes from the lower integration boundary, where the
dependence of the matrix elements on $P^2$ is small.

Both, the $y_c$-dependence test and the comparison with the results from
\cite{6e, 6f}, give us confidence that our computer program for the
calculation of jet cross sections in electron-proton scattering yields
reliable results. It is interesting now to study the scale dependences
of the LO and NLO cross sections. The relevant scales are the
renormalization scale $\mu$, the factorization scale for the virtual
photon $M_\g$ and that for the proton $M_p$. Since the dependence on
these scales should vanish in an all-order calculation, we expect the
dependences to be reduced by going from LO to NLO.

\begin{figure}[hhh]
  \unitlength1mm
  \begin{picture}(122,115)
    \put(8,-3){\epsfig{file=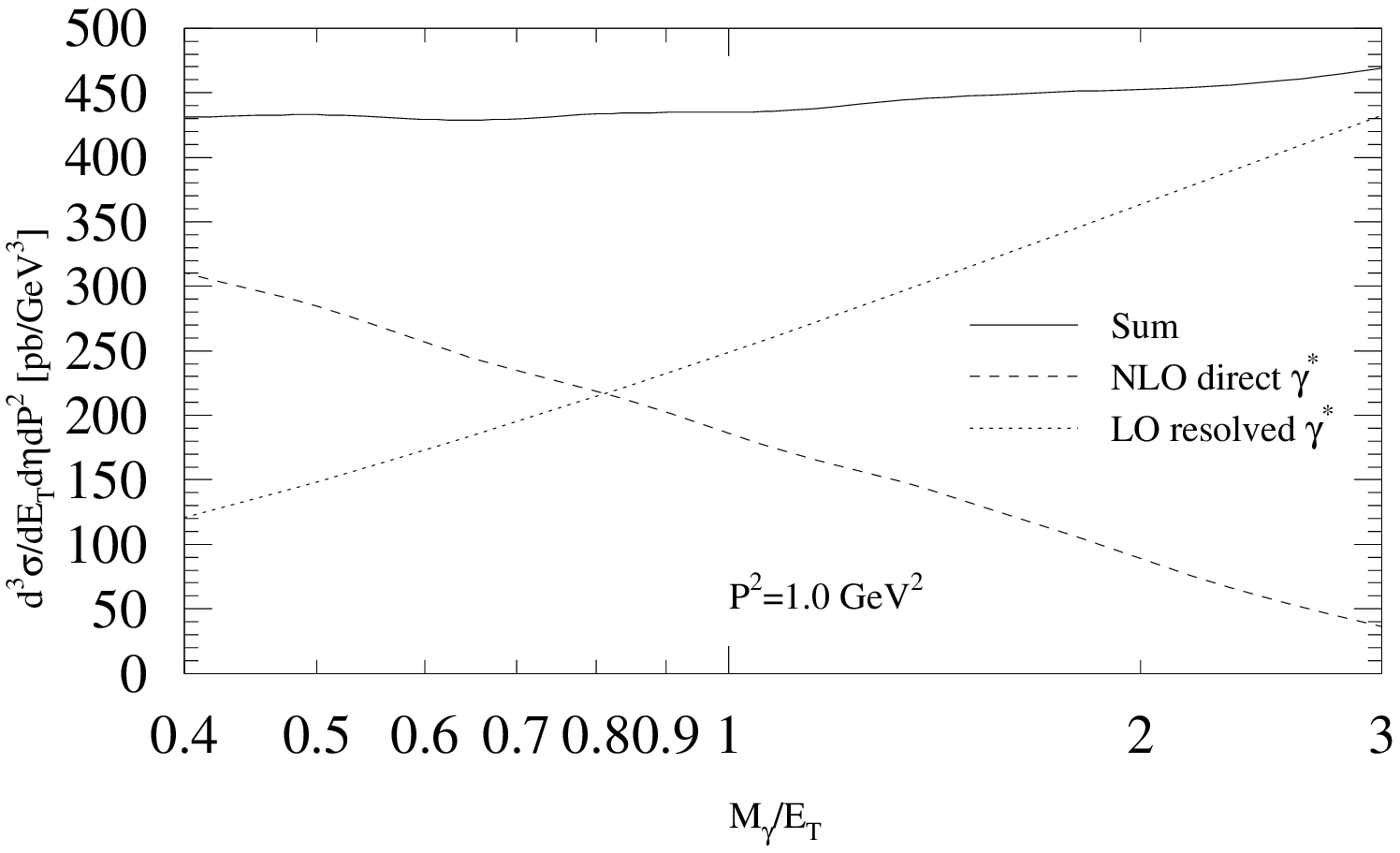,width=11cm}}
    \put(8,-61){\epsfig{file=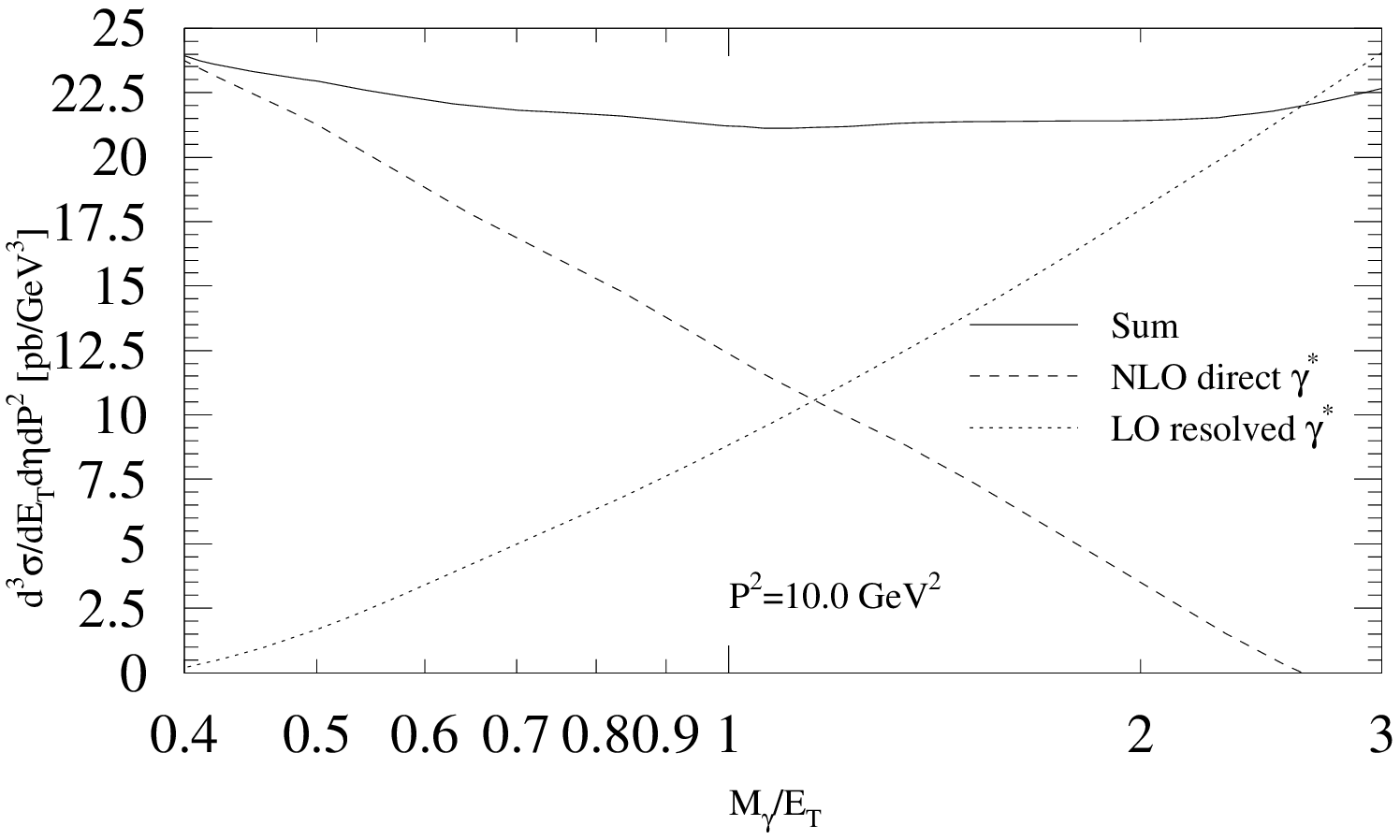,width=11cm}}
    \put(62,108){\footnotesize (a)}
    \put(62,49){\footnotesize (b)}
  \end{picture}
\caption{\label{mgam}(a) Single-jet
        inclusive cross section integrated over $\eta =[-1.875,1.125]$
	for $E_T=7$ GeV and for the virtuality $P^2=1.0$ GeV$^2$ as
	a function of the scale parameter $M_\g/E_T$. The
	$\overline{\mbox{MS}}$-SaS1M parametrization with $N_f=4$ is
	chosen. The solid line gives the sum of the NLO direct and LO
	resolved virtual photon cross sections; (b) $P^2=10.0$ GeV$^2$.}
\end{figure}

Of special interest is the dependence of the cross section on the
factorization scale $M_\g$, which comes from the factorization of the
photon initial state singularities. The dependence is logarithmic,
since terms proportional to $\ln (M_\g^2/P^2)$ have been
subtracted from the NLO cross section for the direct virtual photon,
as indicated in equation (\ref{llog}). The dependence of
the NLO direct single-jet inclusive cross section, integrated over the
region $\eta\in [-1.875,1.125]$ for $E_T=7$ GeV, on the parameter
$M_\g /E_T$ for two different values of $P^2$ is shown in Fig.\
\ref{mgam} a, b as the dashed curve. It is compared to the resolved
virtual photon contribution in LO, which gives the dotted curve. 
It is sufficient to use the LO matrix elements, since the main $M_\g$
dependence of the resolved contribution stems from the 
dependence of the photon PDF on the negative logarithm $-\ln
(M_\g^2/P^2)$. For the comparison we used the SaS1M
parametrization of the virtual photon, which is given in the
$\overline{\mbox{MS}}$ scheme.  As one can see from the
Fig. \ref{mgam} a, b, the dependences on the logarithms of the direct
and the resolved contributions cancel rather well in the sum.

\begin{figure}
  \unitlength1mm
  \begin{picture}(122,115)
    \put(8,-3){\epsfig{file=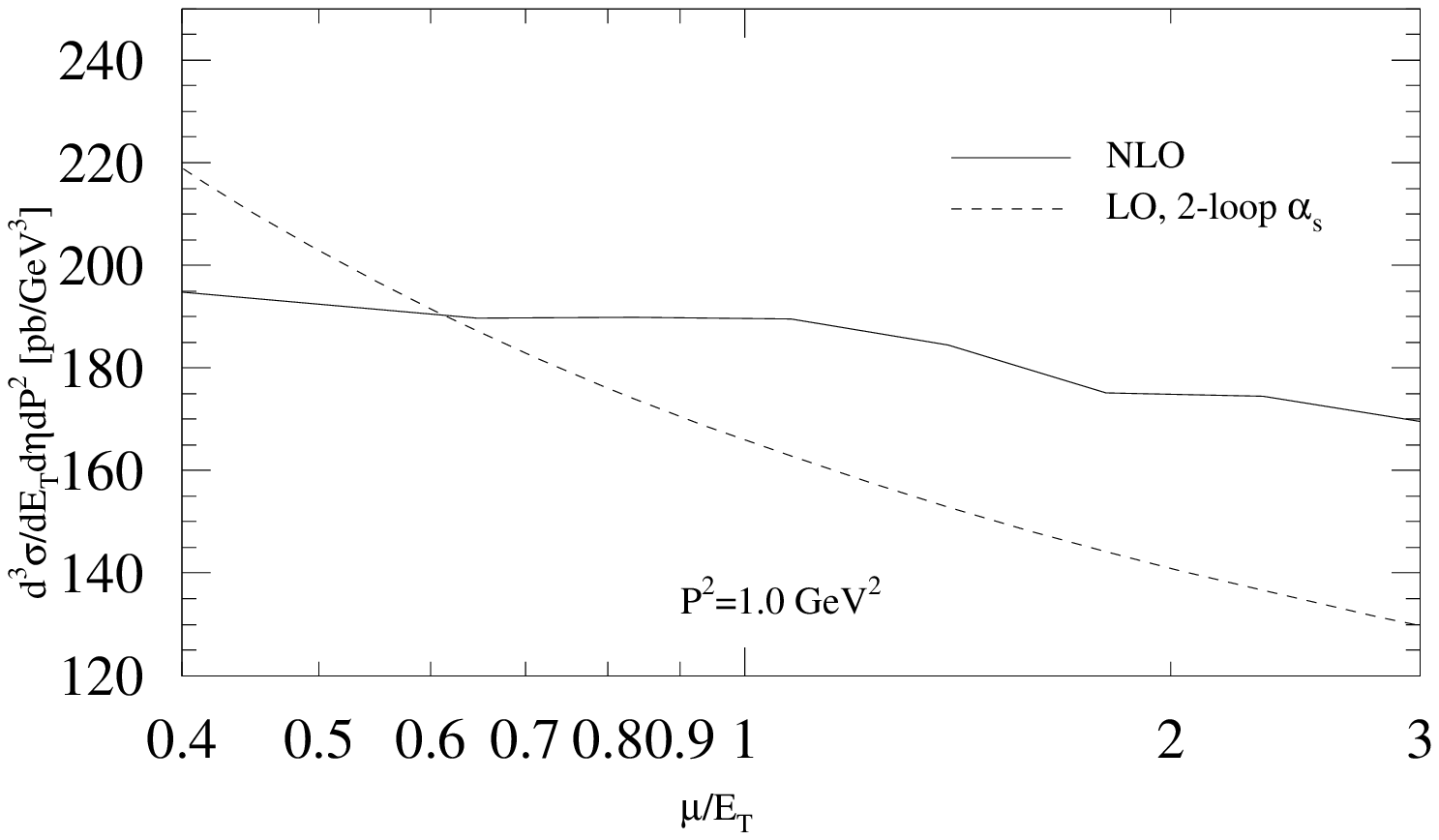,width=11cm}}
    \put(8,-61){\epsfig{file=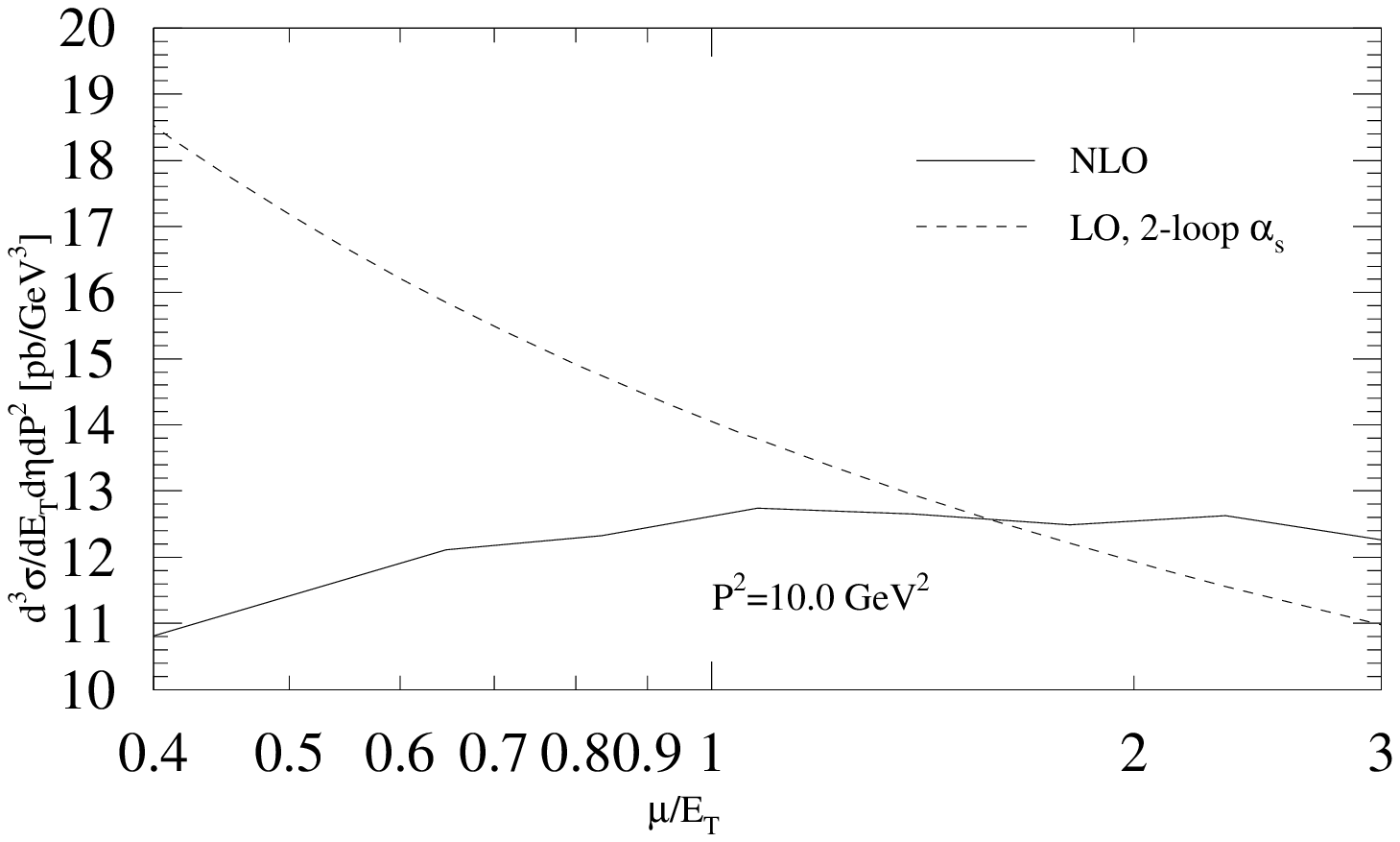,width=11cm}}
    \put(62,108){\footnotesize (a)}
    \put(62,49){\footnotesize (b)}
  \end{picture}
\caption{\label{mm}(a) Single-jet
        inclusive cross section integrated over $\eta =[-1.875,1.125]$
	for $E_T=7$ GeV and for the virtuality $P^2=1.0$ GeV$^2$ as
	a function of the scale parameter $\mu/E_T$. The
	$\overline{\mbox{MS}}$-SaS1M parametrization with $N_f=4$ is
	chosen. The solid line gives the NLO direct prediction,
	whereas the dashed curve shows the LO cross
	section; (b) $P^2=10.0$ GeV$^2$.}
\end{figure}

Finally, we study the renormalization scale dependence of the NLO
corrections as compared to the LO cross section. We consider only the
SR contributions, since the DR have been tested and shown to have a
reduced scale dependence in \cite{6f, 18}. In Fig.\ \ref{mm} a, b we have
plotted the one-jet inclusive cross section integrated over 
$\eta\in [-1.875,1.125]$ for $E_T=7$ GeV and $P^2=1, 10$ GeV$^2$. 
We have used the two-loop formula for $\al_s$ also in the LO
calculation for better comparison. For $P^2=1$ GeV$^2$ the scale
dependence of the NLO curve is reduced considerably. At $P^2=10$ GeV$^2$
though, the NLO cross section falls off slightly for the smaller
scales below $\mu\simeq E_T$. This could be attributed to the fact,
that the scale and the virtuality are of the same order in this region
and the condition $P^2\ll Q^2$ required in the construction of the
virtual photon PDF begins to become violated.  Actually, for a
full test of the renormalization scale dependence it would be
necessary to vary all scales $\mu =M_\g =M_p$ simultaneously, since
also the structure functions are renormalization scale dependent. It
is an empirical fact, that the scale dependence of the proton structure
function is small. The dependence of the photon structure function on
the renormalization scale is large, but this can only be accounted for
by the resolved contributions. The reduced renormalization scale
dependence of the sum of the direct and resolved contributions has
already been demonstrated in \cite{6f, 18}.

\subsection{Single-Jet Inclusive Cross Sections}

In this section, we present numerical results for inclusive one-jet
cross sections as a function of the virtuality $P^2$. We choose the
following notation of the curves as to make the discussion
clearer: the SR cross sections shall be denoted as {\em Dir} (reminding
of the direct character of the virtual photon), whereas the DR cross
sections are labeled {\em Res}. In addition, the sum of the SR and DR
contributions is shown and labeled {\em Sum}.  Note, that in the case
of electron-proton scattering the D component does not exist. As has
been calculated in section 3, large logarithmic contributions occur
for small photon virtualities for the, direct virtual photon that can
be subtracted and absorbed into the PDF of the virtual photon. The SR
cross sections, where these logarithmic terms have been subtracted,
are specified by the index $s$, giving the contributions {\em Dir$_s$}.
All plots in this section are taken from ref.\ \cite{10}.

\begin{figure}
  \unitlength1mm
  \begin{picture}(122,170)
    \put(8,55){\epsfig{file=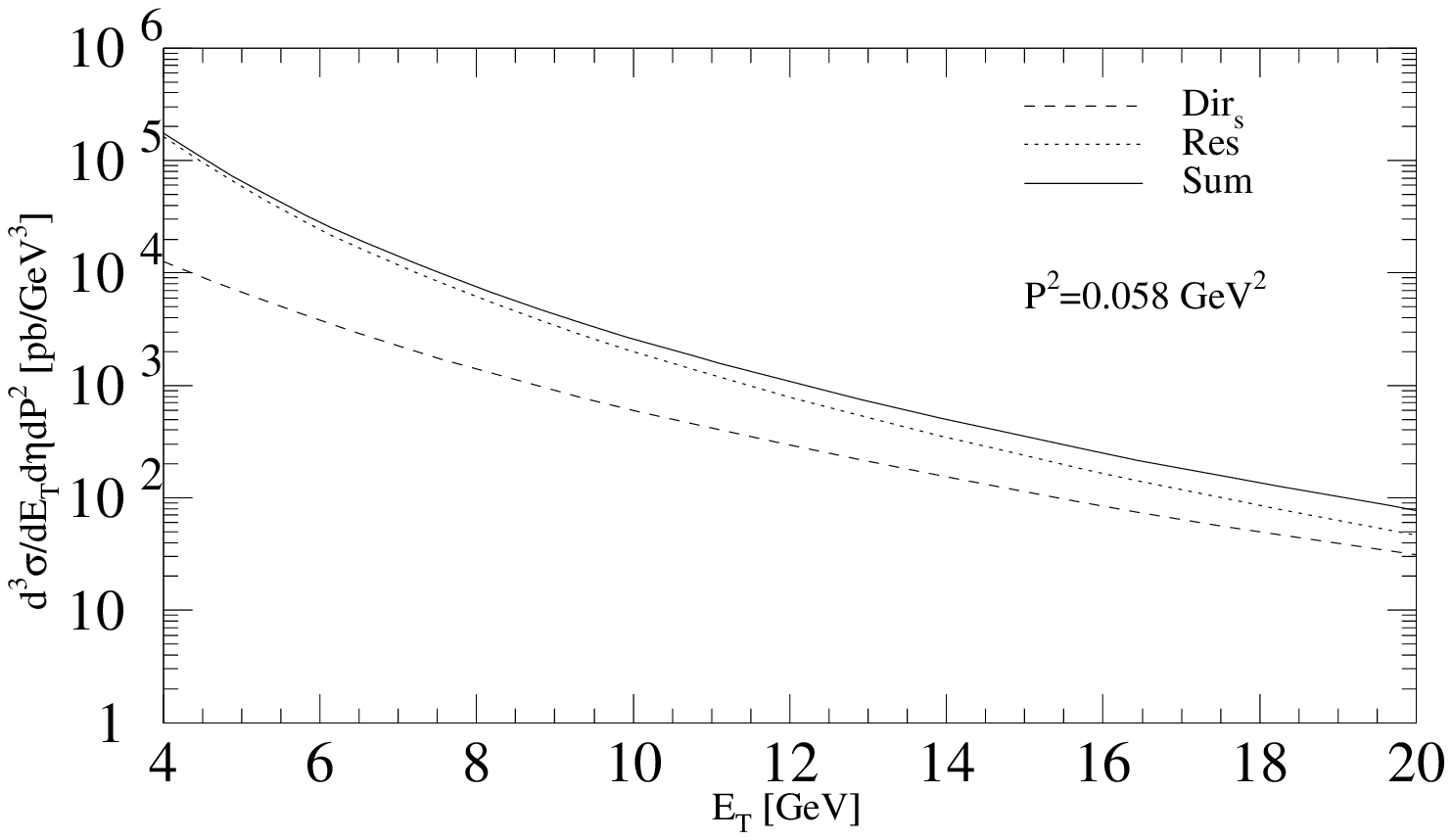,width=11cm}}
    \put(8,-3){\epsfig{file=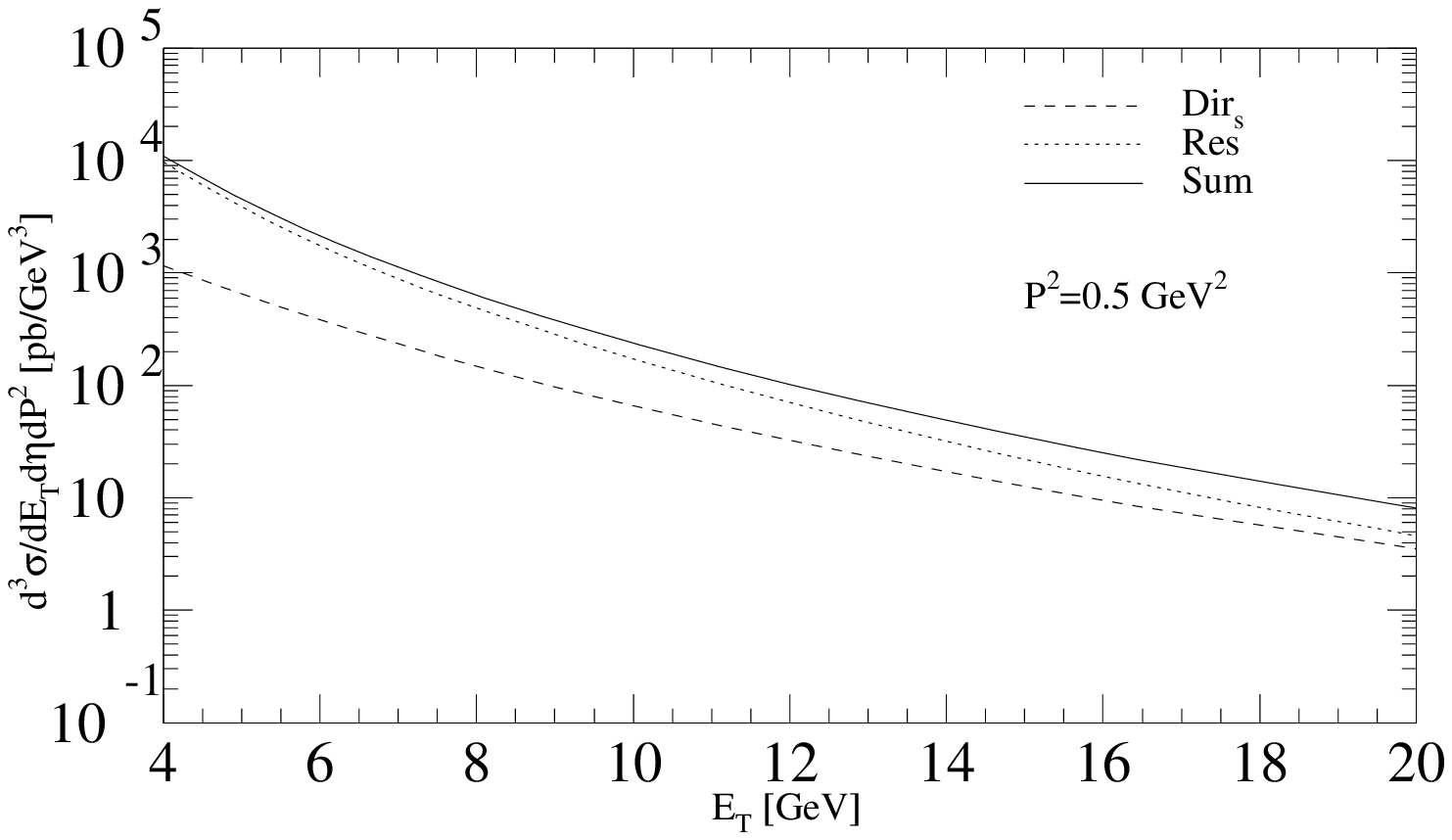,width=11cm}}
    \put(8,-61){\epsfig{file=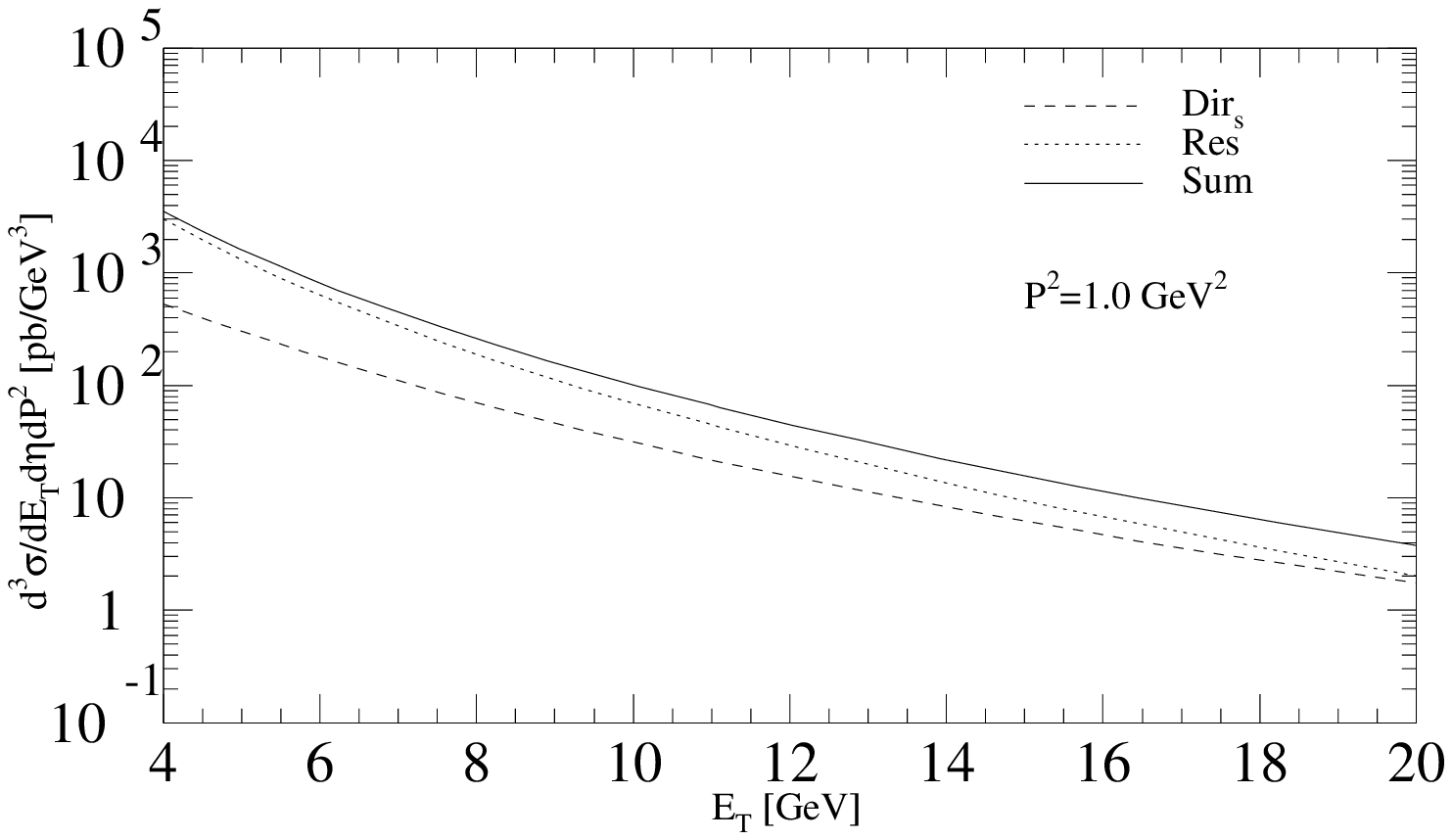,width=11cm}}
    \put(62,165){\footnotesize (a)}
    \put(62,108){\footnotesize (b)}
    \put(62,49){\footnotesize (c)}
  \end{picture}
\caption{\label{1}(a) Single-jet
        inclusive cross section integrated over $\eta \in
        [-1.125,1.875]$ for the virtuality $P^2=0.058$ GeV$^2$. The
        $\overline{\mbox{MS}}$-GRS
        parametrization with $N_f=3$ is chosen. The solid line gives
        the sum of the subtracted direct and the resolved term; (b)
        $P^2=0.5$ GeV$^2$; (c) $P^2=1.0$ GeV$^2$}
\end{figure}

We first concentrate on predictions with the PDF's of GRS. In Fig.\
\ref{1} a, b, c, the results for $d^3\si /dE_Td\eta dP^2$ are shown
as a function of $E_T$ integrated over $\eta$ in the interval $-1.125
\le \eta \le 1.875$, which are the boundaries employed in the ZEUS
analysis \cite{7}. We show results for the the three values of 
$P^2 =0.058,0.5$ and $1$ GeV$^2$. For all three $P^2$ the cross
section is dominated by the Res component at small $E_T$. Near
$E_T=20$ GeV the Dir$_s$ contribution is of the same magnitude as the
Res cross section. The sum of the cross sections as a function of
$P^2$ falls off nearly uniformly in the considered $E_T$ range with
increasing $P^2$. This decrease is stronger for smaller $E_T$.

\begin{figure}
  \unitlength1mm
  \begin{picture}(122,170)
    \put(8,55){\epsfig{file=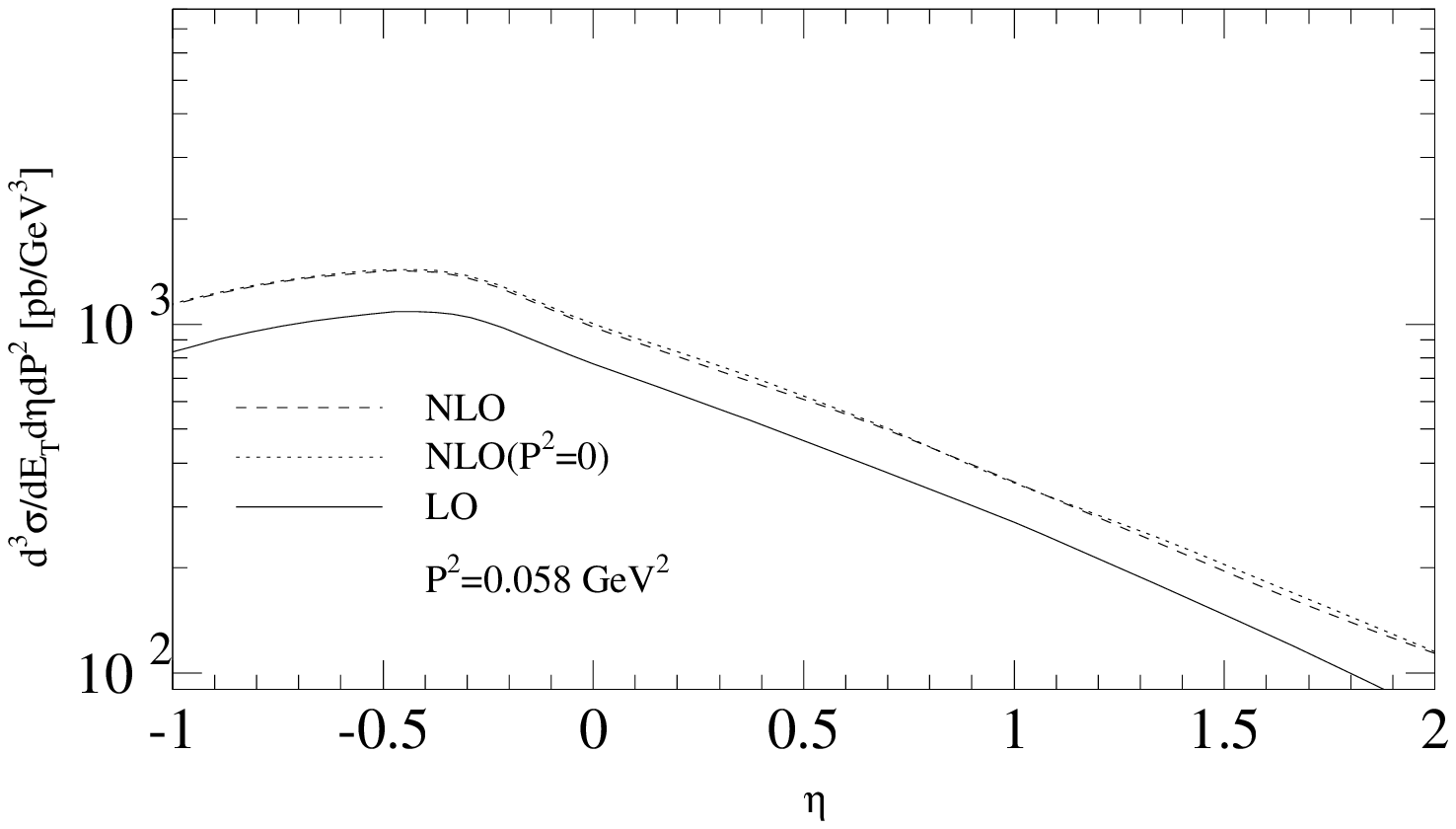,width=11cm}}
    \put(8,-3){\epsfig{file=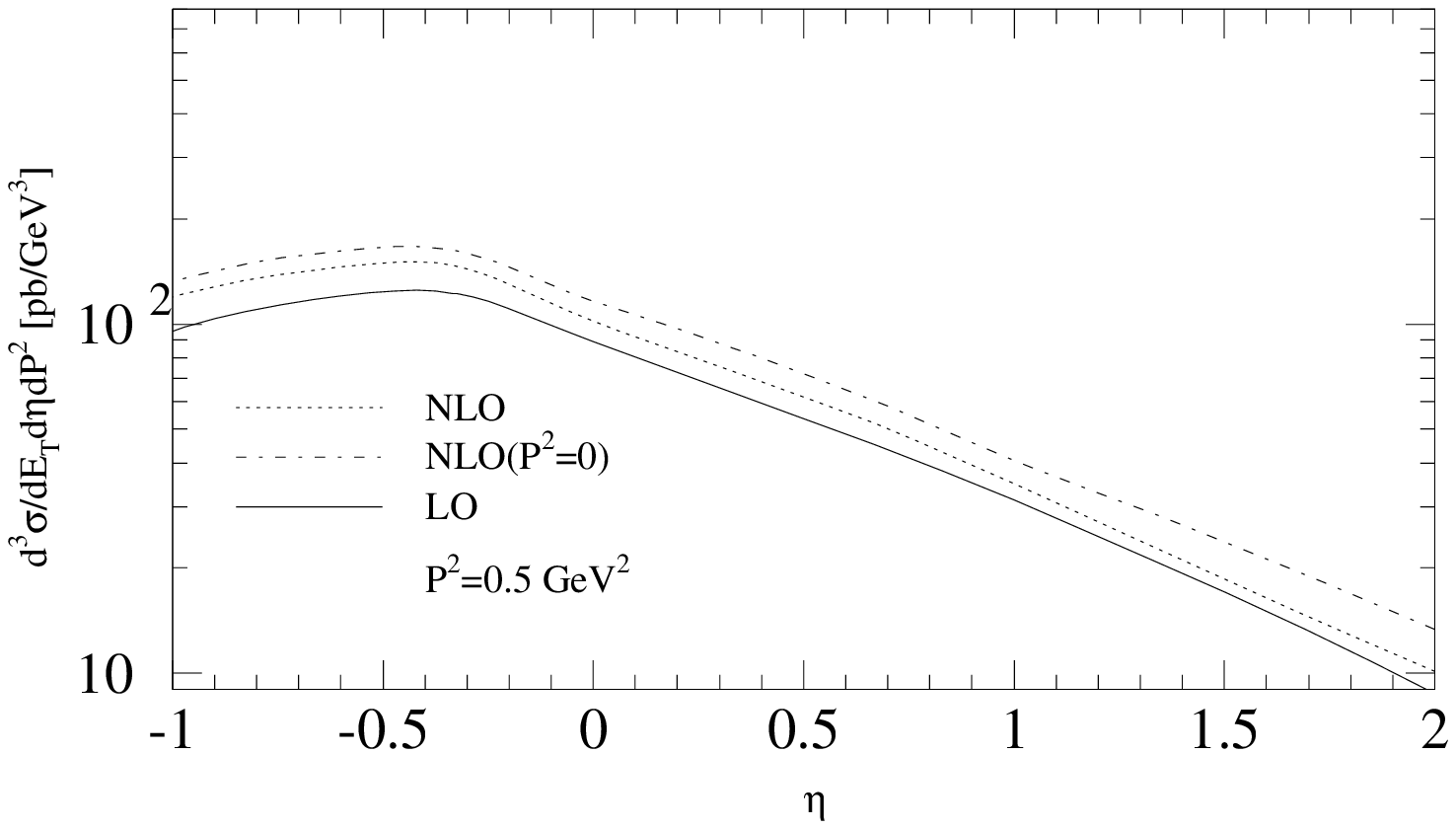,width=11cm}}
    \put(8,-61){\epsfig{file=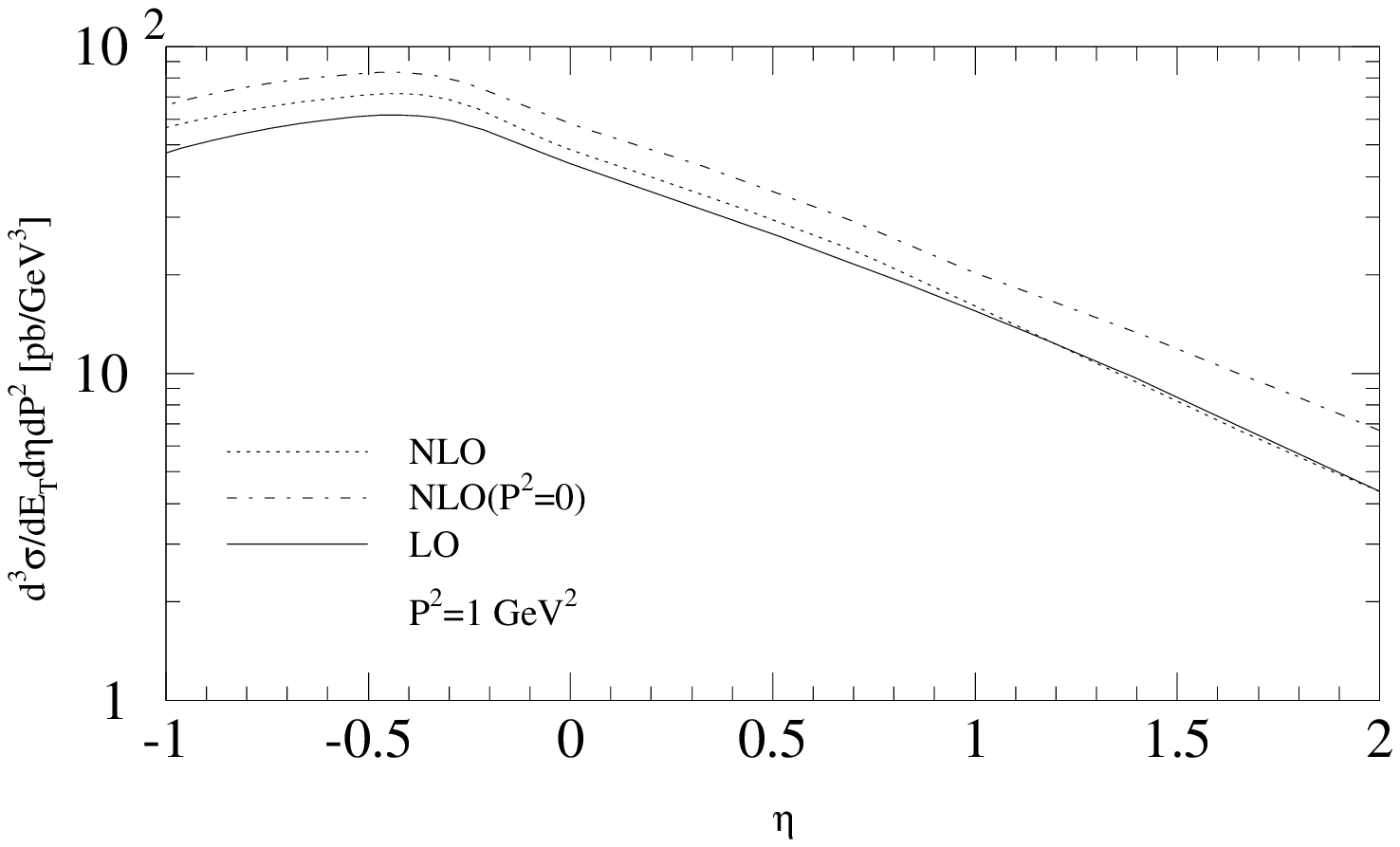,width=11cm}}
    \put(62,165){\footnotesize (a)}
    \put(62,108){\footnotesize (b)}
    \put(62,49){\footnotesize (c)}
  \end{picture}
\caption{\label{2}(a) Single-jet
        inclusive cross sections for $E_T=7$ GeV and $P^2=0.058$
        GeV$^2$. The $\overline{\mbox{MS}}$-GRS parametrization with
        $N_f=3$ is chosen. Only the SR part with subtraction
        (Dir$_s$) is  plotted. The solid line gives the LO
        contribution. The dashed curve is the full NLO cross section,
        whereas the dotted curve gives the NLO cross section,
        where the NLO matrix elements have no $P^2$-dependence;
	(b) $P^2=0.5$ GeV$^2$; (c) $P^2=1.0$ GeV$^2$.}
\end{figure}

Next, we studied the $\eta$ distribution of the Dir$_s$
contribution at fixed $E_T=7$ GeV and the same $P^2$ values as in
Fig.\ \ref{1}. The results are shown in Fig.\ \ref{2} a, b, c, where two
approximations are shown, namely the LO cross section and the NLO
cross section from \cite{6f}. There, $P^2=0$ everywhere, except for the
unintegrated Weizs\"acker-Williams approximation, which leads to a 
$1/P^2$ dependence. Obviously this approximation is good for
$P^2=0.058$ GeV$^2$. At the larger $P^2$ however it overestimates the
cross section and should not be used. This means that the $P^2$
dependence of the Dir$_s$ part, although the strongest logarithmic $P^2$
dependence has been subtracted, should be taken into account. In the
sum of the Dir$_s$ and Res cross sections the difference is small as long
as the Res part dominates. This holds for the smaller $E_T$'s. The LO
prediction is evaluated with the same structure functions and $\al_s$
value as the NLO result. It is smaller than the NLO result, which
it approaches with increasing $P^2$. Of course, this finding
depends on the chosen value of $R$ because the NLO cross section
depends on $R$, whereas the LO curve does not, as we have already
noted before. Estimates of the inclusive cross section with LO
calculations can therefore only be trusted for large cone radii. 

\begin{figure}
  \unitlength1mm
  \begin{picture}(122,170)
    \put(8,55){\epsfig{file=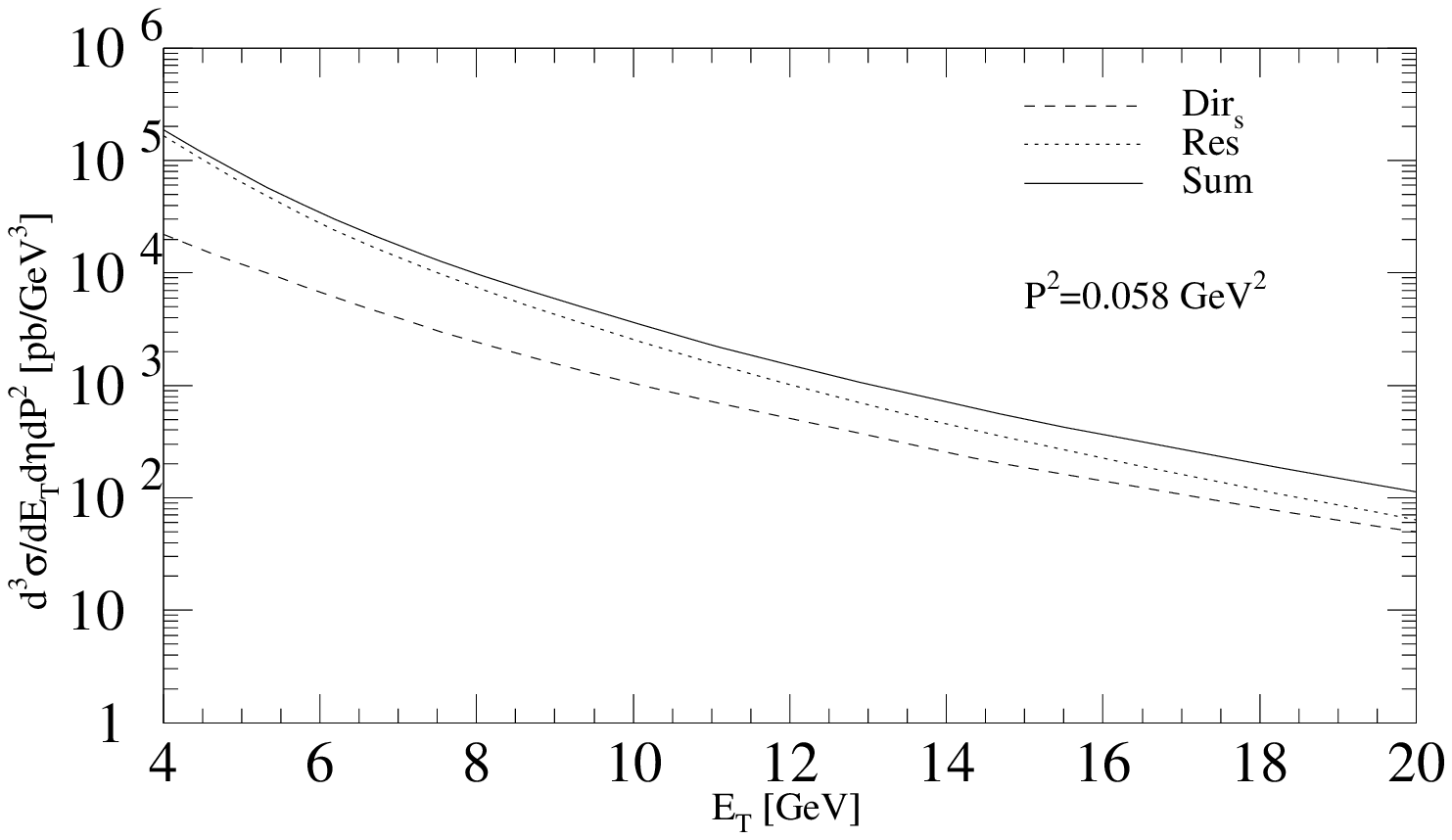,width=11cm}}
    \put(8,-3){\epsfig{file=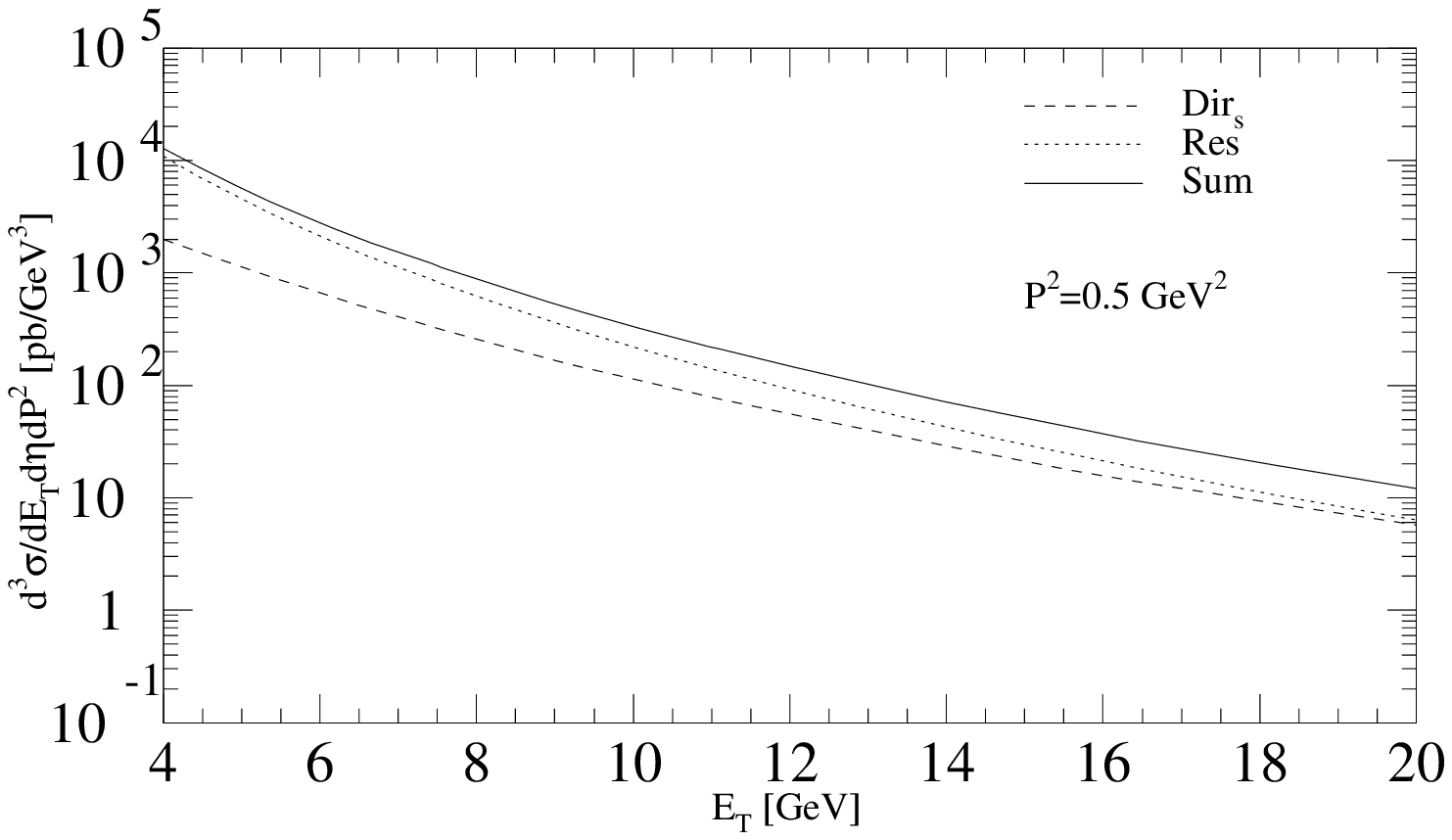,width=11cm}}
    \put(8,-61){\epsfig{file=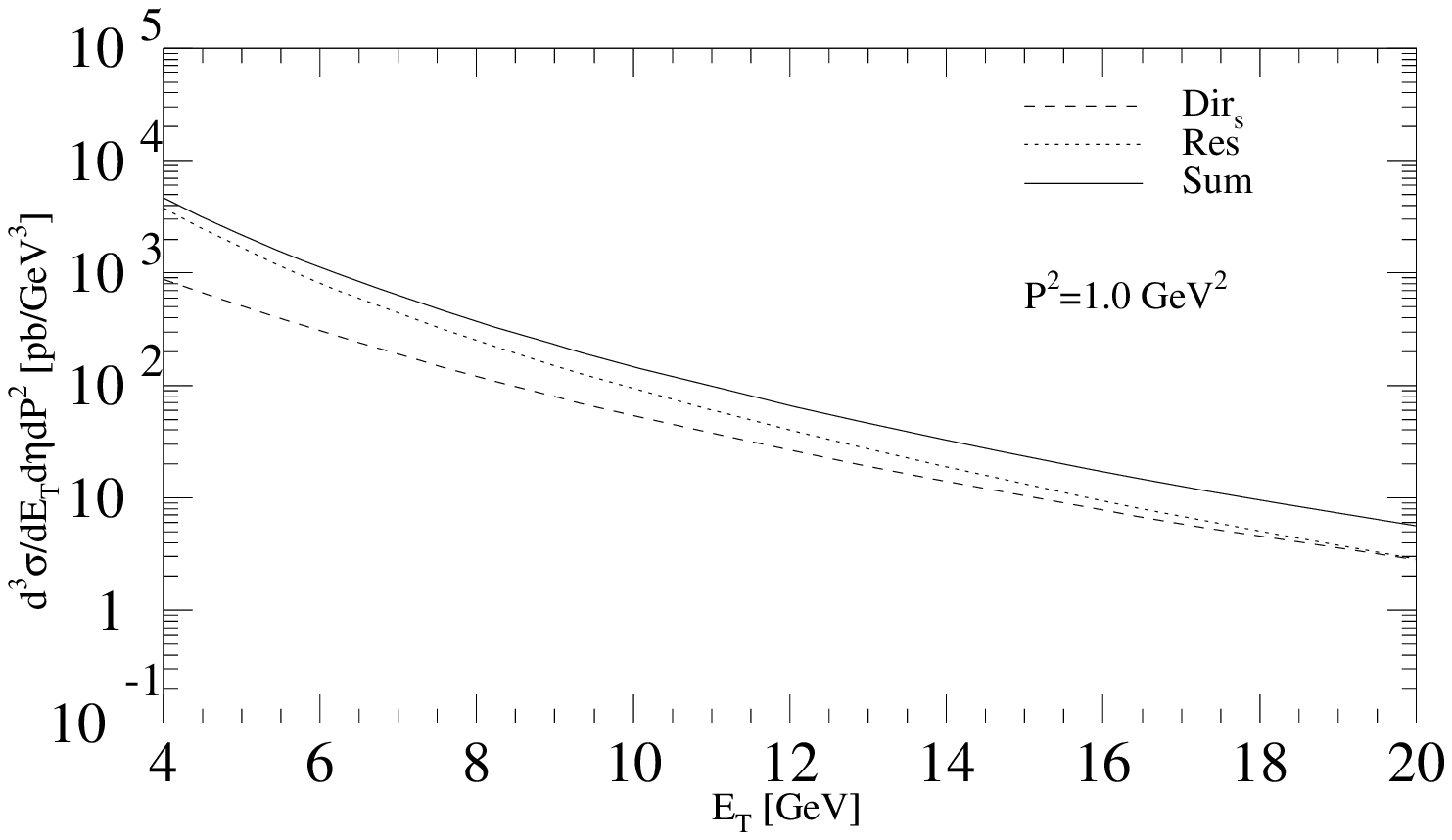,width=11cm}}
    \put(62,165){\footnotesize (a)}
    \put(62,108){\footnotesize (b)}
    \put(62,49){\footnotesize (c)}
  \end{picture}
\caption{\label{4}(a) Single-jet
        inclusive cross section integrated over $\eta \in
        [-1.125,1.875]$ for the virtuality $P^2=0.058$ GeV$^2$. The
        $\overline{\mbox{MS}}$-SaS1M
        parametrization with $N_f=4$ is chosen. The solid line gives
        the sum of the subtracted direct and the resolved term; 
	(b) $P^2=0.5$ GeV$^2$; (c) $P^2=1.0$ GeV$^2$.}
\end{figure}

The results shown so far are for a model with three flavors only and
therefore should not be compared to the experimental data except when
the contribution from the charm quark is added at least in LO. A more
realistic approach is to use the photon PDF's SaS1M \cite{22}
which are constructed for four flavors. In Fig.\ \ref{4} a, b, c results
are presented for $d^3\si /dE_Td\eta dP^2$ integrated over 
$\eta\in [-1.125,1.875]$ as a function of $E_T$ for $P^2=0.058, 0.5$
and $1.0$ GeV$^2$. We can compare these curves with the results in
Fig.\ \ref{1}a, b, c obtained with the PDF of GRS, where $N_f=3$. The sum of
the Dir$_s$ and Res contributions changes by 10\% to 30\% in the small
$E_T$ region and approximately 50\% in the large $E_T$ region. The
larger cross section for $N_f=4$ results mainly from the Dir$_s$
contribution. The direct component is more important for larger $E_T$ than
for smaller $E_T$. Therefore the increase is stronger in the large
$E_T$ region. 

\begin{figure}[hhh]
  \unitlength1mm
  \begin{picture}(122,110)
    \put(-3,-1){\epsfig{file=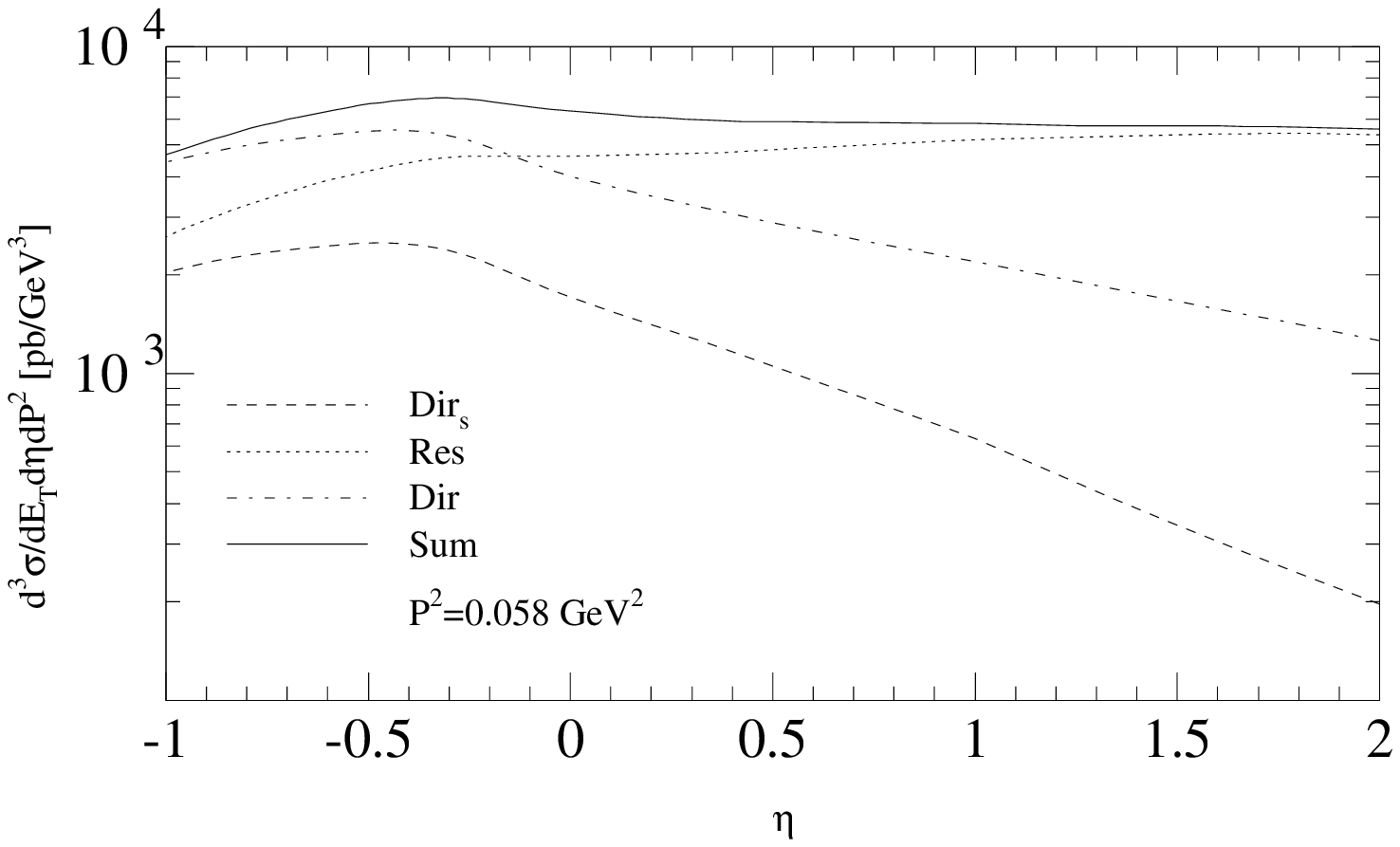,width=7cm,height=12cm}}
    \put(-3,-59){\epsfig{file=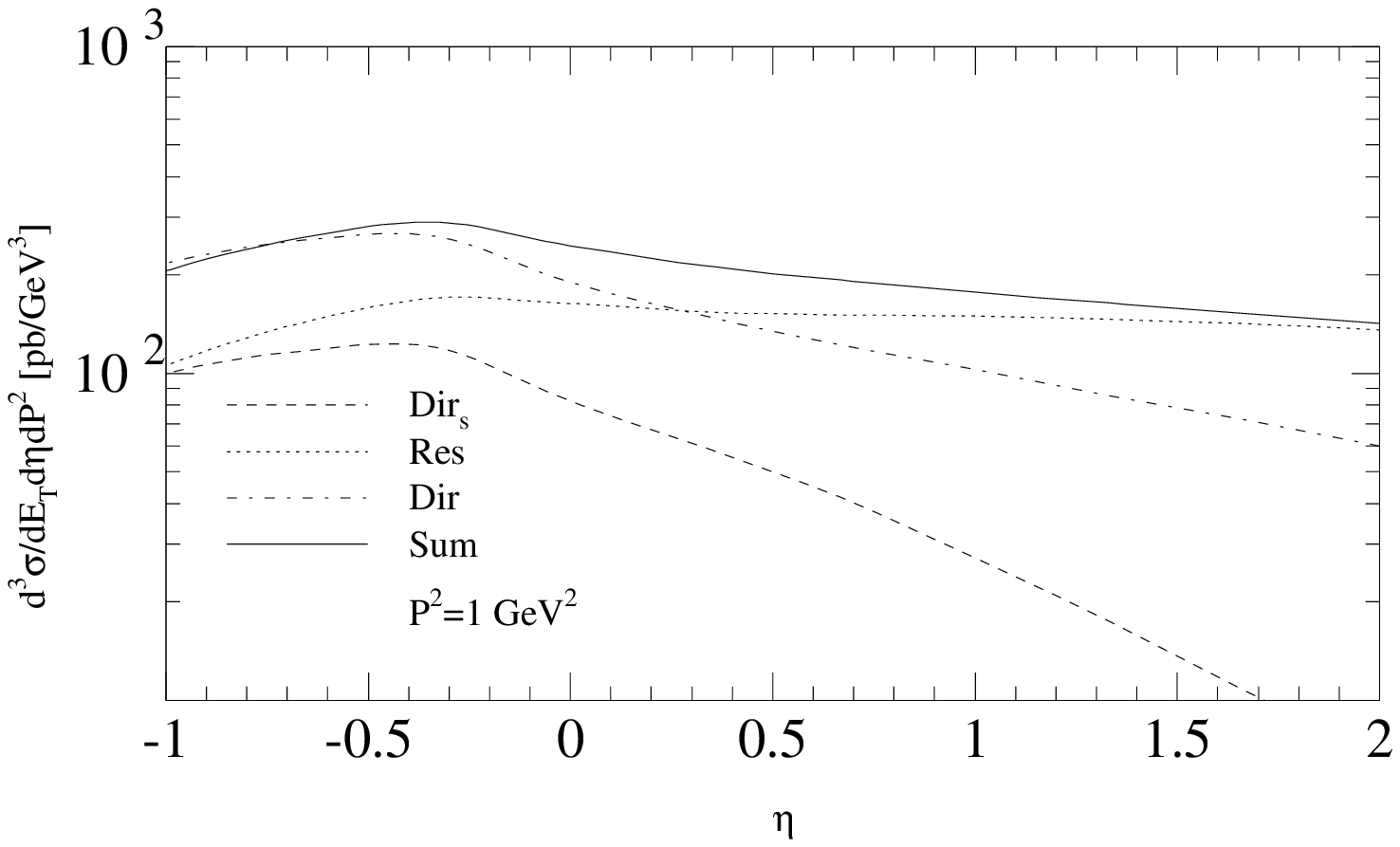,width=7cm,height=12cm}}
    \put(60,-1){\epsfig{file=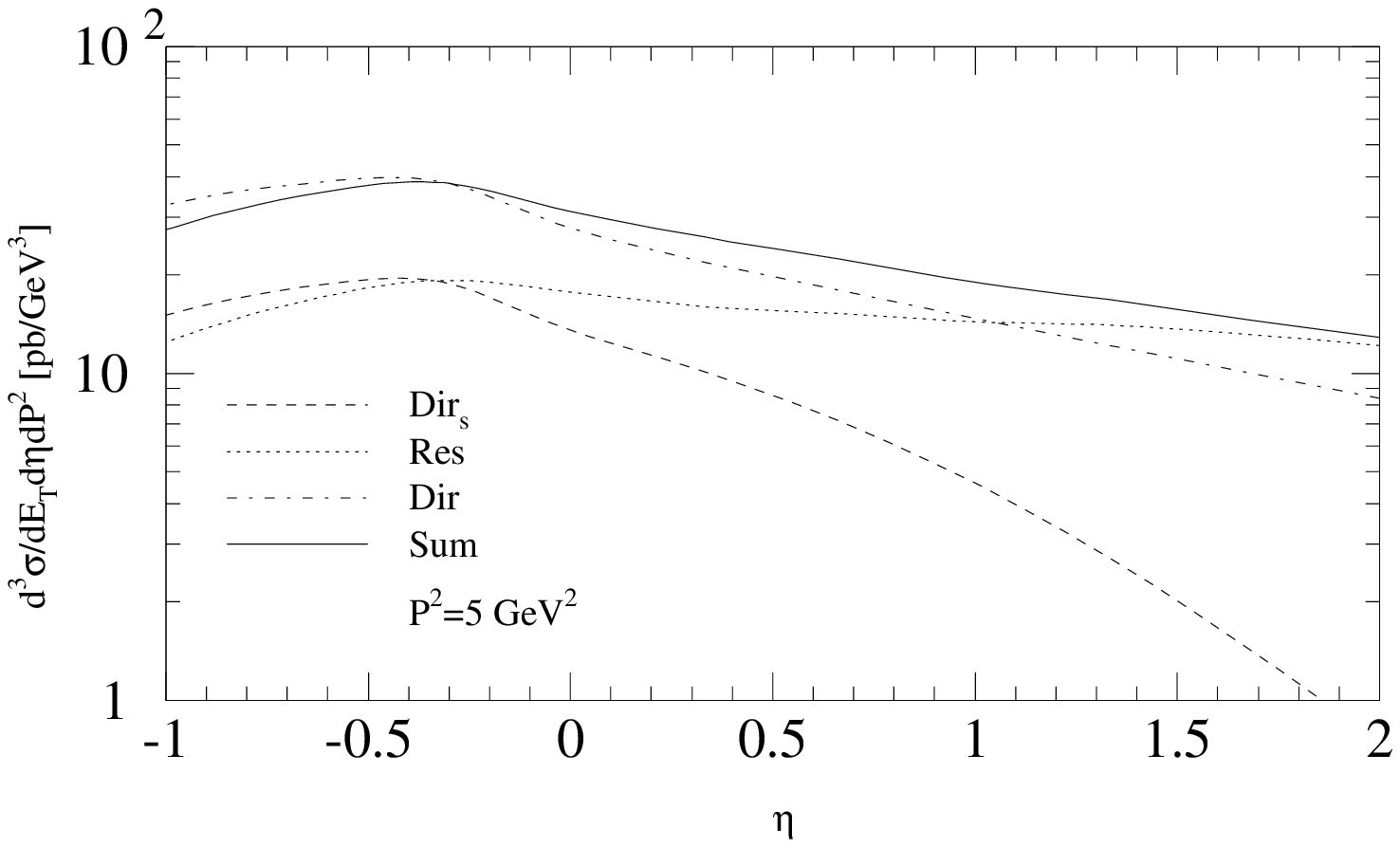,width=7cm,height=12cm}}
    \put(60,-59){\epsfig{file=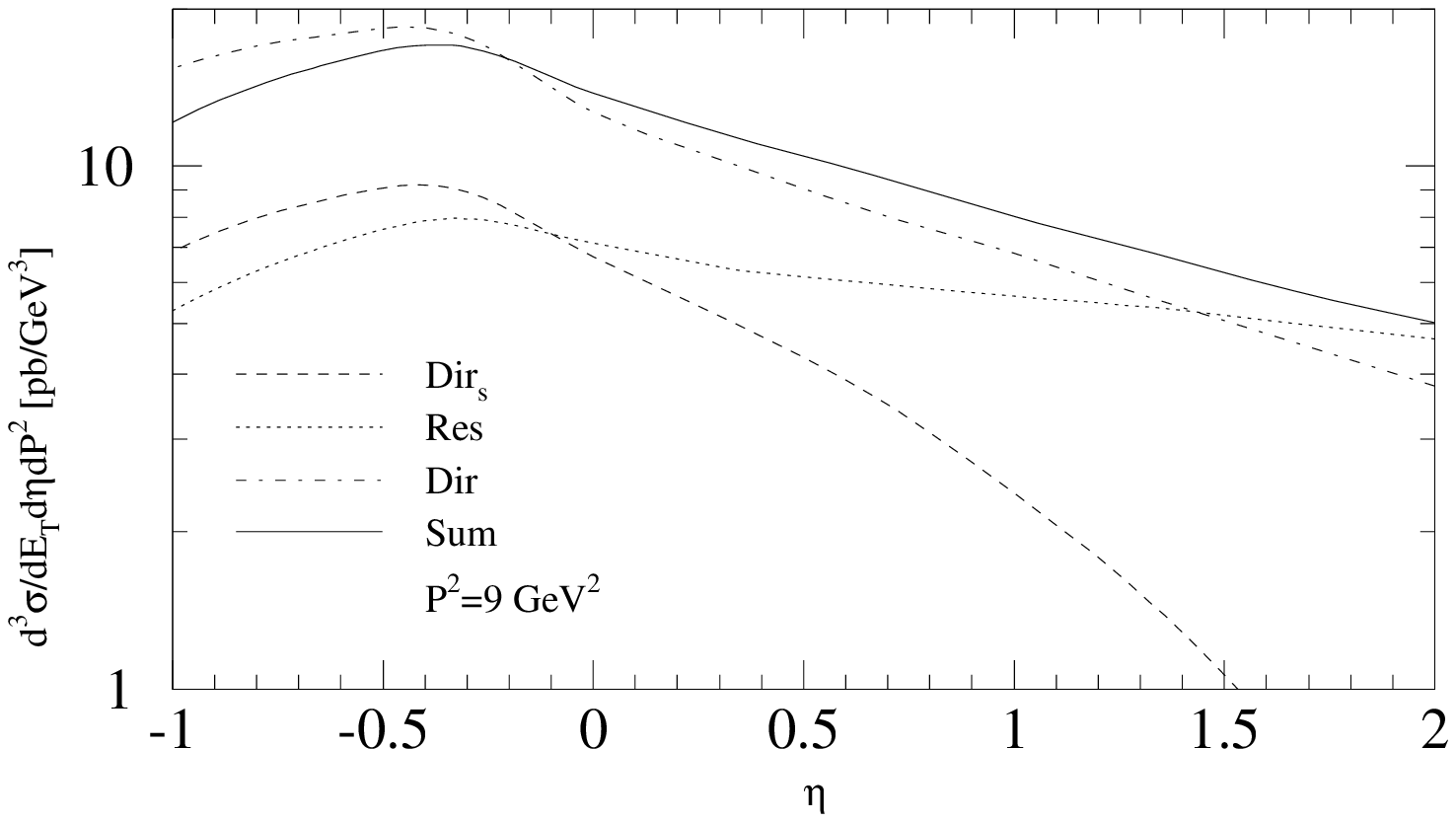,width=7cm,height=12cm}}
    \put(32,103){\footnotesize (a)}
    \put(32,44){\footnotesize (b)}
    \put(92,103){\footnotesize (c)}
    \put(92,44){\footnotesize (d)}
  \end{picture}
\caption{\label{5}(a) Comparisons of
        single-jet inclusive cross sections for $E_T=7$ GeV and
        the virtuality $P^2=0.058$ GeV$^2$. The
        $\overline{\mbox{MS}}$-SaS1M parametrization with $N_f=4$ is 
        chosen. The solid line gives the sum of the subtracted direct
        and the resolved term. The dash dotted curve is the direct
        contribution without subtraction; (b) $P^2=1$ GeV$^2$; 
	(c) $P^2=5$ GeV$^2$; (c) $P^2=9$ GeV$^2$.}
\end{figure}

Of interest are also the rapidity distributions for fixed $E_T$. These
are shown for $E_T=7$ GeV as a function of $\eta$ between $-1\le \eta
\le 2$ choosing $P^2=0.058,1,5$ and $9$ GeV$^2$ in Fig.\ \ref{5} a, b, c, d. 
We show the subtracted Dir$_s$ cross section, the Res cross section
and their sum. The Res component has its maximum shifted to positive
$\eta$'s in contrast to the Dir$_s$ component, as expected. The
Dir$_s$ component decreases quite rapidly with increasing $\eta$. This
stems from the subtraction of the $(\ln P^2/M_\g^2)$ terms as can be
seen by comparison with the unsubtracted cross section, denoted Dir, in
Fig.\ \ref{5} a, b, c, d. The sum of the resolved and subtracted direct cross
section Dir$_s$ is more or less constant for the smaller $P^2$ values and
decreases with increasing $\eta$ for $P^2=5$ and $9$ GeV$^2$. 

In section 4.3 we have checked by a direct comparison, that the 
subtraction term (\ref{llog}) approximates the PDF of the photon
rather well for large enough $P^2$. Thus, for these large virtualities
we expect the unsubtracted cross section (Dir) to be the correct one,
rather than the sum of the subtracted direct Dir$_s$ and the resolved
contributions, at least for small $\eta$. The larger $P^2$, the closer
does the full direct cross section Dir approach the sum Res+Dir$_s$,
as can be observed in Fig.\ \ref{5} a, b, c, d. As we have seen in
section 4.3, there still is a deviation of the pure perturbative
contribution from the evolved PDF in the small $x$ region. This
corresponds to the kinematic region of large $\eta$, which is the
forward direction of the proton. This deviation is evident in the
Fig.\ \ref{5} a, b, c, d as well; at $P^2=9$ GeV$^2$ the two cross
sections differ at $\eta =2$ by approximately 30 \%. Another difference
shows up in the backward direction at $\eta =-1$. In this region,
which corresponds to the region in the photon PDF where the
perturbative component dominates, no deviation is expected. The cause
might be the neglection of the transverse momentum $q_T$ of the
virtual photon in the calculation of the Dir and Res cross sections,
which becomes especially important for larger $P^2$. Actually, looking
at Fig.\ \ref{u-dis} b from section 4, the purely perturbative curve,
which will occur in the unsubtracted Dir component, overestimates the
up-distribution at $x=1$, which corresponds to the backward $\eta$
region. So, also in this region, the sum of the Dir$_s$ and the Res
components could still be a better estimate of the cross section, than
the Dir component alone. 

It is clear that the resolved and the direct cross sections decrease
with increasing $P^2$ for fixed $\eta$ and $E_T$. It is of interest to
know how the ratio of Res to the Dir cross section behaves
as a function of $P^2$. This has been analyzed in \cite{10}. Apart
from the fact that the ratios cannot be measured directly, we found a
strong dependence of the ratio on the scheme chosen for the 
photon PDF and very large corrections when going from LO to NLO. As
one can deduce from these results, it is not very sensible to compare
the Dir and Res contributions directly. Rather one has to introduce a
parameter that experimentally separates Dir and Res contributions. We
will introduce this parameter in the following section.

\subsection{Dijet Inclusive Cross Sections}

In comparison to single-jet cross sections, dijet cross sections
provide a much stronger test of QCD, since they depend on one variable
more. We will now present inclusive dijet cross
sections $d^4\si/dE_Td\eta_1d\eta_2dP^2$ as a function of $P^2$. The
variable $E_T$ is defined according to \cite{6, 6f} to be the
transverse momentum of the measured (trigger) jet, which has rapidity
$\eta_1$. The second rapidity $\eta_2$ is associated with the second
jet, where the two measured jets are those with highest $E_T$ in the
three-jet sample, i.e.\ $E_{T_1}, E_{T_2}>E_{T_3}$. 

\begin{figure}
  \unitlength1mm
  \begin{picture}(122,170)
    \put(8,55){\epsfig{file=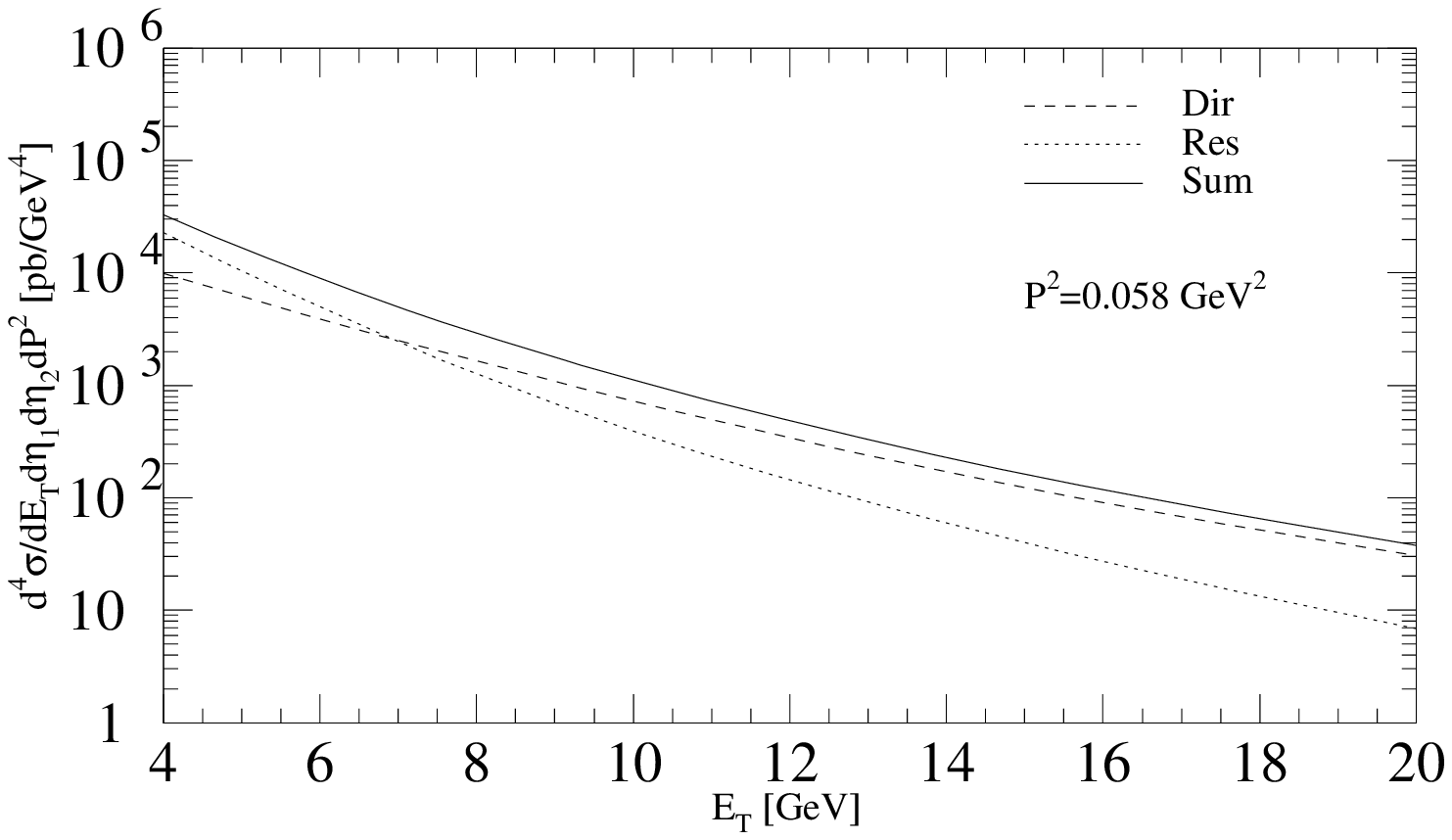,width=11cm}}
    \put(8,-3){\epsfig{file=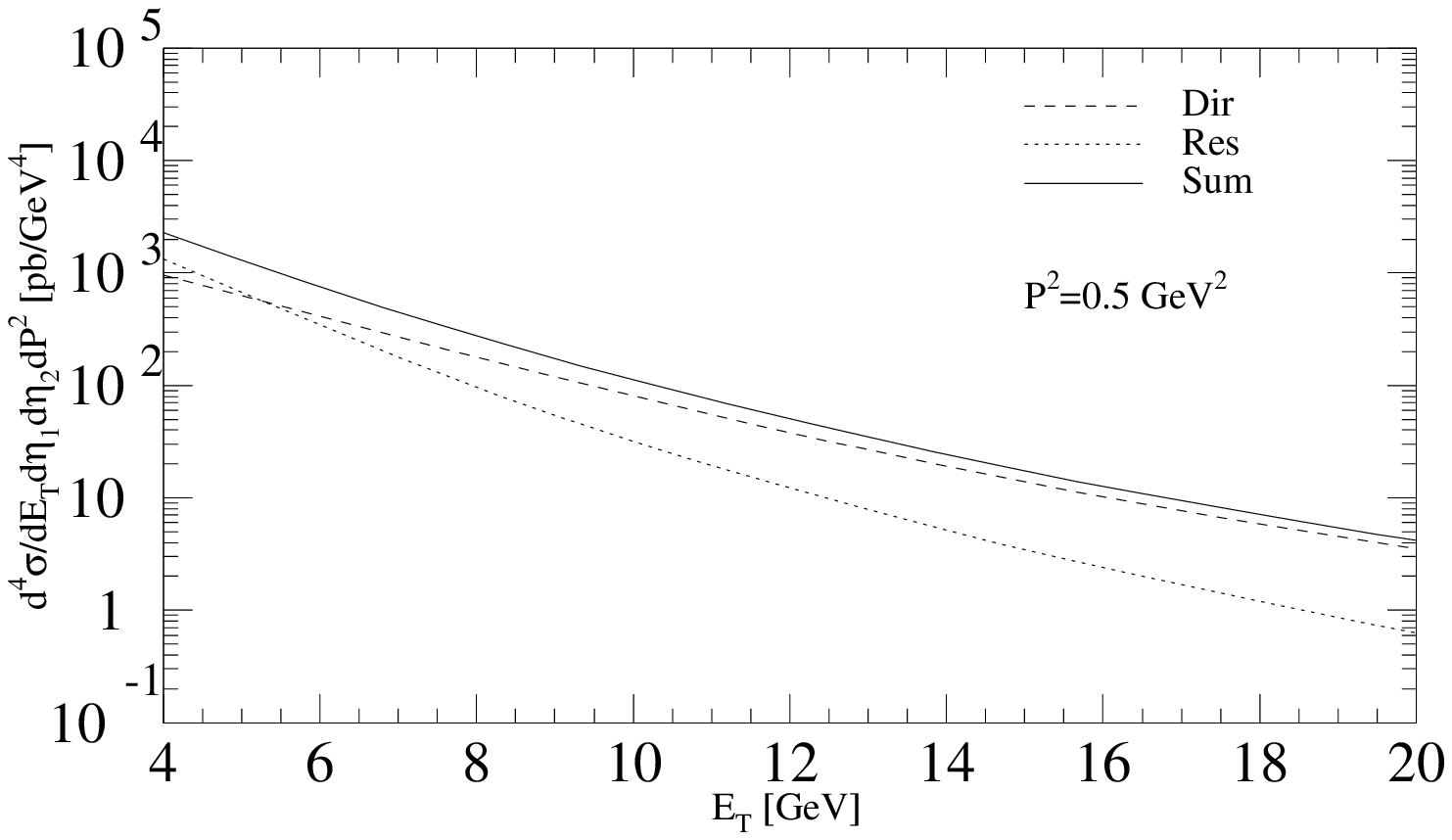,width=11cm}}
    \put(8,-61){\epsfig{file=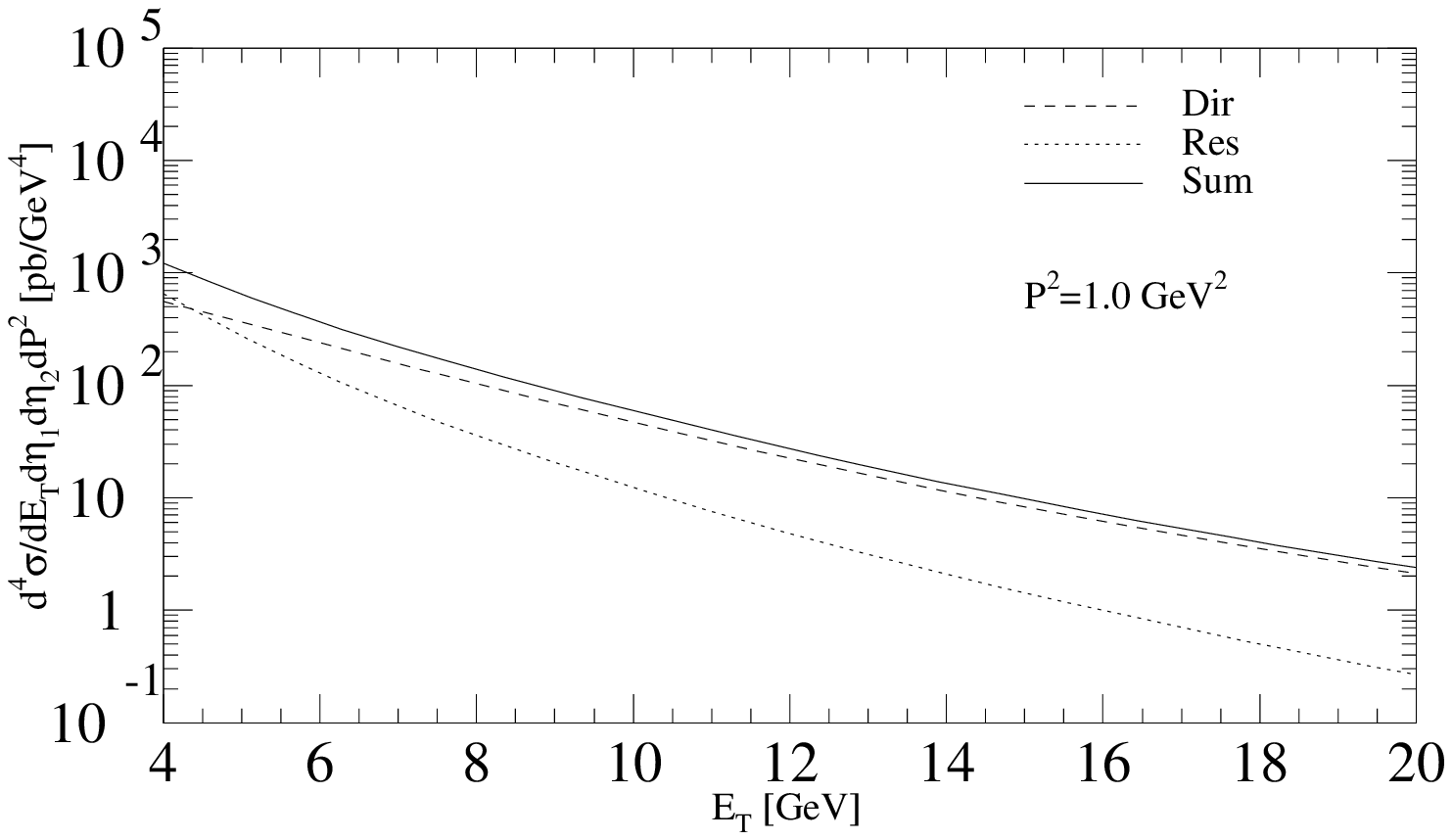,width=11cm}}
    \put(62,165){\footnotesize (a)}
    \put(62,108){\footnotesize (b)}
    \put(62,49){\footnotesize (c)}
  \end{picture}
\caption{\label{7}(a) Dijet inclusive
        cross section integrated over $\eta_1,\eta_2 \in
        [-1.125,1.875]$ for the virtuality $P^2=0.058$ GeV$^2$. The
        $\overline{\mbox{MS}}$-GRS parametrization with $N_f=3$ is
        chosen. The solid line is the sum of the direct and the
        resolved contribution. The dashed line is the direct-enriched
        contribution with $x_\gamma^{obs}>0.75$ and the dotted curve is
        the resolved enriched contribution with
        $x_\gamma^{obs}<0.75$; (b) $P^2=0.5$ GeV$^2$; (c) $P^2=1.0$
        GeV$^2$.} 
\end{figure}

In principle we could predict $\eta$ distributions
similar to those in \cite{6}. Since experimental data on these
distributions are not expected in the near future because of limited
statistics, we refrain from showing such plots here and present only the
$E_T$ distributions integrated over the interval
$-1.125<\eta_1,\eta_2<1.875$ following the constraints of the ZEUS
analysis \cite{7}. The results for $P^2=0.058, 0.5$ and $1.0$ GeV$^2$
are shown in Fig.\ \ref{7} a, b, c, where the full curve is given by 
the cross section 
$d^4\si/dE_Td\eta_1d\eta_2dP^2$ as a function of $E_T$ integrated over
$\eta_1$ and $\eta_2$ in the specified interval and for
$0.2<y<0.8$ (the plots in this section are taken from ref.\
\cite{10}). The functional dependence on $E_T$ does not change as a
function of $P^2$, only the absolute value of the cross section
decreases with increasing $P^2$.

Furthermore we show the so-called {\em enriched} direct and resolved
cross section in Fig.\ \ref{7}. These two contributions are defined
with a cut on the variable $x_\g^{obs}$, which is given by  
\equ{}{ x_\gamma^{obs} \equiv \frac{\sum_i E_{T_i}e^{-\eta_i}}{2yE_e}
  \quad , }
where the sum runs over the two highest $E_T$ jets. The variable
$x_\g^{obs}$ gives the fraction of the photon energy going into the two
measured jets. It is a good estimate of the theoretically defined
variable $x_a$, that defines the fraction of the photon momentum 
participating in the hard interaction, see Fig.\ \ref{fac1}. For
$x_a=1$, the photon couples directly to the subprocess, whereas for
$x_a<1$ some of the photon energy goes into the production of a
remnant jet, leading to resolved processes. Note, that in LO 
$x_a = x_\g^{obs}$. For experimental considerations, one defines the
direct enriched contribution for  $x_\g^{obs}>0.75$, whereas the
resolved enriched component has $x_\g^{obs}<0.75$. Both enriched cross
sections contain contributions from the direct and the resolved
part. In Fig.\ \ref{7} a, b, c the sum of the Dir and Res curves is
equal to the full cross section $d^4\si/dE_Td\eta_1d\eta_2dP^2$ with
no cut on $x_\gamma^{obs}$.  The curves in Fig.\ \ref{7} are for the GRS 
parton distributions in the $\overline{\mbox{MS}}$ scheme. As to be 
expected, with increasing $P^2$ the full cross section is more and
more dominated by the Dir component, in particular at the larger
$E_T$. This means that the cross section in $x_\gamma^{obs}<0.75$
decreases stronger with $P^2$ than in the $x_\gamma^{obs}>0.75$
region. This could be studied experimentally by measuring the ratio of
the two cross sections as a function of $P^2$ for fixed $E_T$. This
has not been done yet.

\begin{figure}[hhh]
  \unitlength1mm
  \begin{picture}(122,60)
    \put(8,-61){\epsfig{file=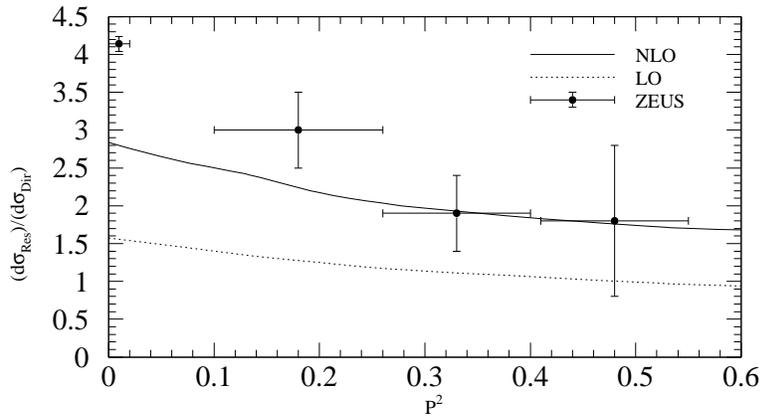,width=11cm}}
  \end{picture}
\caption{\label{77}The ratio of the resolved-enriched to the direct-enriched
        contributions as calculated in Fig.\ \ref{7} a, b, c, integrated over
        $E_{T_1},E_{T_2}>4$ GeV in LO (dotted) and NLO (full) for the
        SaS1M parametrization with $N_f=4$ compared with ZEUS data.}
\end{figure}

Instead, the ZEUS collaboration \cite{7} has presented data on the
ratio $r=\mbox{Res}/\mbox{Dir}$, where Dir and Res refer to the 
enriched direct and resolved cross sections. The ZEUS data in \cite{7}
has actually been obtained by integrating the 
transverse momenta of the two-jet cross sections over the region
$E_{T_1},E_{T_2}\ge 4$ GeV and the rapidities in the range 
$-1.125<\eta_1, \eta_2 < 1.875$ for various $P^2$-bins. 
With the integration cut on the transverse momenta of the two hardest
jets, the transverse momentum of the unobserved jet can vanish, which
is not IR safe in NLO QCD. We therefore allow the second jet to have
less than $4$ GeV if the third unobserved particle is soft (i.e.\ has a
transverse momentum of less than $1$ GeV) \cite{6}. Through this
procedure, a $y_c$ dependence is avoided. We calculated the ratio $r$
as a function of $P^2$ up to $P^2=0.6$ GeV$^2$ and compared it with the 
ZEUS \cite{7} data in Fig.\ \ref{77} in LO (dotted curve) and NLO (full
curve), using the SaS1M photon PDF with $N_f=4$ flavors. We find
quite good agreement of the NLO prediction with the data points for
$P^2\ge 0.25$ GeV$^2$. The curve deviates from the data for 
$P^2\simeq 0.2$ GeV$^2$, though, and even more for the point $P^2\simeq 0$,
which lies about 30\% above the prediction. Surely the photoproduction
data is much more precise then the other points shown in Fig.\
\ref{77}. For photoproduction it has been shown in \cite{6}, that the
measured enriched resolved component is larger than the predicted one
for a small cut in the transverse momentum. This has been attributed
to additional contributions from multiple interactions with the proton
remnant jet in the resolved cross section, which have not been
included in the NLO calculations. This underlying event contribution
is reduced for larger $E_T^{min}$ and for cone radii smaller than 1. These
problems must be present in the comparison shown here for the smallest
$P^2$ value as well. As it seems, the underlying event contribution is
also reduced by going to higher values of $P^2$. This could be studied
more directly by measuring rapidity distributions for the enriched
resolved $\g^*$ sample as was done for the photoproduction case \cite{6}.

\section{Photon-Photon Collisions}

In this section we predict inclusive jet rates for the case of
photon-photon collisions in kinematic regions that will become
available in LEP2 experiments. We first introduce the notation
and kinematics for the hadronic cross section and then discuss some
numerical results.

\subsection{Jet Production Cross Section and Kinematics}

To obtain a close correspondence between $ep$ scattering discussed in the
previous section and $\g\g^*$ scattering to be discussed here, we use 
similar notations. Thus, we deviate somewhat from the notation used in
section 4. We start from electron-positron scattering for
two-jet production, which may be written as
\equ{}{ e^+(k_1) + e^-(k_2) \to e^+(k_1')+e^-(k_2') +
    \mbox{Jet}_1(E_{T_1},\eta_1) +  
    \mbox{Jet}_2(E_{T_2},\eta_2) + X \quad .}
We assume, that the interaction of the electrons is processed via the
interaction of one quasi-real and one virtual photon, that are
radiated by the electron and positron, respectively. Thus, we consider
the subprocess
\equ{}{ \g^*_a(p) + \g_b(q) \to \sum\mbox(\mbox{Jet})_i + X \quad , }
with $p\equiv k_1-k_1'$ and $q\equiv k_2-k_2'$. The electron-positron
c.m.\ energy is given by $s_H=(k_1+k_2)^2$, whereas the $\g\g^*$
c.m.\ energy is given by $W^2=(p+q)^2$. Since both, the real and the
virtual photon, acquire some hadronic substructure through the resolved
processes, it is not quite clear which is the probing and which is the
probed photon. We therefore will not speak about target and probing
photon, but simply of the real and the virtual photon. We define the 
virtuality of the real photon as $Q^2\equiv -q^2$ with $Q^2\simeq 0$
and that of the virtual photon as $P^2\equiv -p^2$. Next, the
momentum fractions of the photon in the electron and positron
$y_a$  and $y_b$ have to be defined, which are given by 
\equ{}{ y_a\equiv \frac{pk_2}{k_1k_2} \qquad \mbox{and} \qquad 
	y_b\equiv \frac{qk_1}{k_1k_2}\simeq \frac{E_\g}{E_e} \quad . }
Here, $E_\g$ is the energy of the real photon and $E_e$ is the
electron energy in the $e^+e^-$ c.m.\ system. 

\begin{figure}[hhh]
\unitlength1mm
\begin{picture}(121,40)
\put(-25,-18){\psfig{file=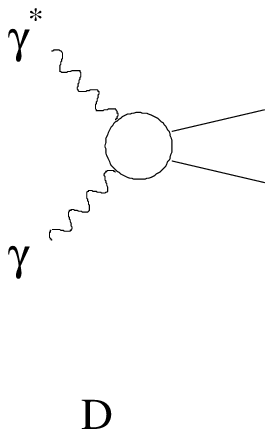,width=8cm}}
\put(5,-18){\psfig{file=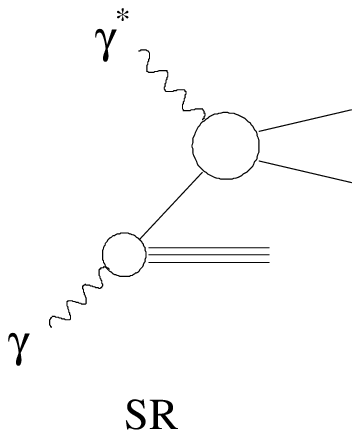,width=8cm}}
\put(40,-18){\psfig{file=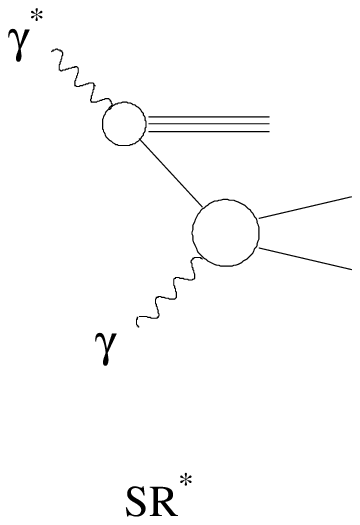,width=8cm}}
\put(75,-18){\psfig{file=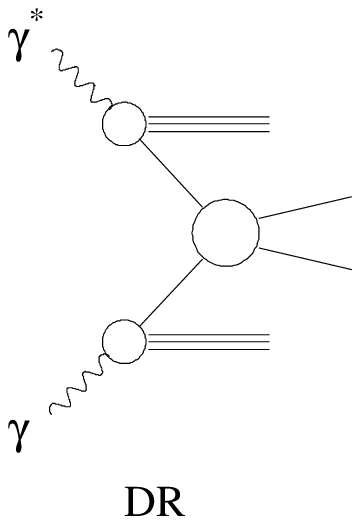,width=8cm}}
\end{picture}
\caption{\label{dir-res}The different components 
	contributing in $\g\g^*$ scattering.}
\end{figure}

Considering jet production, the interaction of the photons have three
different parts \cite{16, 17, 18}. First in the direct (D)
contribution, both photons can interact directly, which yields the QPM
box diagram in LO. The NLO QCD corrections consist of the radiation
of one additional gluon in the final state and the virtual
corrections. Next, the single-resolved (SR) components have to be
considered, which result from the  hadronic structure of either of one
of the photons. Since the resolved  real and the resolved virtual
photon have different hadronic structures, described by different
PDF's, we will specify the SR contribution of a resolved virtual
photon as SR$^*$, whereas the single resolved real photon contribution
is simply denoted SR. Finally, the contribution from two resolved
photons is called the double-resolved (DR) contribution. The different
components are pictured in Fig.\ \ref{dir-res}.

The resolved photons are considered as sources for partons, which
afterwards interact in a subprocess. The factorization of hard and
soft regions in the $e^+e^-$ cross section is given by
\begin{eqnarray}
  d\si (e^+e^-\to \mbox{jets}) = \sum_{k,l} \int \! dx_a dx_b \ 
  f_{\g^*/l}(x_a) d\si_{k/l} f_{k/\g}(x_b) \nonumber \\ 
   df_{\g^*/e}(y_a) f_{\g /e}(y_b) dy_ady_b \quad . 
\end{eqnarray}
It is written as a convolution of the PDF's of the virtual and the real
photons $f_{\g^* /l}(x_a)$ and $f_{k/\g}(x_b)$, respectively, with the
hard partonic cross section $d\si_{ij}$ and the spectra of the
photons, that are described by the Weizs\"acker-Williams
approximation. The spectrum of the real photon is integrated over the
low $Q^2$ region from $Q^2_{min}=\frac{m_e^2x^2}{1-x}$ to 
$Q^2_{max}=4$ GeV$^2$, giving 
\equ{real g}{ f_{\g /e}(y_b) =
  \frac{\al}{2\pi} \frac{1+(1-y_b)^2}{y_b} \ln\left( 
	\frac{Q^2_{max}}{Q^2_{min}} \right) \quad .} 
The function $f_{\g^*/e}$ is given by equation (\ref{1/p2}). 
The kinematics can be described most easily in the c.m.\
system of the virtual photon and the electron that radiates $\g_b$. We
start from the D case, for which from energy-momentum conservation one has
\begin{eqnarray}
  W &=& E_T(e^{-\eta_1}+e^{-\eta_2}) \quad , \\
  y_a &=& \frac{W^2+P^2}{s_H} \quad , \\
  y_b &=& 1 + \frac{2W}{W^2+P^2}\ E_T (\sinh\eta_1 + \sinh\eta_2) \quad . 
\end{eqnarray}
Both variables $y_a$ and $y_b$ are integrated out and are included in
the phase space for convenience, which gives 
\equ{}{ d\mbox{PS}^{(2)}dy_ady_b = \frac{W^2}{W^2+P^2}
  \frac{E_T}{2s_H} \frac{dE_T}{(2\pi )^2} \ d\eta_1d\eta_2 \quad . } 
In the SR and DR cases, the additional variables $x_a$ and $x_b$ have
to be introduced. They give the momentum fractions of the partons in
the resolved photons. We neglect the transverse momentum of the
incoming virtual photon $q_T$ as we have done in the case of
electron-proton scattering, so the partons are traveling in the
direction of the incoming photons. Thus the energies of the partons
$p_a$ and $p_b$ are given by $x_aE_{\g^*}$ and $x_bE_\g$,
respectively. The rapidities are boosted by 
\equ{}{ \eta_i' = \eta_i + \frac12 \ln (x_ax_b) \quad . }
In the DR case, $x_a\ne 1$ and $x_b\ne 1$, whereas in the SR case only
one of the variables $x_a$ or $x_b$ is less then 1.

\subsection{Predictions for Inclusive Jet Rates}

We now come to a presentation of numerical results for the scattering
of a virtual on a real photon. First we note that the factorization
and renormalization scale dependences of the D contribution is tested
indirectly with the tests done in section 5.4, since the matrix
elements for the D case are proportional to the abelian color class
for the subprocess $\g g\to q\bar{q}g$ which is included in the SR
case. All tests hold for each color class separately, since e.g.\ the
cancellation and factorization of singularities holds for each color
class separately. The $M_\g$ dependence of the D and the SR$^*$
contribution compensate each other, just as for the SR and DR
components, as tested in section 5.4. 

For producing our plots we assume kinematical conditions that
will be encountered at LEP2, where the photons are emitted by
colliding electrons and positrons, both having the energy of
$E_e=83.25$ GeV. We choose the configuration, where the virtual photon
travels in the positive $z$-direction. We
focus on one-jet cross sections and do not present results on dijet
rates, since the studies here have only exploratory character. For the
same reason we have used only the $\overline{\mbox{MS}}$-GRS \cite{21}
parametrization of the photon PDF, for obtaining our results and do not
consider the SaS PDF's \cite{22}. We have implemented the PDF of GRS
for both, the real and the virtual photon, since the GRS
parametrization goes over into the GRV \cite{20} parametrization for
the real photon, when choosing $Q^2=0$. The real photon will be
integrated over $Q^2$ using the Weizs\"acker-Williams approximation
for the region described before equation (\ref{real g}), whereas the
virtual photon will have fixed $P^2$-values. Because of the high c.m.\
energies encountered at LEP2, we have set the number of flavors to
$N_f=4$, adding the contributions from photon-gluon fusion by fixed
order perturbation theory. We took the value
$\Lambda^{(4)}_{\overline{MS}}=239$ MeV for the QCD scale, which is
also used in the $\al_s$ two-loop formula, for which $\mu =E_T$. The 
factorization scales are set equal, as in the case of electron-proton 
scattering, with $M_\g =M_{\g^*}=E_T$. The Snowmass jet definition 
\cite{44} is used as explained in section 5. 

The D and SR curves presented in the following are the NLO
contributions for the direct virtual photon, where the large
logarithm has been subtracted and should therefore be denoted D$_s$
and SR$_s$ in accordance with the notation in section 5. Since we do
not present curves for the unsubtracted cross sections, we suppress
the index $s$ to simplify the notation.

\begin{figure}
  \unitlength1mm
  \begin{picture}(122,170)
    \put(8,55){\epsfig{file=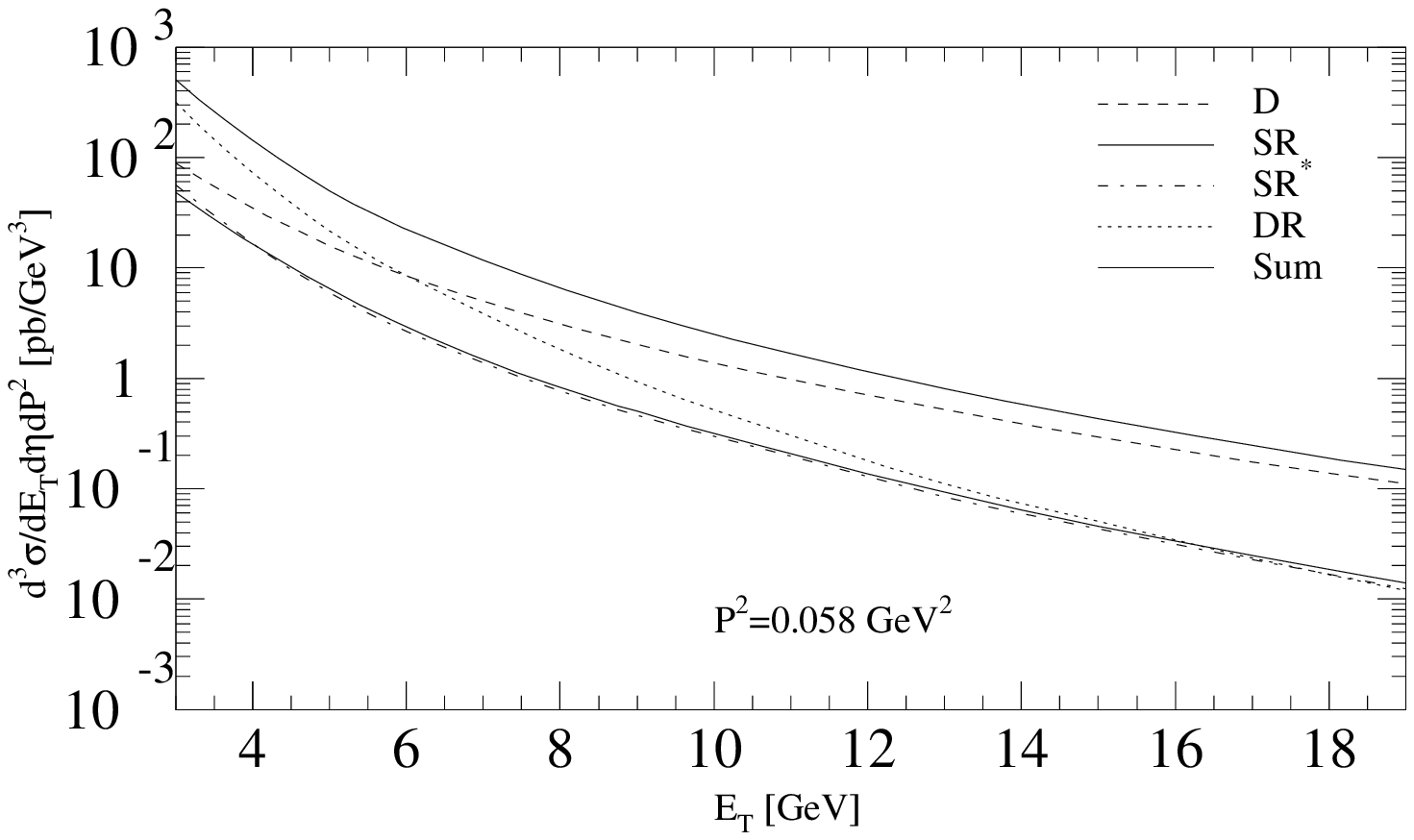,width=11cm}}
    \put(8,-3){\epsfig{file=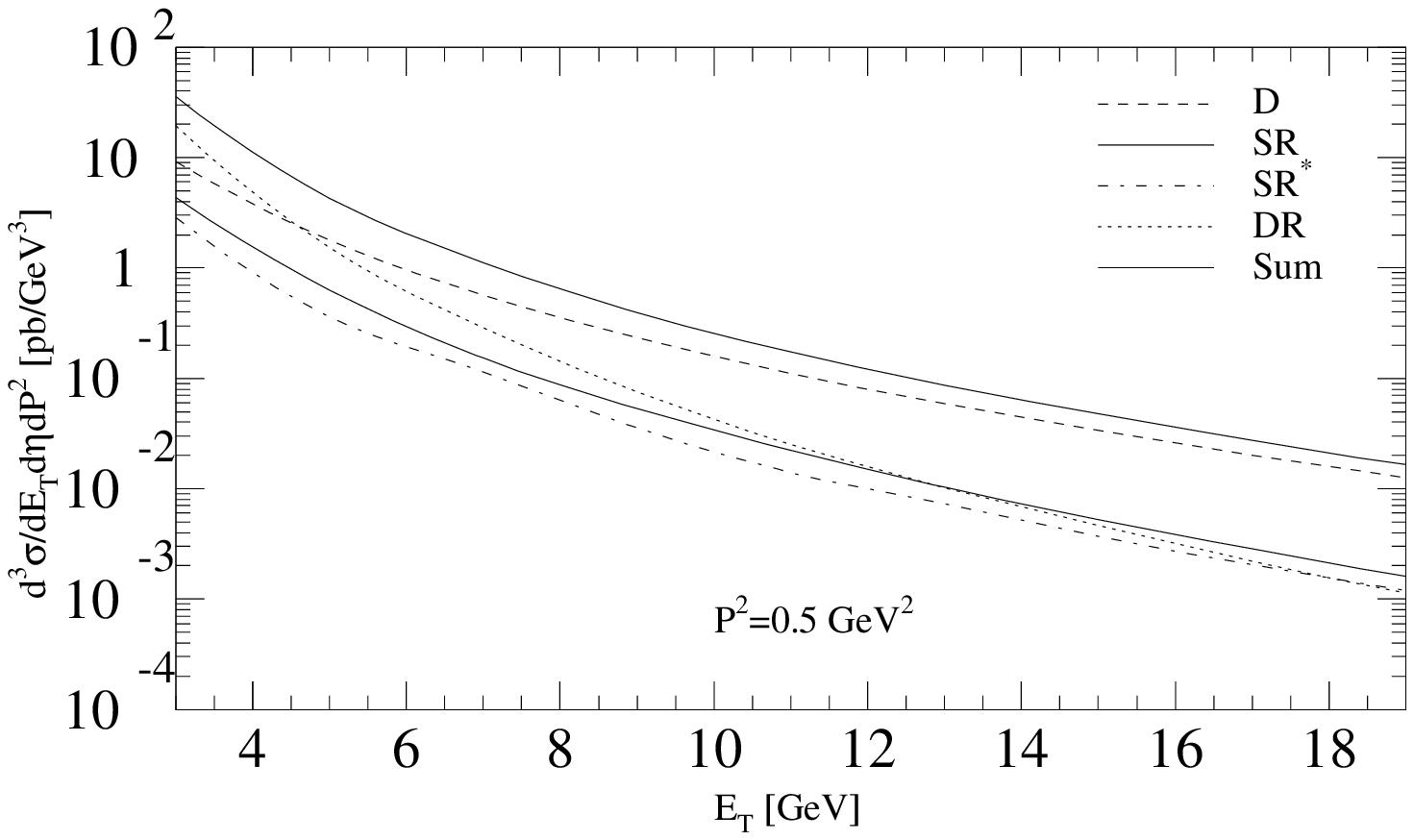,width=11cm}}
    \put(8,-61){\epsfig{file=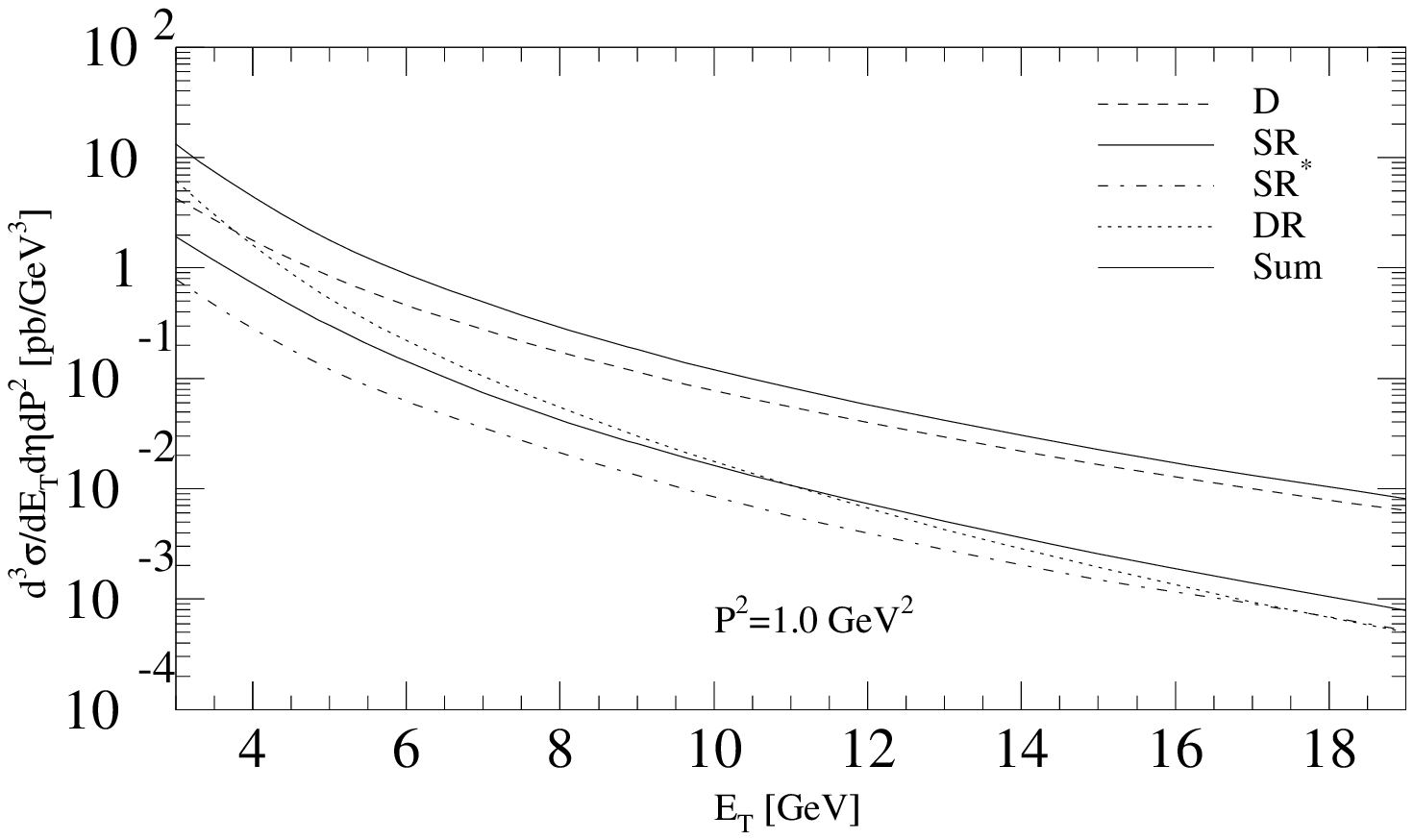,width=11cm}}
    \put(62,165){\footnotesize (a)}
    \put(62,108){\footnotesize (b)}
    \put(62,49){\footnotesize (c)}
  \end{picture}
\caption{\label{10}(a) Single-jet
        inclusive cross section integrated over $\eta \in
        [-2,2]$ for the virtuality $P^2=0.058$ GeV$^2$. The
        $\overline{\mbox{MS}}$-GRS parametrization with $N_f=4$ is
	chosen. The upper full curve is the sum of the D, SR,
	SR$^*$ and the DR components; (b) $P^2=0.5$ GeV$^2$; (c)
        $P^2=1.0$ GeV$^2$.} 
\end{figure}

\begin{figure}[hhh]
  \unitlength1mm
  \begin{picture}(122,110)
    \put(-3,-1){\epsfig{file=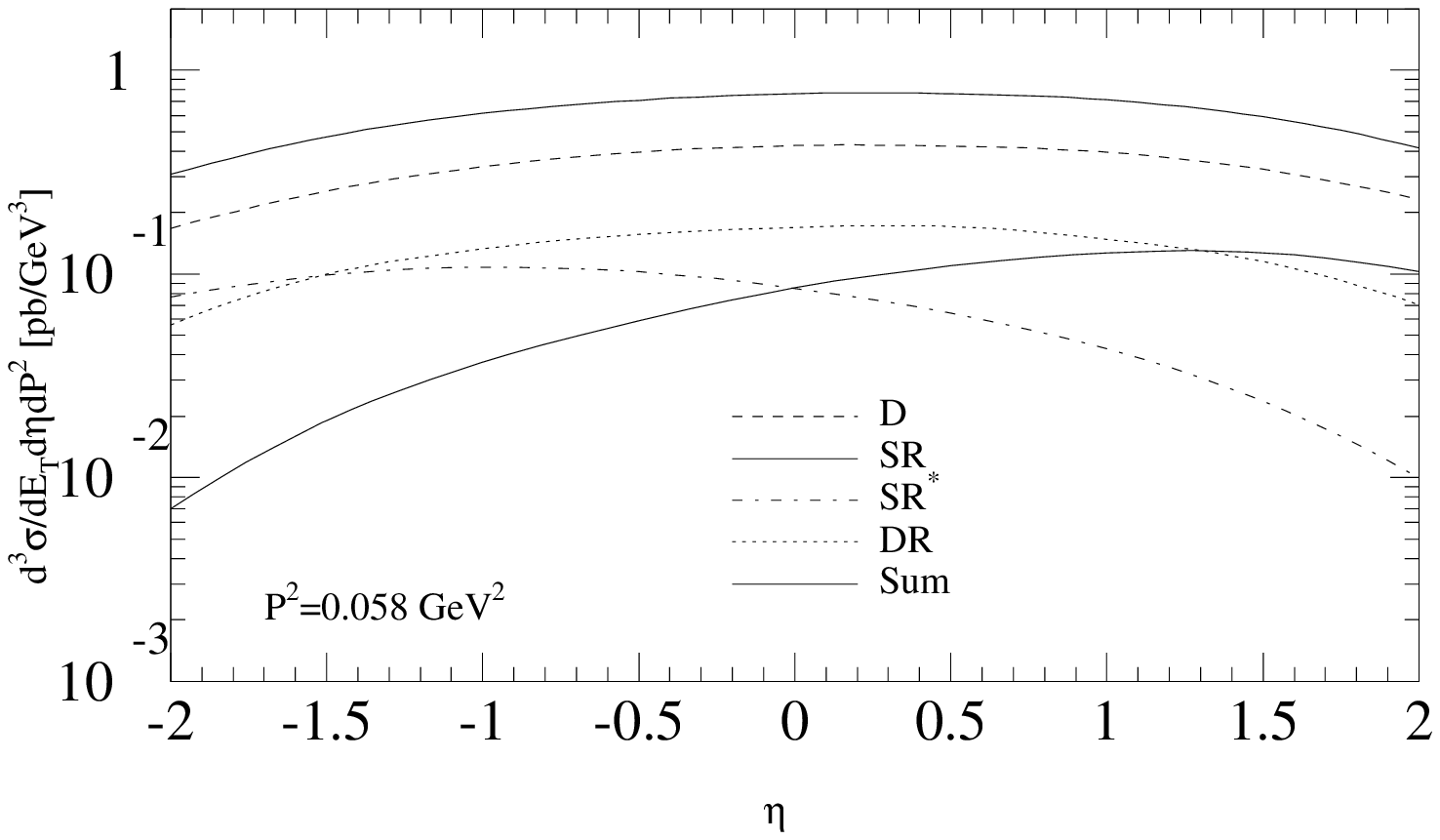,width=7cm,height=12cm}}
    \put(-3,-59){\epsfig{file=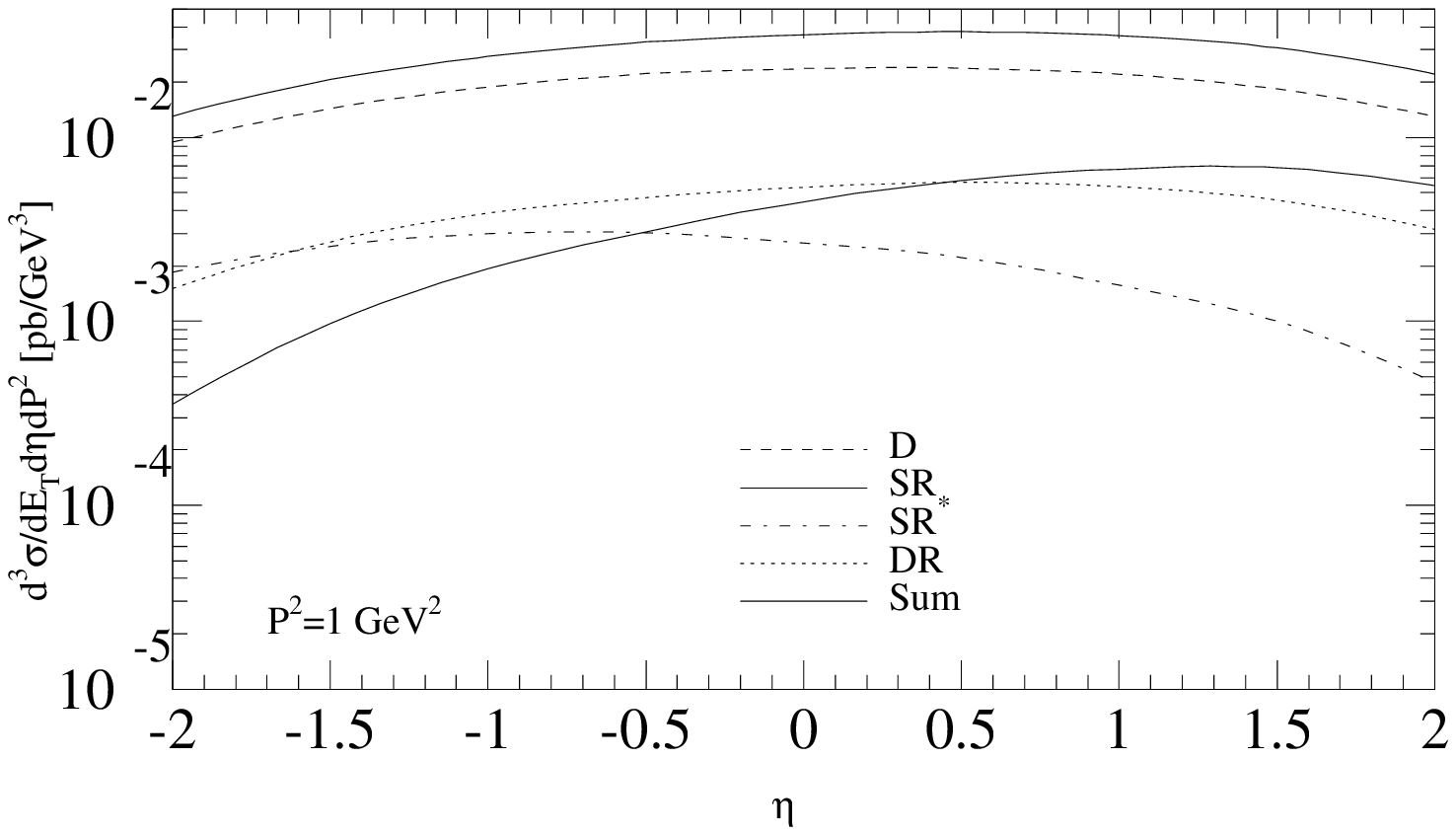,width=7cm,height=12cm}}
    \put(60,-1){\epsfig{file=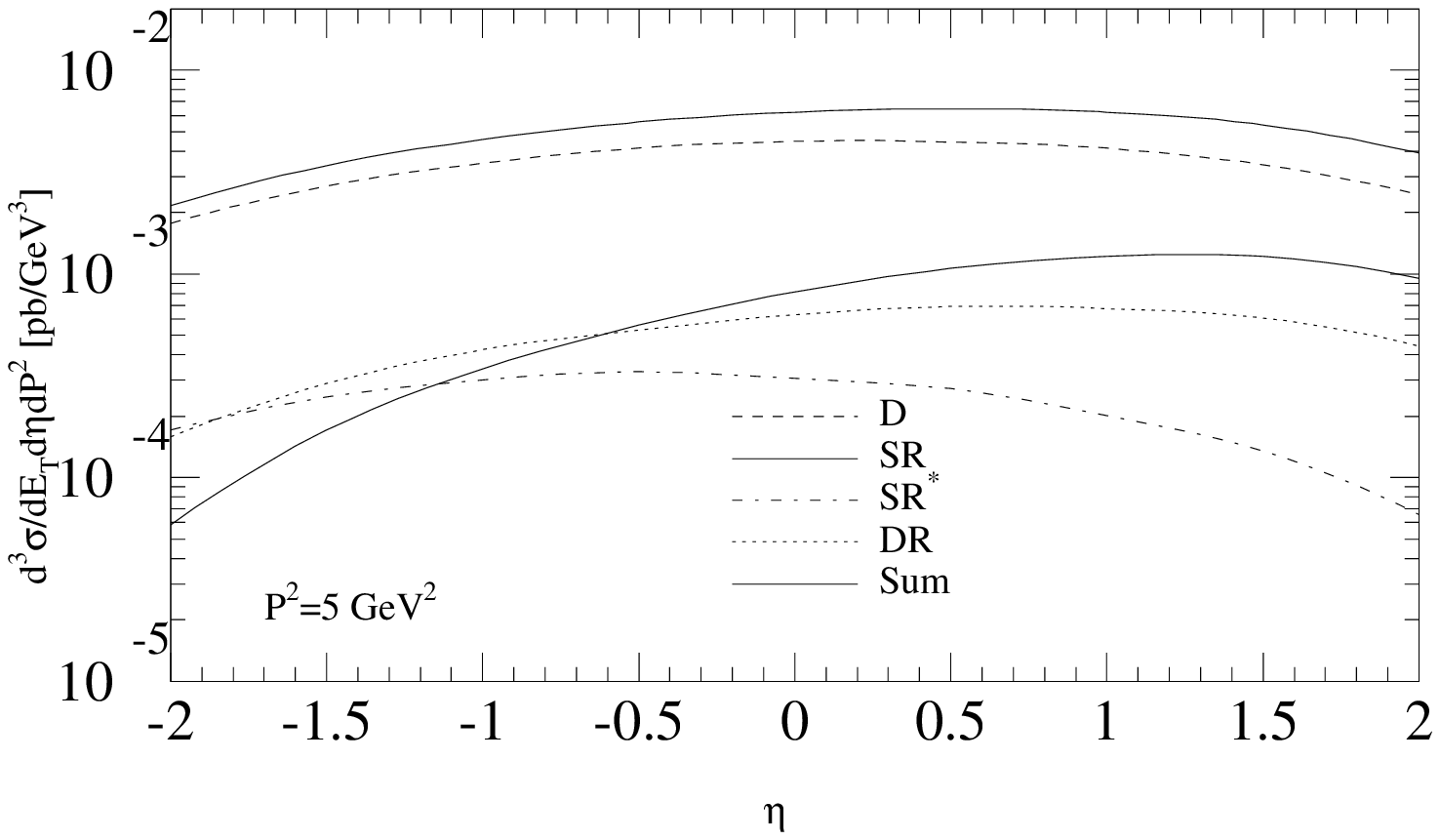,width=7cm,height=12cm}}
    \put(60,-59){\epsfig{file=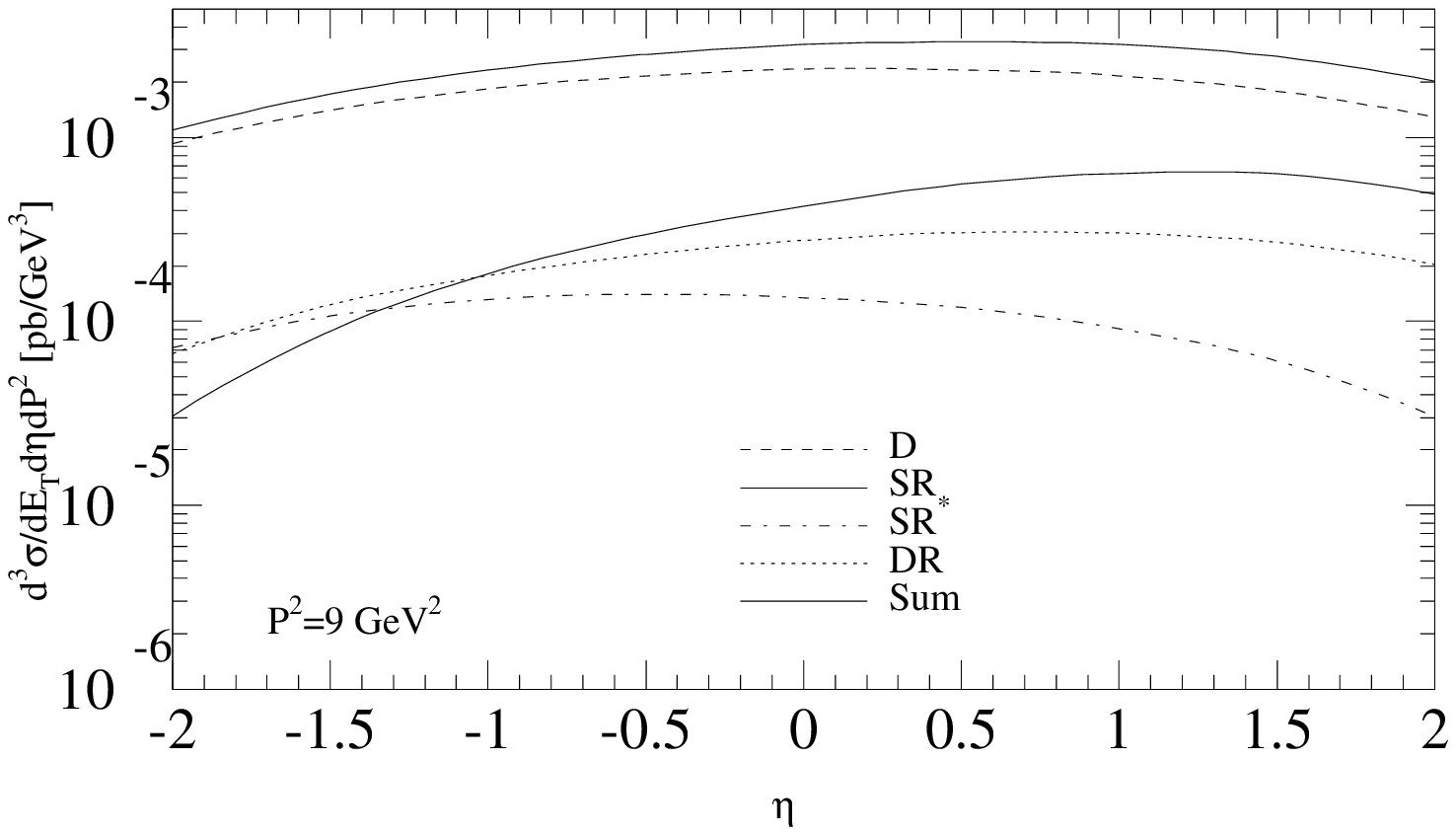,width=7cm,height=12cm}}
    \put(32,103){\footnotesize (a)}
    \put(32,44){\footnotesize (b)}
    \put(92,103){\footnotesize (c)}
    \put(92,44){\footnotesize (d)}
  \end{picture}
\caption{\label{11}(a) Single-jet
        inclusive cross section as a function of $\eta$ for fixed
	$E_T=10$ GeV and virtuality $P^2=0.058$ GeV$^2$. The 
        $\overline{\mbox{MS}}$-GRS parametrization with $N_f=4$ is
	chosen. The upper full curve is the sum of the D, SR, SR$^*$ and
	the DR components; (b) $P^2=1$ GeV$^2$; (c) $P^2=5$ GeV$^2$;
        (d) $P^2=9$ GeV$^2$.}
\end{figure}

In Fig.\ \ref{10} a, b, c the $E_T$ spectra\footnote{I thank T.\
Kleinwort for the consent in using his computer program for producing
the SR$^*$ and DR curves.} for the virtualities
$P^2=0.058, 0.5$ and $1.0$ GeV$^2$ for the cross section
$d^3\si /dE_Td\eta dP^2$ are shown, integrated over the interval
$-2\le \eta \le 2$, which are the boundaries being presently used at
LEP1.5. As explained in section 5, the value $P^2_{eff}=0.058$ GeV$^2$
is chosen as to reproduce the $P^2\simeq 0$ case. As one can see, the
SR (lower full) and SR$^*$ (dash-dotted) curves coincide in Fig.\
\ref{10} a, where the real photon is approximated by the integrated
Weizs\"acker-Williams formula and the virtual photon has the value
of $P^2_{eff}$. The full cross section (upper full curve) is dominated
by the DR component only in the small $E_T$ range for small
$P^2$ values. For $P^2=0.5$ and $1.0$ GeV$^2$, the DR and D
contributions are of the same order around $E_T=4$ GeV, but the DR component
falls off quickly for the higher $E_T$'s, leaving the D component as
the dominant contribution. This is expected, as the resolved
virtual photon is important for smaller $P^2$ and suppressed for the larger
virtualities. For the same reason, the SR$^*$ contribution falls below
the SR curve when going to higher values of $P^2$ (remember that the
resolved real photon is not suppressed by $P^2$). In all
curves, both SR contributions do not play an important role for the
full cross section. Of course, all contributions decrease with
increasing $P^2$, so that the full cross section falls of with
increasing $P^2$. 

We turn to the $\eta$-distribution of the single-jet cross section for
fixed $E_T=10$ GeV between $-2\le \eta \le 2$ for the virtualities
$P^2=0.058, 1, 5$ and $9$ GeV$^2$. As one sees in Fig.~\ref{11} a--d,
the D and DR distributions 
for the lowest virtuality $P^2_{eff}$ are almost symmetric, because of
the identical energies of the incoming leptons. The SR curve falls off
for negative $\eta$, whereas the SR$^*$ component is suppressed for
positive $\eta$. Going to higher $P^2$ values, the D contribution
stays more or less symmetric and dominates the full cross section, as
we have already seen in Fig.\ \ref{10} a, b, c for the larger $E_T$
values. The components containing contributions from the resolved
virtual photon DR and SR$^*$ fall of in the region of negative $\eta$
so that they become more and more asymmetric. This is clear, 
since we have chosen the virtual photon to be incoming from the positive
$z$-direction and the resolved virtual photon is falling off for
higher virtualities. To observe the compensation of the 
$\ln (P^2/M_\g^2)$ term, subtracted  from the direct virtual photon,
with the similar but negative behavior of the resolved virtual
photon, one has to compare the DR and SR, and the D and SR$^*$
components (see Fig.\ \ref{dir-res}). The DR and SR contributions are
of the same magnitude in the negative $\eta$ region and the DR component
is dominant for the larger $\eta$ values, where the resolved photon is
more important. We have observed these findings already for $ep$
scattering. The same holds for the D and SR$^*$ distributions in the
negative $\eta$ region, only here the D component is far more dominant
then the SR$^*$ one in the whole $\eta$ region.

\section{Summary and Outlook}

We have calculated single- and dijet inclusive jet
cross sections for $\g^*p$ and $\g^*\g$ scattering through a consistent
extension of methods used in the calculation of $\g\g$ scattering and
of photoproduction in $ep$ scattering. The partonic cross sections for
the considered reactions were calculated in NLO QCD using the
phase-space slicing method, where a technical cut-off $y_c$ is
introduced to separate singular and finite regions of phase space. The
spectrum of the real photon was approximated with the integrated,
whereas the spectrum of the virtual photon was approximated by the
differential Weizs\"acker-Williams formula. For the resolved virtual
photon contribution we used the parton distributions of the virtual
photon in a LO parametrization. 

For the hard cross section, we have in particular worked out the
subtraction of singularities that appear when integrating the phase
space over the collinear region of the virtual photon. Contrary to
real photons, the singularity appearing for the virtual photon is not
regulated in the dimensional regularization scheme but by the
virtuality $P^2$ of the photon. This leads to a logarithm depending on
$P^2$, which is absorbed into the PDF of the virtual photon. Through
this procedure, the PDF becomes scheme and scale dependent. The terms
remaining in the subtracted cross section are constructed in such a
way, that the corresponding real photon term is obtained in the limit
$P^2\to 0$ in the $\overline{\mbox{MS}}$ scheme.

We have presented several tests of the numerical program \cite{jv} for the
evaluation of the cross sections. We have shown, that the dependence
on the slicing parameter $y_c$ vanishes when the regulated singular
and finite contributions are added. Furthermore, the factorization
scale dependences of the NLO direct and the LO resolved contributions
cancel to a large extend. The renormalization scale dependence was
shown to be reduced in NLO compared to LO. 

The jet cross sections for $\g^*p$ scattering were computed under HERA
conditions using the Snowmass jet definition. We presented
distributions in the transverse energy and rapidity of the observed
jet. For very small $P^2$ we found good numerical agreement between
real and virtual photoproduction. For the larger $P^2$ values, the
unsubtracted direct contribution corresponding to the case of deep
inelastic scattering approximates the sum of the subtracted direct and
resolved contribution rather well, at least for not too large
rapidities. This is in accordance with the result that the
perturbatively calculable subtracted term agrees quite well with the
evolved quark distributions of the virtual photon PDF in the larger
$x$ range. Differences between the unsubtracted direct and the sum of
the subtracted direct and resolved components can be attributed to
small differences in the subtraction term and the quark distribution
and to effects from neglecting the transverse momentum of the incoming
virtual photon. 

Furthermore we have calculated distributions in the transverse energy
for inclusive dijet cross sections. For experimental considerations the
variable $x_\g^{obs}$ has been introduced, which is used for a
separation of the direct and resolved contributions. The 
resolved part was defined for $x_\g^{obs}<0.75$, whereas the direct
part was given for $x_\g^{obs}>0.75$. In this way, both contributions 
contained non-negligible direct and resolved parts. The sum of the
enriched direct and resolved curves of course showed to be independent
of the value of $x_\g^{obs}$. As an application we have
calculated the ratio of the resolved to the direct enriched cross
sections, that could be compared to ZEUS data. The ratio shows
significant NLO effects and is in good agreement with ZEUS data for
$P^2>0.2$ GeV$^2$. For smaller virtualities the experimental data
points lie above the theoretical prediction, which can be attributed
to, e.g., multiple scattering between the photon and proton remnants.

The jet cross sections for $\g^*\g$ scattering were evaluated for
conditions to be met at LEP2. As for $ep$ scattering we used the
Snowmass jet definition. We showed distributions in the transverse
energy and the rapidity only for single inclusive cross sections with
one parametrization of the virtual photon. In contrast to the
$ep$ scattering, for $\g^*\g$ scattering one additional subprocess is
encountered, which is the direct interaction of the real with the
virtual photon. The singularities of the real photon were regularized
in the dimensional regularization scheme, whereas the virtual photon
singularities have been handled as described above by subtracting
the large logarithm. The direct component was shown to be the most
dominant one for larger $E_T$. The resolved virtual photon
contributions were suppressed for larger values of $P^2$ due to the
suppression of the virtual photon PDF for larger virtualities. 

Future investigations on virtual photoproduction will require more data on
single inclusive jet production at large transverse energy. A
detailed dijet analysis of an infrared safe cross section such as
$d^4\sigma/dE_Td\eta_1 d\eta_2dP^2$, where the transverse energies of
the two jets are not cut at exactly the same value, will provide an
improved insight into the structure of the virtual photon. 
Furthermore, choosing a $k_T$-cluster-like jet definition with
smaller cone radii will reduce both the uncertainties in the jet
algorithm and in the underlying event. On the theoretical side, one
possible improvement is the correct treatment of the transverse
momentum of the incoming photon for larger $P^2$ including a correct
transformation from the photonic c.m.~frame to the HERA or LEP laboratory
systems. For a consistent NLO treatment, the inclusion of NLO parton
densities for the photon is necessary. These are, however, needed in a
parametrized form and should also be studied in correlation with deep
inelastic $e\g^*$ scattering data.

\subsection*{Acknowledgements}

I thank G.~Kramer for his capability to formulate clear goals for
interesting research, which led to this paper. He was always supportive
and helpful concerning any questions and problems I encountered during
my work. Furthermore I thank the group at the II.\ Institute of Theoretical
Physics, namely J.~Binnewies, P.~B\"uttner, M.~Klasen
and T.~Kleinwort for a very pleasant working atmosphere. I am indebted
to M.~Klasen and J.~Binnewies for a careful reading of this manuscript
and I thank M.~Klasen for his collaboration. T.~Kleinwort has
given me useful advice concerning details of the computer code
developed during this work.

\begin{appendix}

\section{General Definitions}

To simplify the notation, we state some definitions here that will be
used throughout the appendix. The Mandelstam variables $s,t$ and $u$
are defined in the usual way. The scale $M^2$ that appears in the 
virtual, initial and final state corrections is normally set equal to
the virtuality of photon $P^2$, only in the photoproduction limit
$P^2\to 0$  we set $M^2=s$. The Born terms that appear throughout this
work are given by 
\begin{eqnarray}
 T_\g (s,t,u) &=& (1-\e ) \left( \frac{t}{u} + \frac{u}{t} \right)
 - 2P^2\ \frac{s}{ut} - 2\e \quad , \\
 T_1(s,t,u) &=& 4N_CC_F  \left( \frac{s^2+u^2}{t^2} -\e \right) \quad , \\
 T_2(s,t,u) &=& -8C_F (1-\e )\left( \frac{s^2}{ut} -\e\right) \quad , \\
 T_3(s,t,u) &=& 4C_F (1-\e )\left(
 \frac{2N_CC_F}{ut}-\frac{2N_C^2}{s^2} \right)(t^2+u^2-\e s^2) \quad , \\
 T_4(s,t,u) &=& 32 N_C^3C_F(1-\e )^2\left( 3 - \frac{ut}{s^2} -
 \frac{us}{t^2} - \frac{st}{u^2} \right) \quad . 
\end{eqnarray}
For the initial state corrections the plus distribution function 
\equ{}{ R_+(x,z) :=  \left( \frac{\ln \left( x\left(
    \frac{1-z}{z}\right)^2\right)}{1-z}\right)_+ }
is needed. As the integration over $z$ in the initial state
singularities runs from $z_{min}$ to $1$, the plus distribution 
function is defined as
\equ{D12}{ R_+[g] = \int\limits_{z_{min}}^1 \!\! dz\ R(x,z) g(z)
  -\int\limits_0^1 dz\  R(x,z) g(1) , } 
for any regular function $g(z)$. This leads to additional terms not 
given here explicitly when (\ref{D12}) is transformed so that both 
integrals are calculated in the range $[z_{min},1]$. The singular terms 
in the initial state corrections are proportional to the Altarelli-Parisi
splitting functions
\begin{eqnarray}
  P_{q\leftarrow \g}(z) &=& N_C \left( 1 + 2z(1-z) \right) \quad , \\
  P_{q\leftarrow q} (z) &=& C_F \left[ \frac{1+z^2}{(1-z)_+} +
  \frac{3}{2} \delta (1-z) \right] , \\
   P_{g\leftarrow q} (z) & = &
  C_F \left[ \frac{1+(1-z)^2}{z} \right] , \\
  P_{g\leftarrow g} (z) & = &
  2 N_C \left[ \frac{1}{(1-z)_+}+\frac{1}{z}+z(1-z)-2 \right]
  \nonumber \\
  &+& \left[ \frac{11}{6}N_C-\frac{N_f}{3} \right] \delta (1-z),\\
  P_{q\leftarrow g} (z) & = &
  \frac{1}{2} \left[ z^2+(1-z)^2 \right] . 
\end{eqnarray}
The plus functions appearing here are defined, in contrary to equation 
(\ref{D12}), in the limits $[0,1]$. In the virtual corrections 
the function  
\begin{eqnarray}
  L(x,y) &=& \ln \left|\frac{x}{P^2}\right|\ln\left|\frac{y}{P^2}\right|
 - \ln \left|\frac{x}{P^2}\right|\ln \left|1-\frac{x}{P^2}\right| 
 - \ln \left|\frac{y}{P^2}\right|\ln \left|1-\frac{y}{P^2}\right|
 \nonumber \\
 &-& \lim_{\eta\to 0} \mbox{Re}\left[ {\cal L}_2 \left( \frac{x}{P^2}
 + i\eta \right) + {\cal L}_2\left( \frac{y}{P^2}+i\eta \right) 
 \right]  + \frac{\pi^2}{6} 
\end{eqnarray}
appears \cite{23}, where ${\cal L}_2(x)$ is the Dilogarithm function. In the
limiting case $P^2\to 0$ one finds for ${\cal L}_2(x)$
\equ{}{ {\cal L}_2\left( \frac{x}{P^2}\right) = -\frac{\pi^2}{6}
  - \frac{1}{2} \ln^2\left( \frac{x}{P^2}\right) \quad . }
The square of the logarithm has two different values according to the
sign of $x$:
\equ{}{ \ln^2\left( \frac{x}{P^2}\right) = \left\{
  \begin{array}{c@{\mbox{\ \ for\ \ }}c} \ln^2(-x/P^2)-\pi^2 & x<0 \\  
    \ln^2(x/P^2) & x>0 \end{array} \right. \quad . }
Therefore, ones obtains the following three cases for the function
$L(x,y)$ appearing in the virtual corrections of the real photon with
$P^2=0$: 
\begin{eqnarray}
  L(x,y) &=& \frac{\pi^2}{2}-\frac{1}{2}\ln^2\left(\frac{x}{y}\right)
  \qquad \mbox{for} \qquad x>0, y>0 \quad ,   \\
  L(x,y) &=& \frac{3\pi^2}{2}-\frac{1}{2}\ln^2\left(\frac{x}{y}\right)
  \qquad \mbox{for} \qquad x<0, y<0 \quad ,   \\
  L(x,y) &=& \pi^2 -\frac{1}{2}\ln^2\left(-\frac{x}{y}\right)
  \qquad \mbox{for} \qquad \left\{ {x>0, y<0 \atop x<0,y>0} \right. \quad .
\end{eqnarray}

\section{Virtual Corrections}

In this subsection we give the explicit expressions for the virtual
corrections that arise from the interference of the LO Born processes
for $\g^*\g\to q\bar{q}$, $\g^*q\to gq$ and $\g^*g\to q\bar{q}$ with
the corresponding one-loop amplitudes. The expressions can be found in
\cite{23}. The results depend on the two-body variables $s, t$ and $u$:
\begin{eqnarray}
  E_1 &=& \left[ -\frac{2}{\e^2} + \frac{1}{\e}(2\ln
  \frac{-u}{M^2}-3) - \frac{\pi^2}{3} -8 -\ln^2 \frac{-u}{M^2}
   \right] T_\g (s,t,u) \nonumber \\
   &-& 4\ln \frac{-u}{M^2} \left( \frac{2u}{s+t} + \frac{u^2}{(s+t)^2}
 \right)  \nonumber \\
 &-& \ln \frac{s}{M^2} \left( \frac{4u+2s}{u+t} - \frac{st}{(u+t)^2} 
  \right) - \ln \frac{-t}{M^2} \left( \frac{4u+2t}{u+s} -
  \frac{st}{(u+s)^2} \right) \nonumber \\
 &+& 2L(-u,-s) \frac{u^2+(u+t)^2}{st} +
 2L(-u,-t)\frac{u^2+(u+s)^2}{st} \nonumber \\
 &-& \left( \frac{4u}{s+t} + \frac{u}{s+u} + \frac{u}{u+t} \right) 
 + \left( \frac{u}{s}+\frac{u}{t}+\frac{s}{t}+\frac{t}{s} \right) \\
  E_2 &=& \left[ \frac{2}{\e^2} + \frac{2}{\e}\left(
  \ln\frac{-u}{M^2} - \ln\frac{s}{M^2} - \ln\frac{-t}{M^2} \right)
  \right]T_\g (s,t,u) \nonumber \\
  &+& \left[ - \frac{\pi^2}{3} -\ln^2\frac{-u}{M^2} +
  \ln^2\frac{s}{M^2} +\ln^2\frac{-t}{M^2} + 2L(-s,-t) \right]T_\g (s,u,t)
  \nonumber \\ 
   &-& 4\ln \frac{-u}{M^2} \left( \frac{2u}{s+t} + \frac{u^2}{(s+t)^2}
 \right) 
 + \ln \frac{s}{M^2} \frac{2s}{u+t} + \ln \frac{-t}{M^2} \frac{2t}{u+s} 
  \nonumber \\ &+& 2L(-u,-s) \frac{u^2+(u+t)^2}{st} +
 2L(-u,-t)\frac{u^2+(u+s)^2}{st} \nonumber \\
 &-& 2 \left( \frac{2u}{s+t}
 -\frac{u}{s}-\frac{u}{t}-\frac{s}{t}-\frac{t}{s} \right)   \\
  E_3 &=& \left[ -\frac{2}{\e^2} + \frac{1}{\e}(2\ln
  \frac{s}{M^2}-3) + \frac{2\pi^2}{3} -8 -\ln^2 \frac{s}{M^2}
   \right] T_\g (s,u,t) \nonumber \\
   &+& 4\ln \frac{s}{M^2} \left( \frac{2s}{u+t} + \frac{s^2}{(u+t)^2}
 \right)  \nonumber \\
 &+& \ln \frac{-u}{M^2} \left( \frac{4s+2u}{s+t} - \frac{ut}{(s+t)^2} 
  \right) + \ln \frac{-t}{M^2} \left( \frac{4s+2t}{s+u} -
  \frac{ut}{(s+u)^2} \right) \nonumber \\
 &-& 2L(-s,-u) \frac{s^2+(s+t)^2}{ut} -
 2L(-s,-t)\frac{s^2+(s+u)^2}{ut} \nonumber \\
 &+& \left( \frac{4s}{u+t} + \frac{s}{u+s} + \frac{s}{s+t} \right) 
 - \left( \frac{s}{u}+\frac{s}{t}+\frac{u}{t}+\frac{t}{u} \right) \\
  E_4 &=& \left[ \frac{2}{\e^2} + \frac{2}{\e}\left(
  \ln\frac{s}{M^2} - \ln\frac{-u}{M^2} - \ln\frac{-t}{M^2} \right)
  \right]T_\g (s,u,t) \nonumber \\
  &+& \left[ \frac{4\pi^2}{3} +\ln^2\frac{-u}{M^2} -
  \ln^2\frac{s}{M^2} +\ln^2\frac{-t}{M^2} + 2L(-u,-t) \right]T_\g (s,t,u)
  \nonumber \\ 
   &+& 4\ln \frac{s}{M^2} \left( \frac{2s}{u+t} + \frac{s^2}{(u+t)^2}
 \right)  
 - \ln \frac{-u}{M^2} \frac{2u}{s+t} - \ln \frac{-t}{M^2} \frac{2t}{u+s} 
  \nonumber \\ &-& 2L(-s,-u) \frac{s^2+(s+t)^2}{ut} -
 2L(-s,-t)\frac{s^2+(u+s)^2}{ut} \nonumber \\
 &+& 2 \left( \frac{2s}{u+t}
 -\frac{s}{u}-\frac{s}{t}-\frac{u}{t}-\frac{t}{u} \right)  \quad . 
\end{eqnarray}

\section{Final State Corrections}

In the following we give the real final state corrections that
appear when the $2\to 3$ matrix elements are integrated over the
singular region of phase space. The expressions depend on the
invariant mass cut-off $y_F$ and on the two-body variables $s, t$ and
$u$. Terms of order ${\cal O}(\e )$ have been neglected. The
contributions $F_2,\ldots F_5$ can be found in \cite{23}.

\begin{eqnarray}
 F_1 &=& \al^2\al_s 4N_CC_FQ_i^4T_\g (s,t,u) \bigg\{ \frac{1}{\e^2} + 
  \frac{1}{\e} \bigg(\frac{3}{2} -\ln \frac{s}{M^2} \bigg) 
   \nonumber \\  &+& \frac{7}{2} - \frac{3}{2}\ln
  \frac{-y_F(t+u)}{M^2}  - \ln^2\frac{-y_F
    (t+u)}{s} + \frac{1}{2}\ln^2\frac{s}{M^2} -\frac{\pi^2}{3} \bigg\} \\
 F_2 &=& \al\al_s^2 C_FQ_i^2T_\g (t,s,u) \bigg\{ C_F\bigg[ \frac{1}{\e^2} 
  + \frac{1}{\e} \bigg(\frac{3}{2} -\ln \frac{-t}{M^2} \bigg) \nonumber \\  
  &+& \frac{7}{2} - \frac{3}{2}\ln \frac{-y_F (t+u)}{M^2} -
  \ln^2\frac{y_F (t+u)}{t} + \frac{1}{2}\ln^2\frac{-t}{M^2}
  -\frac{\pi^2}{3} \bigg] \nonumber \\ &+& 
 \frac{1}{2}N_C \bigg[ \frac{1}{\e^2} -
 \frac{1}{\e} \bigg(-2 +\ln \frac{-st}{Q^4} + \ln\frac{-s}{t}
 -\frac{5}{3} \bigg)  \nonumber \\
  &+& \ln\frac{-(t+u)}{M^2}\ln\frac{-s}{t} + \ln^2\frac{y_F (t+u)}{t} 
  - \ln^2\frac{-y_F (t+u)}{s} \nonumber \\ &-& \frac{1}{2}\ln^2\frac{t+u}{t} 
 + \frac12 \ln^2\frac{-(t+u)}{s} - 2\ln\frac{-y_F (t+u)}{M^2}
  \nonumber \\
  &+& \ln\frac{-(t+u)}{M^2}\ln\frac{-st}{(t+u)^2} +
   \ln^2\frac{-(t+u)}{M^2}  \nonumber \\ 
  &-& \ln^2\frac{y_F (t+u)}{t} - \ln^2\frac{-y_F (t+u)}{s} +
  \frac12 \ln^2\frac{-t-u}{s} + \frac12  \ln^2\frac{t+u}{t} \nonumber \\  
  &-& \frac{5}{3}\ln\frac{-y_F (t+u)}{M^2} 
  +\frac{67}{9} - \frac{2\pi^2}{3} \bigg]\bigg\} \\
 F_3 &=& \al\al_s^2 (N_f-1)C_FQ_i^2T_\g (t,s,u)\bigg\{ 
  -\frac{1}{3}\frac{1}{\e} +
  \frac{1}{3}\ln\frac{-y_F (t+u)}{M^2} - \frac{5}{9} \bigg\} \\
 F_4 &=& \al\al_s^2 C_FQ_i^2T_\g (t,s,u)\bigg\{ -\frac{1}{3}\frac{1}{\e} +
  \frac{1}{3}\ln\frac{-y_F (t+u)}{M^2} - \frac{5}{9} \bigg\} \\
 F_5 &=& \al\al_s^2 Q_i^2T_\g (s,t,u) \bigg\{ C_F \bigg[ \frac{1}{\e^2} 
   + \frac{1}{\e} \bigg(\frac{3}{2} -\ln \frac{s}{M^2} \bigg)  
  + \frac{7}{2} - \frac{3}{2}\ln \frac{-y_F (t+u)}{M^2}
    \nonumber \\  &-&    \ln^2\frac{-y_F
    (t+u)}{s} + \frac{1}{2}\ln^2\frac{s}{M^2} -\frac{\pi^2}{3}
 \bigg] -\frac{1}{4}N_C \bigg[ \frac{1}{\e}\ln\frac{-t}{s}
  \nonumber \\ &-& \ln\frac{-(t+u)}{M^2}\ln\frac{-t}{s} 
 + \ln^2\frac{y_F (t+u)}{t}  - \ln^2\frac{-y_F (t+u)}{s}  \nonumber \\
  &+&  \frac{1}{2}\bigg( \ln^2\frac{-(t+u)}{s}
 -\ln^2\frac{t+u}{t} \bigg)  \bigg] \bigg\} + (t\leftrightarrow u)
\end{eqnarray}

\section{Initial State Corrections for Massles Partons}

Here, we state the parton initial state singularities as functions of
the invariants $s, t$ and $u$, the cut-off parameter $y_J$ and the
additional variable of integration $z_b$. Again, terms of 
${\cal O}(\e )$ have been neglect. The contributions $I^b_2,\ldots
I^b_5$ can be found in \cite{23}.

\begin{eqnarray} I_1^b &=& \al^2\al_s Q_i^4\bigg[ -\frac{1}{\e} 
  P_{q\leftarrow \g}(z_b)
  +N_C\bigg\{ (1-2z_b+2z_b^2) \ln\bigg( \frac{-(t+u)}{M^2}
  \frac{1-z_b}{z_b}y_I \bigg) \nonumber \\ 
  &+& 2z_b(1-z_b) \bigg\} \bigg] (1-\e )C_FT_\g(s,t,u) \quad , \\
  I_2^b &=& \al\al_s^2Q_i^2 C_FT_\g (s,t,u)\bigg\{ C_F \bigg[
  -\frac{1}{\e}\frac{1}{C_F} P_{q\leftarrow q}(z_b)
    \nonumber \\  
  &+& \delta (1-z_b) \bigg( \frac{1}{\e^2} 
  +\frac{1}{\e} \bigg( -\ln \frac{-u}{M^2} +\frac{3}{2} \bigg)
  +\frac{1}{2}\ln^2\frac{-u}{M^2}+\pi^2\bigg) \nonumber \\  
  &+& (1-z_b)\bigg( 1+ \ln
  \bigg(\frac{-(t+u)}{M^2}\frac{1-z_b}{z_b}y_I \bigg)\bigg) 
   +2R_+\bigg(\frac{-u}{M^2},z_b \bigg)    \nonumber \\
   &-&  2\ln\bigg(\frac{-u}{M^2}\bigg(\frac{1-z_b}{z_b}
  \bigg)^2\bigg) -  \frac{2z_b}{1-z_b}\ln\bigg( 1+\frac{u}{t+u}
   \frac{1-z_b}{y_I z_b}\bigg) \bigg]  \nonumber \\
   &-& \frac12 N_C \bigg[ \delta (1-z_b) \bigg( \frac{1}{\e} \ln\frac{t}{u}
   +\frac{1}{2}\ln^2\frac{-u}{M^2}-\frac{1}{2}\ln^2\frac{-t}{M^2}\bigg)
   +2R_+\bigg(\frac{-u}{M^2},z_b\bigg)  \nonumber \\ 
   &-&2R_+\bigg(\frac{-t}{M^2},z_b\bigg) -2\ln\bigg( \frac{-u}{M^2}
  \bigg(\frac{1-z_b}{z_b}\bigg) ^2\bigg)  
   +2\ln\bigg(\frac{-t}{M^2}\bigg(\frac{1-z_b}{z_b}\bigg)
  ^2\bigg)\nonumber \\   
  &-&   \frac{2z_b}{1-z_b}\ln\bigg(
  1+\frac{u}{t+u} \frac{1-z_b}{y_I z_b}\bigg) 
   + \frac{2z_b}{1-z_b}\ln\bigg( 1+\frac{t}{t+u}
   \frac{1-z_b}{y_I z_b} \bigg)
   \bigg] \bigg\} , \\
   I_3^b &=& \al\al_s^2Q_i^2 \bigg[ -\frac{1}{\e}\frac{1}{C_F}
   P_{g\leftarrow q}(z_b)
   +\frac{1}{C_F}P_{g\leftarrow q}(z_b)\ln\bigg( \frac{-(t+u)}{M^2}
   \frac{1-z_b}{z_b}y_I \bigg)  \nonumber \\
   & & {}  -2\frac{1-z_b}{z_b} \bigg] \frac{C_F}{2}T_\g (s,u,t) , \\
   I_4^b &=& (N_f-1)I_3^b , \\
   I_5^b &=& \al\al_s^2Q_i^2 \bigg[ -\frac{2}{\e} P_{q\leftarrow g}(z_b)
   \nonumber \\ &+& 2P_{q\leftarrow g}(z_b)\ln\bigg( \frac{-(t+u)}{M^2}
   \frac{1-z_b}{z_b}y_I\bigg) +1
   \bigg] C_FT_\g (s,u,t)   \nonumber \\
  &+& \bigg[ \frac{2}{\e}\frac{1}{N_C}P_{g\leftarrow g}(z_b)
  +\delta(1-z_b)
  \bigg( -\frac{2}{\e^2}+\frac{1}{\e} \bigg( \ln\frac{tu}{Q^4}
    -\frac{11}{3} +\frac{2N_f}{3N_C} \bigg) -2\pi^2
   \nonumber \\ 
    &-&  \frac{1}{2}\ln^2\frac{-u}{M^2}-\frac{1}{2}\ln^2
    \frac{-t}{M^2} \bigg)
    -2R_+\bigg(\frac{-u}{M^2},z_b\bigg)-2R_+\bigg(\frac{-t}{M^2},z_b 
  \bigg) \nonumber \\ 
    &+& 2\ln\bigg(\frac{-u}{M^2}\bigg(\frac{1-z_b}{z_b}\bigg)^2\bigg)
    +2\ln\bigg(\frac{-t}{M^2}\bigg(\frac{1-z_b}{z_b}\bigg)^2\bigg)
    \nonumber \\ 
    &+& \frac{2z_b}{1-z_b}\ln\bigg( 1+\frac{u}{t+u}
    \frac{1-z_b}{y_I z_b}\bigg) 
    + \frac{2z_b}{1-z_b}\ln\bigg( 1+\frac{t}{t+u}
    \frac{1-z_b}{y_I z_b}\bigg)\nonumber \\
    &-&  4(1+z_b^2)\frac{1-z_b}{z_b}
    \ln\bigg( \frac{-(t+u)}{M^2}\frac{1-z_b}{z_b}y_I\bigg)
  \bigg] \bigg( -\frac{N_C}{4}\bigg) T_\g (s,t,u) .
\end{eqnarray}

\section{Initial State Corrections for the Virtual Photon}

A virtual photon can decay into a $q\bar{q}$-pair. After the
integration over the collinear region of phase space the Born matrix
elements factorize. There are three types of Born matrix elements,
namely the processes $qq'\to qq'$ (denoted by $T_1$), 
$qq\to qq$ (denoted by $T_1$ and $T_2$) and $q\bar{q}\to gg$
(denoted by $T_3$). The divergence is regularized by the virtuality
of the photon $P^2$. The corrections depend on the two-body variables
$s, t$ and $u$, on the cut-off parameter $y_J$ and on the additional
variable of integration $z_a$.
\begin{eqnarray}
  I_1^a &=&  -2\al^2\al_sQ_i^4 M \bigg[ T_\g(t,s,u) +T_\g(u,s,t) 
             \bigg] \ , \\
  I_2^a &=&  2C_F\al\al_s^2Q_i^2 M T_1(s,t,u) \ , \\
  I_3^a &=&  -2\al\al_s^2Q_i^2 M \bigg[ T_1(t,s,u) +T_1(u,s,t) \bigg]
             \ , \\
  I_4^a &=&  4C_F\al\al_s^2Q_i^2M \bigg[ T_2(s,t,u) + T_2(t,s,u) \nonumber \\ 
  &+& {\scriptstyle \frac{1}{2}} (T_3(s,t,u) + \mbox{zycl.\
  permutations in $s,t,u$}) \bigg] \ , \\
  I_5^a &=&  2C_F\al\al_s^2Q_i^2 M \bigg[ T_3(s,t,u) + \mbox{zycl.\ 
        permutations in $s,t,u$} \bigg]  \ , 
\end{eqnarray}
where
\equ{}{ M = \frac{1}{2N_C} P_{q\leftarrow\g}(z_a) \ln\left(
  1 + \frac{y_Js}{z_aP^2} \right) \quad . } 
\end{appendix}


\end{document}